\documentclass[12pt]{article}
\pdfoutput=1
\usepackage{amsmath,amssymb,mathrsfs,amsbsy,latexsym,amsfonts,amsthm}
\usepackage[backref=page]{hyperref}%added backref
\hypersetup{
    colorlinks=true,
    linkcolor=blue,
    filecolor=magenta,      
    urlcolor=magenta,%cyan,
    citecolor=magenta,
    pdftitle={Overleaf Example},
    pdfpagemode=FullScreen,
    }
\usepackage[utf8]{inputenc}
\usepackage{authblk}
\usepackage{color}
\usepackage{ascmac}
\usepackage{here}
\usepackage{bm}
\usepackage{comment} 
\usepackage[top=30truemm,bottom=30truemm,left=27truemm,right=27truemm]{geometry}
\usepackage{wrapfig}
\usepackage{graphicx}
\usepackage{caption}
\usepackage{enumerate}
\usepackage[italicdiff]{physics}
\usepackage{fancybox}
\usepackage{slashed}
\usepackage{tensor}
\usepackage{ulem} %Comment out for diff file

\newcommand{\be}{\begin{equation}}
\newcommand{\ee}{\end{equation}}

\newcommand{\bfig}{\begin{figure}[htb!] \centering}
\newcommand{\efig}{\end{figure}}
\newcommand{\ra}{\rightarrow}

\newcommand{\trc}[1]{\underset{#1}{\operatorname{Tr}}}

\newcommand{\rmas}{\mathrm{as}}
\newcommand{\rmsoft}{\mathrm{soft}}
\newcommand{\rmhard}{\mathrm{hard}}
\newcommand{\rmin}{\mathrm{in}}
\newcommand{\rmout}{\mathrm{out}}
\newcommand{\wC}{\widetilde{C}}
\newcommand{\wD}{\widetilde{D}}

\newcommand{\bN}{\mathbf{N}}
\newcommand{\bC}{\mathbf{C}}
\newcommand{\bD}{\mathbf{D}}
\newcommand{\bS}{\mathbf{S}}
\newcommand{\bGamma}{\mathbf{\Gamma}}

\newcommand{\ins}[2]{(#1, #2)_\mathrm{s}}

\DeclareMathOperator*{\SumInt}{%
\mathchoice%
  {\ooalign{$\displaystyle\sum$\cr\hidewidth$\displaystyle\int$\hidewidth\cr}}
  {\ooalign{\raisebox{.14\height}{\scalebox{.7}{$\textstyle\sum$}}\cr\hidewidth$\textstyle\int$\hidewidth\cr}}
  {\ooalign{\raisebox{.2\height}{\scalebox{.6}{$\scriptstyle\sum$}}\cr$\scriptstyle\int$\cr}}
  {\ooalign{\raisebox{.2\height}{\scalebox{.6}{$\scriptstyle\sum$}}\cr$\scriptstyle\int$\cr}}
}

\renewcommand*{\backref}[1]{}
\renewcommand*{\backrefalt}[4]{%
  \ifcase #1 %
    \relax % use \relax if you do not want the "No citations" message
  \or
    (page~#2).%
  \else
    (pages~#2).%
  \fi%
}
\newcommand{\nn}{\nonumber\\}

\newcommand{\intx}{\int \!\! d^3 x}

\newcommand{\intpw}{\int \widetilde{d^3p}\,}
\newcommand{\intqw}{\int \widetilde{d^3q}\,}
\newcommand{\intp}{\int \!\! \frac{d^3 p}{(2\pi)^3(2 E_p)}}

\newcommand{\vx}{\vec{x}}

\newcommand{\vk}{\vec{k}}

\newcommand{\vp}{\vec{p}}
\newcommand{\vq}{\vec{q}}
\newcommand{\vpp}{\vec{p'}}

\numberwithin{equation}{section}

\begin{document}
%%%%%%%%%%%%%%%%%%%% TITLE %%%%%%%%%%%%%%%%%%%%
\title{
Dress code for infrared safe scattering in QED}
\author[1]{
	Hayato~Hirai\thanks{\tt hirai(at)n.kisarazu.ac.jp }
}
\author[2, 3]{
	Sotaro~Sugishita\thanks{
	\tt sotaro.sugishita(at)yukawa.kyoto-u.ac.jp}
	\vspace{5mm}
}
\affil[1]{\it\normalsize Natural Science Education, National Institute of Technology, Kisarazu College, 2-11-1 Kiyomidai-Higashi, Kisarazu,
Chiba 292-0041, Japan}
\affil[2]{\it\normalsize Institute for Advanced Research, 
Nagoya University,
Nagoya, Aichi 464-8601, Japan}
\affil[3]{\it\normalsize Department of Physics, 
Nagoya University,
Nagoya, Aichi 464-8602, Japan}
\setcounter{Maxaffil}{0}

\date{ }

\maketitle

\begin{abstract}
We study the $S$-matrix and inclusive cross-section for general dressed states in quantum electrodynamics. 
We obtain an infrared factorization formula of the $S$-matrix elements for general dressed states. 
It enables us to study what dressed states lead to infrared-safe $S$-matrix elements.
The condition for dressed states can be interpreted as the memory effect which is nothing but the conservation law of the asymptotic symmetry.
We derive the generalized soft photon theorem for general dressed states.
We also compute inclusive cross-section using general dressed states. 
It is necessary to use appropriate initial and final dressed states to evaluate interference effects, which cannot be computed correctly by using Fock states due to the infrared divergence.
\end{abstract}

\newpage
\setcounter{tocdepth}{2}
\tableofcontents

\newpage

\section{Introduction and Summary}
\paragraph{Infrared divergence and the problem}\ \\
In quantum electrodynamics (QED), infrared (IR) divergences appear in the computation of the conventional $S$-matrix elements between Fock states due to the long-range nature of QED interaction.
The IR divergences resummed to all orders of perturbative series set all the $S$-matrix elements for any nontrivial  processes to zero  \cite{Weinberg:1965nx, Carney:2017jut}. It can be understood as a consequence of asymptotic symmetry in QED \cite{He:2014cra, Campiglia:2015qka, Kapec:2015ena}, as pointed out in \cite{Kapec:2017tkm}.
The  traditional treatment of IR divergences is to consider the inclusive cross-sections by summing over all the effects of  soft (undetectably low energy) photon emissions  \cite{Bloch:1937pw, Yennie:1961ad, Weinberg:1965nx}.
Although the $S$-matrix is still ill-defined in the above sense, the inclusive cross-section can give the IR finite result.
However the inclusive cross-sections for Fock states cannot predict any non-trivial scatterings when the initial or final state is a continuous wave packet because the cross-sections are set to zero due to the {\it zero-measure problem} caused by infrared divergence, which was pointed out by \cite{Carney:2018ygh}.
Even if the initial state or final state is the superposed state of a discrete set of different momentum eigenstates as $\ket{f^\rmin}=\sum_{\alpha}f_{\alpha}\ket{\alpha}_0,\, \ket{g^\rmout}=\sum_{\beta}g_{\beta}\ket{\beta}_0$, all the interference terms are set to zero by the resummed infrared divergence in the inclusive cross-section of the process $\ket{f^\rmin}\ra \ket{g^\rmout}$.
The situation can be schematically written as
\begin{align}
\begin{split}\label{eq:int_deco}
    \sigma^{\mathrm{inc}}(f^\rmin\ra g^\rmout)
    &=\trc{\gamma: \text{soft}}\left(\left|\bra{g^\rmout\otimes \gamma}S\ket{f^\rmin}\right|^2\right)\\[1em]%\nn[0.5em]
    &=\sum_{\alpha,\alpha'}\sum_{\beta,\beta'}g_{\beta}g_{\beta'}^{\ast}f_\alpha f^\ast_{\alpha'}\,\trc{\gamma:\text{soft}} \left(S^{}_{\beta+\gamma, \alpha}S^{\dagger}_{\alpha',\beta'+\gamma}\right)
    \xrightarrow{\text{IR divergence}}\ \sum_{\alpha,\beta}|g_\beta|^2|f_\alpha|^2|S_{\beta,\alpha}|^2\,,%\nonumber
    \end{split}
\end{align}
where all the interference terms including $f_\alpha f^\ast_{\alpha'}$ or $g_\beta g^\ast_{\beta'}$ with $\alpha\neq\alpha', \beta\neq\beta'$ vanish in the final result.
The decoherence of the initial interference terms including $f_\alpha f^\ast_{\alpha'}$ was  first argued also in \cite{Carney:2018ygh} (see also \cite{Gomez:2017rau, Gomez:2018war}).
The vanishing of the interference terms means the non-cancellation of infrared divergence at each order of the perturbative series.
Then the standard computation of inclusive cross-sections works only when both initial and final states are momentum eigenstates if the interference is observed in reality.
This decoherence can be understood as a consequence of the superselection rule associated with the asymptotic symmetry in QED.
All the problems originating from infrared divergences suggest that using Fock states as asymptotic states is invalid in the case of long-range force. 

\paragraph{Need for reconsideration of dressed states}\ \\
Dressed states were proposed as the appropriate asymptotic states instead of the Fock states \cite{Chung:1965zza, Greco:1967zza, Kibble:1968sfb, Kibble:1969ip, Kibble:1969ep, Kibble:1969kd, Kulish:1970ut, Bagan:1999jf}.\footnote{The dressed states have been reconsidered with the connection to the asymptotic symmetry recently \cite{Mirbabayi:2016axw, Gabai:2016kuf, Carney:2017oxp, Kapec:2017tkm, Dybalski:2017mip, Gomez:2017rau, Gomez:2018war, Carney:2018ygh, Hirai:2019gio, Duch:2019wpf,  Choi:2019rlz, Tomaras:2019sjq, Hirai:2020kzx, Furugori:2020vdl, Arkani-Hamed:2020gyp, Agarwal:2021ais, Rai:2021vvq, Irakleous:2021ggq}.}
%All the above problems can be resolved by using proper {\it dressed} states as asymptotic states in QED.
The dressed states are the coherent states in which charged particles are surrounded by the low-energy (soft) photon fields.  
In particular, in \cite{Kulish:1970ut}, a class of dressed states is obtained by solving the asymptotic dynamics. We call the dressed states \textit{the Kulish-Faddeev dressed states}.

However, in the previous paper \cite{Hirai:2020kzx} by the present authors, it was pointed out that the Kulish-Faddeev dressed states are not enough for the cancellation of IR divergences, and some additional dresses are needed. 
The asymptotic states schematically take the form
\begin{align}
\ket{\psi}=e^{R_{KF}}\ket{\psi}_{GB}\,,
\end{align}
where $R_{KF}$ is the Kulish-Faddeev dress (and we omit the phase term for simplicity). In the original Kulish-Faddeev paper \cite{Kulish:1970ut}, $\ket{\psi}_{GB}$ are supposed to be the Fock states satisfying the free Gupta-Bleuler condition. 
In the previous paper \cite{Hirai:2020kzx}, we showed in a simplified setup that $\ket{\psi}_{GB}$ also have to be coherent states. 
This motivates us to reconsider what additional dressed states $\ket{\psi}_{GB}$ we need for the IR-safe $S$-matrix elements in QED.
To study the IR-safe condition (\textit{dress code}) of dressed states,\footnote{A choice of dresses  satisfying the dress code corresponds to those considered in the previous paper \cite{Hirai:2020kzx}. We will see that more general dresses are allowed in our dress code.} we will consider the following general initial and final dressed states $\ket{\alpha}_{GB}, \ket{\beta}_{GB}$ corresponding to Fock  momentum eigenstates $\ket{\alpha}_0, \ket{\beta}_0$,
\begin{align}\label{eq:int_drstate}
   \ket{\alpha}_{GB}=e^{C_{\alpha}+i \Theta^\rmin_\alpha}\ket{\alpha}_{0}\ ,\quad
   \ket{\beta}_{GB}=e^{D_{\beta}+i \Theta^\rmout_\beta}\ket{\beta}_{0}\,,
   %\qquad \left(\ \ket{\alpha}_{0}\in \mathcal{H}_{\mathrm{Fock}}\ \right)
\end{align}
where $\Theta^\rmin_\alpha, \Theta^\rmout_\beta$ represent arbitrary phases which are real functions depending on $\alpha, \beta$ respectively\footnote{Since $\Theta^\rmin_\alpha, \Theta^\rmout_\beta$ are the pure phase of Fock momentum basis, they are not relevant to any observables and we can set them as arbitrary. Then we can use them so that they cancel the IR divergence in the pure phase of the $S$-matrix and inclusive cross-section.}; $C_\alpha, D_\beta$ are arbitrary anti-Hermitian dressing operators of transverse soft photons given by 
\begin{align}
\begin{split}\label{eq:int_drfunc}
   C_{\alpha}&=
    \int^{\Lambda_s}_{\lambda}\widetilde{d^3k} \left[C^A_\alpha(\vk)  a_{A}(\vk)
    -C^{A\ast}_\alpha(\vk) a_{A}^{\dagger}(\vk)
    \right],\\[1em]
    D_{\beta}&=
    \int^{\Lambda_s}_{\lambda}\widetilde{d^3k} \left[D^A_\beta(\vk)  a_{A}(\vk)
    -D^{A\ast}_\beta(\vk) a_{A}^{\dagger}(\vk)
    \right]\,,
    \end{split}
\end{align}
where $a_{A}(\vk), a^{\dagger}_{A}(\vk)\ (A=1,2)$ are the annihilation and creation operators of transverse photons of polarization $\epsilon^A(\vk)$, and $C_{\alpha}^{A}(\vk), D_{\beta}^{A}(\vk)$ are arbitrary functions of $\vk$ which can depend on hard information $\alpha$ and $\beta$, respectively.
In \eqref{eq:int_drfunc}, the integration ranges of $|\vk|$ are from an infrared cutoff $\lambda$ to an energy resolution of a detector  $\Lambda_s$ with the Lorentz invariant measure $\widetilde{d^3k}:=\frac{d^3 k}{(2\pi)^3(2 \omega)}$.

\paragraph{Main results}\ \\
The first goal of this paper (see Section~\ref{sec:S-dress}) is to study the IR structure of the $S$-matrix elements for general superposed dressed states $\ket{f^\rmin}=\SumInt_{\alpha} f_{\alpha}\ket{\alpha}_{GB}, \ket{g^\rmout}=\SumInt_{\beta} g_{\beta}\ket{\beta}_{GB}$ as initial and final states, which are given by continuous or discrete superposed states of \eqref{eq:int_drstate}. 
We will compute the $S$-matrix elements $S_{g,f}$ by taking care of infrared divergence at all orders of perturbative series.  
This analysis provides us with the \textit{dress codes} on the soft photon dresses that lead to IR safe $S$-matrix elements.

The next goal of this paper (see Section~\ref{sec:reduced}) is to investigate the inclusive cross-sections for the asymptotic states including the general additional dresses.  
We compute the inclusive cross-sections 
$\sigma^{\text{inc}}(f^\rmin\ra g^\rmout)$ for the general superposed dressed states $\ket{f^\rmin}=\SumInt_{\alpha} f_{\alpha}\ket{\alpha}_{GB}, \ket{g^\rmout}=\SumInt_{\beta} g_{\beta}\ket{\beta}_{GB}$.
In particular, we find that the following dress, which we call the {\it generalized Chung dress}, can remove all infrared divergences:
\begin{align}\label{int_gen_Chung}
    \mathbf{C}^A_\alpha (\vk)= -\sum_{n \in \alpha} e_n\frac{p_n \cdot \epsilon^A(\vk)}{p_n\cdot k}+\wC^A_\alpha(\vk)\ ,
    \quad  \mathbf{D}^A_\beta(\vk)= -\sum_{n \in \beta} e_n\frac{p_n \cdot \epsilon^A(\vk)}{p_n\cdot k}+\wD^A_\beta(\vk)
\end{align}
where $\wC^A_{\alpha}(\vk), \wD^A_\beta(\vk)$ are subleading dresses of the order $o(k^{-1})$.\footnote{
In our later analysis, the generalized dresses $\bC^{A}_{\alpha}, \bD^{A}_{\beta}$ can contain additional common terms $C^{A}, D^{A}$, which are arbitrary dresses independent of $\alpha, \beta$, as seen in \eqref{gen_Chung}. However it  turns out that in the final results the common terms appear in only phases which can be absorbed into phases $\Theta^\rmin_\alpha, \Theta^\rmout_\beta$, and we can omit them.}
For the generalized Chung dress \eqref{int_gen_Chung}, we find that the inclusive cross-section, $ \sigma^{\mathrm{inc}}(f^{\text{in}}\rightarrow g^{\text{out}})$, is given by
\begin{align}\label{eq:int_cross_result}
    \sigma^{\mathrm{inc}}(f^{\text{in}}\rightarrow g^{\text{out}})%\nn[0.5em]
&=\SumInt_{\alpha,\alpha'}\SumInt_{\beta,\beta'}g_{\beta}g_{\beta'}^{\ast}f_\alpha f^\ast_{\alpha'}\, S^{\mathrm{hard}}_{\beta, \alpha}S^{\mathrm{hard}\dagger}_{\alpha',\beta'}
    \bS^{\beta,\beta'}_{\alpha,\alpha'}\,,
\end{align}
where $S^{\mathrm{hard}}_{\beta, \alpha}$ is the IR-safe $S$-matrix for $\ket{\alpha}_0\ra \ket{\beta}_0$ with physical effective infrared cutoff $\Lambda_s$, and $\bS^{\beta,\beta'}_{\alpha,\alpha'}$ is the IR-safe correction from the subleading dresses $\wC^A_{\alpha}, \wD^A_\beta$.
For $\wC^A_{\alpha}=\wD^A_\beta=0$, the correction can be neglected, i.e.  $\bS^{\beta,\beta'}_{\alpha,\alpha'}=1$.
In \eqref{eq:int_cross_result}, all infrared divergences are removed at every term at any orders of perturbative expansion, and then all the interference effects can occur unlike the undressed case that predicts the decoherence as \eqref{eq:int_deco}.

Although the conclusion that the Chung dressed states avoid the zero-measure problem is discussed in detail in \cite{Carney:2018ygh} when the initial states are wave-packets, it does not lead to the resolution of the problem if the final states are wave packets. In contrast, our new treatment  resolves the problem for both cases.
The crucial difference is the definition of the trace over the soft sector used in the computation of the inclusive density matrix. 
We have to  carefully  define the partial trace when we compute the inclusive cross-sections by tracing out the final soft photons based on the superselection structure of the Hilbert space.
We will compare the results by our proposed definition and the naive definition, and see that our definition leads to results avoiding any zero-measure problem. 
Thus, our treatment is the first way to compute the inclusive cross-sections without any zero-measure problem in QED.

\paragraph{Organization of this paper}\ \\
In section~\ref{sec:usenonFock}, we discuss the reason why we should use non-Fock states  in relation to the incompleteness of Fock space and the asymptotic symmetry.
In section~\ref{sec:ASphys},  we study the interaction, the time evolution, and the physical states in the asymptotic regime at $t\rightarrow \pm\infty$ to evaluate IR parts of $S$-matrix in QED.
In section~\ref{sec:S-dress}, we study  $S$-matrix elements for general dressed states.
We show an infrared factorization formula for the $S$-matrix elements, which is the generalization of infrared factorization of the conventional undressed $S$-matrix.
The formula gives the condition for dresses to lead to IR-safe $S$-matrix elements.
We find that the generalized Chung dress  leads to IR-safe $S$-matrix elements.
We also derive the soft photon theorem for general dressed states.
In section~\ref{sec:reduced}, we compute the inclusive cross-section for generalized dressed states. 
To compute it, we define the partial trace over the soft sector by taking care of the superselection structure of the Hilbert space.
We then derive the main results presented above.
In section~\ref{sec:discuss}, we end by discussing some future problems.
Various technical calculations are confined in appendices.
%

%%%%%%%%%%%%%%%%%%%%%%%%%%
%%%%%%%%%%%%%%%%%%%%%%%%%%
%%%%%%%%%%%%%%%%%%%%%%%%%%
\section{Why are Fock states not enough?}\label{sec:usenonFock}
In this section, we discuss why we should use non-Fock states in relation to the incompleteness of Fock space and the asymptotic symmetry.
In subsec.~\ref{sec:incomplete} we briefly review the incompleteness of Fock space.
In sebsec.~\ref{sec:sel_assym} we briefly review the several consequences of asymptotic symmetry, and argue that all states in any physical superposition should have the same asymptotic charges.

%%%%%%%%%%%%%%%%%%%%%%%%%%%%%
\subsection{Incompleteness of the finite Fock space}\label{sec:incomplete}
In QED, the $S$-matrix elements between Fock states with an IR cutoff $\lambda$ vanish in the limit removing the cutoff $\lambda$
as
\begin{align}
    \lim_{\lambda\to 0} \tensor[_0]{\bra{\beta}}{} S(\lambda) \ket{\alpha}_0=0
    \label{eq:zeroS}
\end{align}
except for the trivial  process $\alpha=\beta$ \cite{Weinberg:1995mt}.
This seems that non-trivial scattering does not occur in QED. It contradicts our expectations or experimental results. This is the IR problem in QED.

However, it is too early to conclude that there is no non-trivial scattering, because we have restricted the initial and final states to Fock states. 
If we allow more general states as the asymptotic states, we might obtain nonzero  $S$-matrix elements. 
In fact, the Fock space is incomplete in the sense that there exist states orthogonal to any finite Fock states. 
Here the \textit{finite} Fock states mean that the numbers of Fock particles are finite.
We denote the Fock space consisting of the finite Fock states by $\mathcal{H}_{\mathrm{Fock}}$.
To illustrate the incompleteness of $\mathcal{H}_{\mathrm{Fock}}$, let us consider the following normalized coherent state of transverse photons (where indices $A, B$ are used for the label of the transverse directions and the more detailed definition is given in appendix~\ref{app:transv})
\begin{align}
    \ket{\psi}=\exp\left(\int^\Lambda_\lambda \!\! \frac{d^3 k}{(2\pi)^3(2 \omega)} \frac{ e p^A}{p\cdot k} \left[a_{A}(\vk)
    -a_A^{\dagger}(\vk)
    \right]\right)\ket{\alpha}_0, \qquad (\ \omega:=|\vk|\ ) 
    \label{cohe_st}
\end{align}
where we have introduced cutoffs $\lambda, \Lambda$, and $\ket{\alpha}_0$ is an arbitrary Fock state with a finite number of photons. 
One can confirm that the coherent state $\ket{\psi}$ is orthogonal to any finite Fock states of photons $\ket{\beta}_0$ 
in the limit removing the IR cutoff $\lambda$ as
\begin{align}
\label{zeroOL}
   \lim_{\lambda \to 0} \tensor[_0]{\braket{\beta}{\psi}}{}=0.
\end{align}
In order to have nonzero overlaps with $\ket{\psi}$, we need to consider more general states instead of finite Fock states $\ket{\beta}_0$. 
The need for the non-Fock spaces is stressed by Kibble in \cite{Kibble:1968sfb, Kibble:1969ip, Kibble:1969ep, Kibble:1969kd}, and also emphasized in \cite{Gomez:2018war}.
Recently, it is also proposed in \cite{Prabhu:2022mcj} that it is better to use an algebraic approach to discuss IR safe scattering, especially in massless QED and gravity. 

Equation \eqref{zeroOL} is essentially the same as \eqref{eq:zeroS} because $S(\lambda) \ket{\alpha}_0$ in \eqref{eq:zeroS} is roughly the coherent state of the form \eqref{cohe_st} at least for the soft photon sector \cite{Hirai:2020kzx}. 
Thus, we have to prepare a final state instead of finite Fock states $\ket{\beta}_0$  in order to have nonzero $S$-matrix elements.
The candidates of such final states are  dressed states involving an infinite number of soft photons (in the limit removing the IR cutoff). 
We might think that the conventional inclusive formalism with Fock states \cite{Bloch:1937pw} is another way to avoid the above incompleteness problem 
because in the treatment we use states containing arbitrarily large numbers of soft photons.
However, we will show that infrared divergences in the interference effect cannot be removed in the conventional treatment.

%%%%%%%%%%%%%%%%%%%%%%%%%%%%
\subsection{IR divergences and asymptotic symmetry}\label{sec:sel_assym}

The necessity for non-Fock states can also be understood from the asymptotic symmetries in QED \cite{Kapec:2017tkm}.
Here we review the relation briefly and comment on the superselection rule associated with the asymptotic symmetry.

The conserved current associated with U(1) gauge symmetry with a gauge parameter $\epsilon(x)$  is given by
\begin{align}\label{eq:gaugeJ}
    J^{\mu}=F^{\mu\nu}\partial_{\nu}\epsilon+j^{\mu}\epsilon=\partial_{\nu}\left(F^{\mu\nu}\epsilon\right)\,,
\end{align} 
where  we have used the equation of motion $\partial_{\nu}F^{\mu\nu}=j^{\mu}$ in the second equality.
The corresponding charges are defined by integrating the conserved current over time slices.
In particular, the asymptotic charges defined on the future infinity  and past infinity  can be written as
\begin{align}\label{eq:ascharges}
    Q^{+}[\epsilon]=\int_{\mathcal{I}^{+}_{-}}\! d^{2}\Omega\,  \epsilon F_{ru}^{(2)}\quad,\quad
    Q^{-}[\epsilon]=\int_{\mathcal{I}^{-}_{+}}\! d^{2}\Omega\,  \epsilon F_{rv}^{(2)}\,,
\end{align}
where $F_{r\nu}^{(2)}=\lim_{r\rightarrow\infty}r^2F_{r\nu}$.
$\mathcal{I}^{+}_{-}$ and $\mathcal{I}^{-}_{+}$ denote the past of future null infinity and  the future of past null infinity, respectively \cite{He:2014cra, Kapec:2015ena,Strominger:2017zoo}.
For appropriate parameters $\epsilon(x)$ which do not decay in the asymptotic regions, the charges $Q^{\pm}[\epsilon]$ can take non-vanishing values, and generate asymptotic symmetries. 
The Ward-Takahashi identity for the asymptotic symmetry is equivalent to the soft photon theorem for Fock states \cite{He:2014cra,Campiglia:2015qka,Kapec:2015ena,Strominger:2017zoo}.
Importantly, the infrared divergence of the conventional $S$-matrix between Fock states can be understood as a consequence of the charge conservation  \cite{Kapec:2017tkm}.
More explicitly, the $S$-matrix elements between Fock states are forced to be zero by the conservation law of the asymptotic charges because Fock states with different momenta generally belong to different sectors of the charges. 
Therefore, to obtain non-zero $S$-matrix elements, we need to use some asymptotic states that are not Fock states and obey the conservation law of asymptotic symmetries.

\paragraph{Superselection rule of asymptotic symmetry}\ \\
Here we stress that the Hilbert space is naively divided into superselection sectors labeled by the asymptotic charges, as usually done by the total electric charge, which corresponds to a special case of asymptotic charge with $\epsilon=\text{const}$.
Since the conserved current \eqref{eq:gaugeJ} is the total derivative, the asymptotic charges can be defined on a sphere with infinite radius like \eqref{eq:ascharges}.
Such charges commute with any quasi-local observables and then divide the Hilbert space into superselection sectors labeled by the charges \cite{Strocchi:1974xh,Gervais:1980bz,Buchholz:1982ea}.
In QED, the Hilbert spaces of asymptotic states defined on the future infinity and past infinity are divided into the superselection sectors labeled by the eigenvalues of $Q^{+}[\epsilon]$ and $Q^{-}[\epsilon]$, respectively.
It follows in general by the superselection rule that states belonging to different superselection sectors cannot interfere with each other  because any quasi-local observables preserve the charge, and any states with different charges are orthogonal to each other. 
It is also important that we cannot prepare the superposition of states belonging to different superselection sectors  because any physical operations and interactions are quasi-local, which cannot change the asymptotic charge of the states.\footnote{If we formally consider such a superposition, the state cannot be distinguished from a classical ensemble of each superposed state because any interference does not occur.}
Therefore it would be reasonable to assume that 
\begin{itemize}
    \item[($\ast$)] all states contained in a physical superposition have the same asymptotic charges.
\end{itemize}
This assumption also requires dressed states because dressing of soft photons is necessary so that states with different momenta have the same asymptotic charges.

However, there is a subtle issue on the superselection rule associated with the asymptotic symmetry when we consider the inclusive quantities.
In fact, the results of the conventional inclusive computations with undressed Fock states seem to contradict the superselection rule in the following sense.  Since each exclusive transition amplitude between Fock states with a finite number of soft photons in the inclusive sum should be zero by the superselection rule, we naively expect the total inclusive amplitude also vanishes.
However the conventional inclusive computation leads to nonzero results.
This is due to the IR cutoff introduced in the conventional computations which breaks the asymptotic symmetry. 
Taking the sum over soft photons before removing the IR cutoff, we can obtain nonzero transition amplitude. 
This example indicates that we have to be careful about the order of the limit removing the cutoff and the sum over an infinite number of photons and that it may be dangerous to trust the superselection rule of the asymptotic symmetry in the inclusive computations. 
Despite this subtlety, we will explicitly show in section \ref{sec:reduced} that we cannot observe the interference among different sectors with respect to the asymptotic charges.
We will also show that we need appropriate dressed states even in the inclusive computation to see possible interference effects among superposed states in the same superselection sector.

%%%%%%%%%%%%%%%%%%%%%%%%%%
%%%%%%%%%%%%%%%%%%%%%%%%%%
%%%%%%%%%%%%%%%%%%%%%%%%%%
\section{Asymptotic physics in QED}
\label{sec:ASphys}
In this section we study the interaction, the time evolution, and the physical states in the asymptotic regime at $t\rightarrow \pm\infty$ to evaluate IR parts of $S$-matrix in QED.
Our analysis of the asymptotic dynamics is similar to the original Kulish-Faddeev one in \cite{Kulish:1970ut}. 
We also use the result of the physical state condition in \cite{Hirai:2019gio}.
Combining them, although we consider the asymptotic interaction, 
we confirm that $S$-matrix with large time approximation is given by the standard $S$-matrix with the Gupta-Bleuler condition, which is used in section~\ref{sec:S-dress}.
Readers who are not interested in the confirmation can skip this section.

In the following, we work in the interaction picture.
For brevity we omit the label $I$ denoting the picture.
The $S$-matrix elements for finite times are defined by
\begin{align}\label{eq:def_Smat}
    S_{\beta,\alpha}(t_f,t_i)
    =\tensor[]{\bra{\beta(t_f)}}{}U(t_f,t_i)\ket{\alpha(t_i)}\,,
\end{align}
where $U(t,t')$ is the time evolution operator in the interaction picture defined by
\begin{align}\label{eq:timeev}
   U(t,t')
    =\mathrm{T}\exp\left( -i\int_{t'}^{t}ds V(s)\right)\,.
\end{align}
The $S$-matrix \eqref{eq:def_Smat} is the transition amplitude from an incoming state prepared at $t=t_i$ to an outgoing state observed at $t=t_f$.
In QED,  the interaction operator $V(t)$ is given by
\begin{align}
    V(t)
    =-\int d^{3}x\, A_{\mu}(t,\vx)j^\mu(t,\vx)
    %=\intk A_{\mu}(t,-\vk)j^\mu(t,\vk)\,,
\end{align}
where $A_{\mu}(x)$ is the $U(1)$ gauge filed and $j^{\mu}(x)=i\bar{\psi}(x)\gamma^{\mu}\psi(x)$ is the current operator of massive\footnote{In this paper, we do not consider massless charged particles which cause collinear singularities and other technical difficulties.} Dirac fields associated with the global U(1) symmetry.
Throughout this paper, we use the Feynman gauge (see \cite{Hirai:2019gio} for details).

\subsection{Asymptotic interaction and time-evolution}\label{subsec:asympt-int}
In the large time limit $t\rightarrow \pm \infty$
the current operator $j^{\mu}(x)$ can be evaluated as (see appendix~\ref{sec:ascurrent} for detail, and also \cite{Hirai:2019gio})
\begin{align}\label{eq:asexcurt}
    j^{\mu}(x)= j^{\mu}_{\rmas}(x)+\text{subleading}\quad \text{as}\quad t\rightarrow \pm \infty
\end{align}
 where $j^\mu_{\rmas}(t,\vx)$ is the ``asymptotic current'' defined by
 \begin{align}
&j^\mu_{\rmas}(t,\vx)=  \intpw\,\rho(\vp)\, \frac{p^\mu}{E_p}\, \delta^{(3)}_{\rmsoft}(\vx -\vp t/E_p) \,,
 \label{eq:pp_current}
 \end{align}
which schematically represents a soft part of the current for classical point particles moving with momenta $p^\mu$.
 In the above equation, $\widetilde{d^3p}:=\frac{d^3p}{(2\pi)^32E_p}$ is the Lorentz invariant measure and $\rho(\vp)$ is the charge density operator defined by
 \begin{align}
&\rho(\vp)=\sum_{n=\text{species}} e_n[b^\dagger_n(\vp)b_n(\vp)-d^{\dagger}_n(\vp)d_n(\vp)]\,,
\end{align}
where $b_n, b^\dagger_n$ (and $d_n, d^{\dagger}_n$) are the annihilation and creation operators of a particle with the charge $e_n$ (and its antiparticle) for each species labeled by $n$.
The $\delta^{(3)}_{\rmsoft}(\vx)$ is defined as
\begin{align}
    \delta^{(3)}_{\rmsoft}(\vx):=\int_{\vec{k}\sim \vec{0}}\frac{d^3k}{(2\pi)^3}e^{i\vk\cdot\vx}\,,
\end{align}
where  ``$\vec{k}\sim\vec{0}$'' denotes the  infrared  region around $\vec{k}=\vec{0}$.\footnote{
In this paper, we do not precisely determine the region ``$\vec{k}\sim\vec{0}$''   because the result of $S$-matrix will turn out to be independent of the detail of the region at least at the leading order of large time approximation.
}
Here we introduce a time scale $t_{\rmas}$ such that  $j^{\mu}_{\rmas}(x)$ becomes dominant in the expansion \eqref{eq:asexcurt} for $|t|\gtrsim t_{\rmas}$ and the following approximation makes sense in the $S$-matrix calculation:
\begin{align}\label{eq:asappcur}
   j^{\mu}(x)\simeq j^{\mu}_{\rmas}(x)\quad\text{for}\quad |t|\gtrsim t_{\rmas}\,.
\end{align}
  We do not make a concrete evaluation of  $t_{\rmas}$ in this paper. 
  This approximation leads to 
  \begin{align}\label{eq:asintapp}
      V(t)\simeq V_{\rmas}(t)\quad \text{for}\quad |t|\gtrsim t_{\rmas}\,,
  \end{align}
  where $V_{\rmas}(t)$ is the ``asymptotic interaction'' defined by
  \begin{align}\label{eq:Vas}
      V_{\rmas}(t)
    =-\int d^{3}x A_{\mu}(t,\vx)j^\mu_{\rmas}(t,\vx)\,.
  \end{align}
  The above asymptotic current and interaction are the same as those used in \cite{Kulish:1970ut}. 
  
The time-evolution at $|t|\gtrsim t_{\rmas}$ can be simplified by using the approximation \eqref{eq:asintapp} as follows, as obtained in \cite{Kulish:1970ut}.
We divide the time evolution operator \eqref{eq:timeev}  as
\begin{align}
  U(T_f,T_i)
  =U(T_{f},t_f)U(t_{f},t_{i})U(t_{i},T_i)
\end{align}
where $T_f>t_f\gg t_{\rmas}$ 
and $-t_{\rmas}\gg t_i>T_i $.
Using the  approximation  \eqref{eq:asintapp}, we obtain
\begin{align}
\begin{split}
\label{eq:def_Uas}
&U(t_{i}, T_i)
    \simeq \mathrm{T}\exp\left( -i\int^{t_{i}}_{T_i}dt V_{\rmas}(t)\right)
    =:U_{\rmas}(t_{i},T_i)\,,\\
    &U(T_f,t_{f})
    \simeq \mathrm{T}\exp\left( -i\int_{t_{f}}^{T_f}dt V_{\rmas}(t)\right)
    =:U_{\rmas}(T_f,t_{f})\,,
    \end{split}
\end{align}
where $U_{\rmas}(t,t')$ is the ``asymptotic'' time evolution operator \cite{Dollard1964,Kulish:1970ut}.
It is known in \cite{Kulish:1970ut} that $U_{\rmas}(t,t')$ takes the following form 
\begin{align}\label{eq:Uas}
U_{\rmas}(t,t')=e^{i\Phi_{KF}\left(t,t'\right)}e^{R_{KF}(t)-R_{KF}(t')},
\end{align}
where\footnote{\quad $\widetilde{d^3k}:=\frac{d^3k}{(2\pi)^32\omega_k}$.}
\begin{align}
\label{def:RKF}
    R_{KF}(t)&:=\intpw\rho(p) \int_{\vec{k}\sim\vec{0}}\widetilde{d^3k}\, \frac{  p^\mu}{p\cdot k} \left[a_{\mu}(\vk)e^{i\frac{p \cdot k}{E_p}t}
    -a_\mu^{\dagger}(\vk)
    e^{-i\frac{p \cdot k}{E_p}t}
    \right]\,,\\[1em]
\label{eq:defphase}
    \Phi_{KF}(t,t')&:=\frac{i}{2}\int^{t}_{t'}dt_1 \int^{t_1}_{t'} dt_2 [V_{\rmas}(t_1),V_{\rmas}(t_2)]\,.
\end{align}
This is the dressing operator introduced by Kulish and Faddeev in \cite{Kulish:1970ut}.
Note that $e^{R_{KF}}$ is a unitary operator, and $e^{i\Phi_{KF}(t_f,t_i)}$ is an oscillating factor with phase $\Phi_{KF}(t_f,t_i)$ because $R_{KF}$ is an anti-Hermitian operator and  $i[V_{\rmas}(t_1),V_{\rmas}(t_2)]$ is Hermitian (e.g. see \cite{Kulish:1970ut, Hirai:2019gio, Hirai:2020kzx}).

Here we apply Feynman's $i\epsilon$ prescription with an infinitesimal positive real constant $\epsilon$, as $p\cdot k\rightarrow p\cdot k-i\eta(t)\epsilon$ where $\eta(t):=\text{sgn}(t)$.\footnote{
Feynman's $i\epsilon$ prescription demands that $p_n\cdot k\rightarrow p_n\cdot k-i\eta_n\epsilon$ where $\eta_n=+1$ if $p_n$ are outgoing momenta and $\eta_n=-1$ if $p_n$ are incoming momenta (e.g. see \cite{Weinberg:1995mt}). This deformation is equivalent to $p\cdot k\rightarrow p\cdot k-i\eta(t)\epsilon$ because $e^{R_{KF}(t)}$  acts on initial states for $t<0$ and finial states for $t>0$. 
}
The $i\epsilon$ prescription deforms \eqref{def:RKF} as
\begin{align}
 \label{def:RKFi}
    R_{KF}(t)&=\intpw\rho(p) \int_{\vec{k}\sim\vec{0}}\widetilde{d^3k} \left[\frac{  p^\mu}{p\cdot k+i\eta(t)\epsilon} a_{\mu}(\vk)e^{i\frac{p \cdot k+i\eta(t)\epsilon}{E_p}t}
    -
    (\mathrm{h.c.})
    \right].
\end{align}
Note that the phase is also deformed because the prefactor $(p\cdot k)^{-1}$ comes from the integration of the phase over time in \eqref{eq:def_Uas}.
This deformation keeps the unitarity of $e^{R_{KF}}$ and leads to the following asymptotic behavior as $t\to\pm\infty$:
\begin{align}\label{eq:RKF_van}
    \lim_{t\to \pm\infty}R_{KF}(t)=0\,.
\end{align}
The above  $i\epsilon$ of the phase can be traced back to the deformation of the asymptotic current given by \eqref{eq:pp_current} as\footnote{
This regularization to the current is the same as used in \cite{Arkani-Hamed:2020gyp}.
}
\begin{align}\label{eq:reg_jas}
    j^{\mu}_{\rmas}(t,\vx)
    =e^{-\epsilon |t|}\intpw\,\rho(\vp)\, \frac{p^\mu}{E_p}\, \delta^{(3)}_{\rmsoft}(\vx -\vp t/E_p) \,.
\end{align}
Using the above regularized asymptotic current \eqref{eq:reg_jas}, we can show (see appendix \ref{sec:ascurrent}) that the phase defined in \eqref{eq:defphase} is simplified as
\begin{align}\label{eq:phase_van}
  \lim_{T_i\to - \infty}\Phi_{KF}(t_i,T_i)=\Phi(t_i)\,,
  \qquad
  \lim_{T_f\to\infty}\Phi_{KF}(T_f,t_i)=-\Phi(t_f)\,,
\end{align}
where the explicit form of $\Phi(t)$ is given by \eqref{phiKF-soft} although it is not relevant for our analysis in this paper.
The equations, \eqref{eq:RKF_van} and \eqref{eq:phase_van}, mean that the KF dress does not play any role at the infinite time under the $i\epsilon$ prescription.
Taking the limits of $T_i\to-\infty, T_f\to \infty$ in \eqref{eq:def_Uas}, we then obtain the following properties:
\begin{align}
U_{\rmas}(t_i,-\infty)
=e^{R_{KF}(t_i)+i\Phi(t_i)}\,, \quad
&U_{\rmas}(\infty,t_f)
=e^{-R_{KF}(t_f)-i\Phi(t_f)}\,.
\label{eq:U_asy}
\end{align}
Thus, using the approximation \eqref{eq:def_Uas}, we have obtained the following approximation of the time-evolution operators for large time:
\begin{align}
\begin{split}
&\lim_{T_i\rightarrow-\infty}U(t_i,T_i)
\simeq \lim_{T_i\rightarrow-\infty}U_{\rmas}(t_i,T_i)
=e^{R_{KF}(t_i)+i\Phi(t_i)}\,,\\
&\lim_{T_f\rightarrow\infty}U(T_f,t_f)
\simeq \lim_{T_f\rightarrow\infty}U_{\rmas}(T_f,t_f)
=e^{-R_{KF}(t_f)-i\Phi(t_f)}\,.
\end{split}
\label{eq:U_large}
\end{align}

\subsection{Asymptotic states and $S$-matrix}
In gauge theories, we have to impose the physical state condition.
The condition in the interaction picture with the Feynman gauge is given by
\begin{align}\label{eq:BRSTcd}
     \left[k^\mu a_\mu(\vk) +e^{i\omega t}j^{0}(t,\vk)\right] \ket{\psi(t)}=0\,,
\end{align}
where $j^\mu(t,\vk)=\intx e^{-i\vec{x}\cdot\vec{k}}j^{\mu}(t,\vec{x})$. 
This condition can be derived as the BRST closed condition (see \cite{Hirai:2019gio}).
For $|t|\gtrsim t_{\rmas}$, the condition \eqref{eq:BRSTcd} can be replaced by using \eqref{eq:asappcur} as
\begin{align}
\label{eq:asBRSRcd}
     \left[k^\mu a_\mu(\vk) +e^{i\omega t}j^{0}_{\rmas}(t,\vk)\right] \ket{\psi(t)}=0
\end{align}
with
\begin{align}
  &j^\mu_{\rmas}(t,\vk)
  =\intx e^{-i\vk\cdot\vx}j^\mu_{\rmas}(t,\vx)
  =\intpw \rho(\vp) \frac{p^\mu}{E_p}e^{-i\frac{\vk\cdot\vp}{E_p}t}\,\widetilde{\delta}^{(3)}_{\rmsoft}(\vk)\,,\label{eq:ascurt_momt}
   \end{align}
where $\widetilde{\delta}^{(3)}_{\rmsoft}(\vk)$ is a function supported only on a soft region around $\vec{k}\sim \vec{0}$ (see appendix \ref{sec:ascurrent} for more detail). 
We note that $\vec{k}\sim \vec{0}$ does not imply  $e^{-i\frac{\vk\cdot\vp}{E_p}t}\sim1$ because $|\vk t| \ll 1$ is not ensured in the large time region $|t| \gtrsim t_{\rmas}$. As explained in appendix~\ref{sec:ascurrent}, $\vec{k}\sim \vec{0}$ means $|\vk| \lesssim 1/t_{\rmas}$.

We write the set of states which satisfy the condition \eqref{eq:asBRSRcd} by $\mathcal{H}_{\rmas}$, and also define $\mathcal{H}_{GB}$ as
\begin{align}\label{eq:freeGB}
   \mathcal{H}_{GB}:=\{\,\ket{\psi}\,|\ k^\mu a_\mu(\vk)\ket{\psi}=0\} \,,
\end{align}
which is the set of states satisfying the conventional Gupta-Bleuler condition \cite{Gupta:1949rh, Bleuler:1950cy} for free Maxwell fields.
Then a class of solutions of \eqref{eq:asBRSRcd} is given by\footnote{
We may absorb the phase $e^{i\Phi(t)}$ to $\ket{\psi}_{GB}$ because $e^{i\Phi(t)}\ket{\psi}_{GB} \in \mathcal{H}_{GB}$.
In the following analysis, we will introduce an arbitrary phase of $\ket{\psi}_{GB}$, and will be convenient to factor the phase $e^{i\Phi(t)}$ as \eqref{eq:KFdressd} because we then have \eqref{I-GB_as}.} 
\begin{align}\label{eq:KFdressd}
    \ket{\psi(t)}=e^{R_{KF}(t)+i\Phi(t)}\ket{\psi}_{GB}\,,
\end{align}
where $\ket{\psi}_{GB}$ is any state in $\mathcal{H}_{GB}$ \cite{Hirai:2019gio}.\footnote{We are using the interaction picture, and $\ket{\psi}_{GB}$ is a state at time $t$ although we do not write the time-dependence explicitly.}
The key point is that $\mathcal{H}_{GB}$ is not restricted to the finite Fock space $\mathcal{H}_{\mathrm{Fock}}$.
That is, $\ket{\psi}_{GB}$ can be dressed states as long as they satisfy $ k^\mu a_\mu(\vk)\ket{\psi}_{GB}=0$.
We will see that this additional dressing is important to obtain IR-safe $S$-matrix elements as argued in \cite{Hirai:2020kzx}.
Using \eqref{eq:U_asy}, we find that the physical states \eqref{eq:KFdressd} can be written as
\begin{align}\label{I-GB_as}
    \ket{\psi(t)}=U_{\rmas}(t,-\infty)\ket{\psi}_{GB}\,.
\end{align}

We then consider $S$-matrix  elements \eqref{eq:def_Smat} between two physical states 
$\ket{\alpha(t_i)},\, \ket{\beta(t_f)} \in \mathcal{H}_{\rmas}$,
\begin{align}
\label{eq:S_ba}
   S_{\beta,\alpha}(t_f,t_i)
    =\bra{\beta(t_f)}U(t_f,t_i)\ket{\alpha(t_i)}. 
\end{align}
Since $t_i$ and $t_f$ are not infinite in real scattering experiments, it is better to suppose that $t_i$ and $t_f$ are large but finite. However, we will see that, when $t_i$ and $t_f$ are so large that we can use the approximation \eqref{eq:asappcur}, the $S$-matrix elements \eqref{eq:S_ba} almost agree with those for infinite time as follows.\footnote{For readers who are interested only in the infinite time $t_i=-\infty, t_f=\infty$, they can skip the following paragraph and $S$-matrix for the infinite time is given by \eqref{eq:conv_Smatrix} without approximation.}

Since the physical states are given by \eqref{eq:KFdressd},
the $S$-matrix elements \eqref{eq:S_ba} become 
\begin{align}
    S_{\beta,\alpha}(t_f,t_i)
    &=\tensor[_{GB}]{\bra{\beta}}{}e^{-R_{KF}(t_f)-i\Phi(t_f)}U(t_f,t_i)e^{R_{KF}(t_i)+i\Phi(t_i)}\ket{\alpha}_{GB}
    \label{eq:realSmat}
    \\[1em]
    &=\tensor[_{GB}]{\bra{\beta}}{}U_{\rmas}(\infty, t_f)U(t_f,t_i)
    U_{\rmas}(t_i,-\infty)\ket{\alpha}_{GB}
    \,,
\end{align}
where we have used \eqref{eq:U_asy}.
Applying the approximation \eqref{eq:U_large}, we then obtain
\begin{align}\label{eq:appSmatrix}
%\tensor[_I]{\bra{\beta(T_f)}}{}U(T_f,T_i)\ket{\alpha(T_i)}_{I}
%=&\tensor[_0]{\bra{\beta}}{}e^{-R_{KF}(T_f)}U(T_f,T_i)e^{R_{KF}(T_i)}\ket{\alpha}_{0}\\[1ex]
S_{\beta,\alpha}(t_f,t_i)\simeq &
\lim_{T_f\rightarrow\infty, T_i\rightarrow-\infty}\tensor[_{GB}]{\bra{\beta}}{}U(T_f,t_f)U(t_f,t_i)U(t_i,T_i)\ket{\alpha}_{GB}
 \\[1ex]
=&\tensor[_{GB}]{\bra{\beta}}{}U(\infty,-\infty)\ket{\alpha}_{GB}\,.
\label{eq:conv_Smatrix}
\end{align}
That is, the finite time $S$-matrix \eqref{eq:realSmat} with the KF dresses for sufficiently large but finite time $|t_i|, t_f \gtrsim t_\mathrm{as}$ agrees with the $S$-matrix \eqref{eq:conv_Smatrix} without the KF dresses for infinite time.
Note that the time evolution from $T_i$ to $t_i$ (and $t_f$ to $T_f$) in \eqref{eq:appSmatrix} comes from the KF dress as \eqref{eq:U_large}.
Thus, we do not need to take care of finite time effects as long as we ignore the subleading corrections to the approximation \eqref{eq:asappcur}. 
Because specifying the subleading corrections is beyond the scope of this paper,  we will henceforth omit $t_i, t_f$  and simply write \eqref{eq:conv_Smatrix} as 
\begin{align}\label{eq:Smatrix}
S_{\beta,\alpha}
=&\tensor[_{GB}]{\bra{\beta}}{}U(\infty,-\infty)\ket{\alpha}_{GB}\,.
\end{align}

%%%%%%%%%%%%%%%%%%%%%%%%%%%%%%%%%%%%%%%%%%%%%%%%%%%%%%%%%%%%%%%%%%%%%%%%%%%%%%%%%%%%%%%%%%%%%%%%
\section{$S$-matrix for dressed states}\label{sec:S-dress}
In this section, we study  $S$-matrix elements \eqref{eq:Smatrix} for general dressed states.
We show an infrared factorization formula for the $S$-matrix elements, which
gives the condition for dresses leading to IR-safe $S$-matrix elements.
We also derive the soft photon theorem for general dressed states.

%%%%%%%%%%%%%%%%
\subsection{General dressed states and the $S$-matrix elements}

We consider the $S$-matrix elements \eqref{eq:Smatrix}, i.e., 
\begin{align}
\label{dressS}
S_{\beta,\alpha}
=&\tensor[_{GB}]{\bra{\beta}}{}U(\infty,-\infty)\ket{\alpha}_{GB}\,
\end{align}
for general states $\ket{\alpha}_{GB}$ and $\ket{\beta}_{GB}$ satisfying the Gupta-Bleuler condition as \eqref{eq:freeGB}.
If we set $\ket{\alpha}_{GB}$ and $\ket{\beta}_{GB}$ as the standard Fock states without any dresses,  the  $S$-matrix elements \eqref{eq:conv_Smatrix} are the conventional $S$-matrix elements which suffer from the IR divergences.
However we can include any dresses in $\ket{\alpha}_{GB}$ and $\ket{\beta}_{GB}$ as long as they satisfy the condition \eqref{eq:freeGB} as argued in \cite{Hirai:2020kzx}.
In fact, we will see that such dresses must be added to Fock states to obtain the IR-safe results.\footnote{It was also discussed in \cite{Hirai:2020kzx} that the asymptotic symmetry requires that $\ket{\alpha}_{GB}$ and $\ket{\beta}_{GB}$ should not be restricted to the Fock states.}
%In \cite{Hirai:2020kzx}, it is shown that the states in $\mathcal{H}_{GB}$ which are consistent with the conservation law of asymptotic charge of QED are the states dressed by \textit{Chung's} dress.
Here we start with the following general initial and final dressed states  in $\mathcal{H}_{GB}$ as
\begin{align}\label{eq:dressdstates}
   \ket{\alpha}_{GB}=e^{C_{\alpha}+i \Theta^\rmin_\alpha}\ket{\alpha}_{0}\ ,\quad
   \ket{\beta}_{GB}=e^{D_{\beta}+i \Theta^\rmout_\beta}\ket{\beta}_{0}\,,
   %\qquad \left(\ \ket{\alpha}_{0}\in \mathcal{H}_{\mathrm{Fock}}\ \right)
\end{align}
where $\ket{\alpha}_{0}, \ket{\beta}_{0}$ are arbitrary Fock states in $\mathcal{H}_{\mathrm{Fock}}$ satisfying
\begin{align}
    k^\mu a_\mu(\vk)\ket{\alpha}_{0}
    =k^\mu a_\mu(\vk)\ket{\beta}_{0}=0\,.
\end{align}
In the above, $C_\alpha, D_\beta$ are arbitrary anti-Hermitian dressing operators of transverse soft photons given by 
%\footnote{
%$C_{\alpha}$ in \eqref{eq:gen_dressC} can be written as
%\begin{align}
    %C=\intpw\rho(p)\int^{\Lambda_s}_{\lambda}\widetilde{d^3k} \left[C^A(\vp,\vk)  a_{A}(\vk)-C^{A\ast}(\vp,\vk) a_{A}^{\dagger}(\vk)\right]
%\end{align}
%}
\begin{align}
\begin{split}
\label{eq:gen_dressC}
   C_{\alpha}&=
    \int^{\Lambda_s}_{\lambda}\widetilde{d^3k} \left[C^A_\alpha(\vk)  a_{A}(\vk)
    -C^{A\ast}_\alpha(\vk) a_{A}^{\dagger}(\vk)
    \right],\\[1em]
    D_{\beta}&=
    \int^{\Lambda_s}_{\lambda}\widetilde{d^3k} \left[D^A_\beta(\vk)  a_{A}(\vk)
    -D^{A\ast}_\beta(\vk) a_{A}^{\dagger}(\vk)
    \right]\,,
    \end{split}
\end{align}
where $a_{A}(\vk), a^{\dagger}_{A}(\vk)\ (A=1,2)$ are the annihilation and creation operators of a transverse photon 
(see appendix~\ref{app:transv} for our convention), and $C_{\alpha}^{A}(\vk), D_{\beta}^{A}(\vk)$ are arbitrary functions of $\vk$ which can depend on hard information $\alpha$ and $\beta$, respectively.
$\Theta^\rmin_\alpha, \Theta^\rmout_\beta$ in \eqref{eq:dressdstates} represent arbitrary phases which are real functions depending on $\alpha, \beta$ respectively.
In \eqref{eq:gen_dressC}, we have introduced two parameters $\lambda, \Lambda_s$ specifying the range of integration over $|\vec{k}|$: $\lambda$ as an IR regulator which goes to zero and  $\Lambda_s$ as an upper cutoff which divides photons into \textit{soft} ones and \textit{hard} ones.
The scale of $\Lambda_s$ depends on a situation. For example, it is fixed by the detector resolution, which we will explain more in section \ref{sec:reduced}.
Since $C_{\alpha}^{A}(\vk)$ and $ D_{\beta}^{A}(\vk)$ only excite the transverse modes of photons, the states \eqref{eq:dressdstates} are annihilated by $k^\mu a_\mu$. Thus, the dressed states  \eqref{eq:dressdstates} are general coherent states of  soft photons in $\mathcal{H}_{GB}$. 
We sometimes write the $k$-integral as 
\begin{align}
    \int^{\Lambda_s}_{\lambda}\widetilde{d^3k}=: \int_\text{soft}\widetilde{d^3k}\,.
\end{align}
We now assume that the Fock states $\ket{\alpha}_0, \ket{\beta}_0$ are momentum eigenstates of (free) charged particles such as $b^\dagger(p_1) d^\dagger(p_2)d^\dagger(p_3)\ket{0}_0$.\footnote{They can also  contain finite numbers of transverse hard photons.}
This assumption does not lose its generality because general Fock states are the superposition of momentum eigenstates. 

Our main task in this section is to see the condition for dressing operators so that $S$-matrix elements \eqref{dressS} are finite in the limit removing the IR regulator $\lambda \to 0$. We call such elements \textit{IR safe}.  In our approach, we explicitly introduce the IR regulator $\lambda$.
We think that the finite photon mass prescription results in the same results at least for the leading IR divergences, although we will not thoroughly examine it in this paper.

In our notation, Chung's dressing,  introduced in \cite{Chung:1965zza}, is obtained by setting
$C^A_{\alpha}=-R^A_{\alpha}$ and $D^A_\beta=-R^A_\beta$ where
\begin{align}
\label{def:RAal}
     R^A_{\alpha}(\vk)&:= \sum_{n \in \alpha} e_n\frac{p_n \cdot \epsilon^A(\hat{k})}{p_n\cdot k}, \quad 
    R^A_{\beta}(\vk):= \sum_{n \in \beta} e_n\frac{p_n \cdot \epsilon^A(\hat{k})}{p_n\cdot k}.
\end{align}
Here, $n\in \alpha$ (and $n\in \beta$) in the sum means that $n$ is a label of the charged hard particles with electric charge $e_n$ and momentum $p_n^\mu$ in the Fock state $\ket{\alpha}_0$ (and $\ket{\beta}_0$). 
That is,  Chung's dressing means that we put the following coherent state of soft photons on each hard charged particle $\ket{\vp}_0$ with electric charge $e$:
\begin{align}
\label{eq:Chung-dress}
    \exp\left(-\int_\text{soft}\widetilde{d^3k}\,e\frac{p \cdot \epsilon^A(\hat{k})}{p\cdot k}[a_{A}(\vk)
    - a_{A}^{\dagger}(\vk)]\right)
    \ket{\vp}_0.
\end{align}
As in \eqref{eq:gen_dressC},
we will represent Chung's dressing operator as $R_{\psi}$ ($\psi = \alpha, \beta$) where
\begin{align}
    R_\psi :=\int_\text{soft}\widetilde{d^3k} \left[R^A_\psi(\vk)  a_{A}(\vk)
    -R^{A\ast}_\psi(\vk) a_{A}^{\dagger}(\vk)
    \right].
\end{align}
 We will soon confirm the result of  \cite{Chung:1965zza} that the states with Chung's dressing give IR-safe $S$-matrix elements by also choosing appropriate phase factors.

%%%%%%%%%%%%%%%%%%%%%%%%%

Plugging \eqref{eq:dressdstates} into \eqref{eq:conv_Smatrix} gives the $S$-matrix element between the two states $\alpha, \beta$ as 
\begin{align}
\label{eq:gen_Smat}
    S_{\beta, \alpha}
    %&=\tensor[_I]{\bra{\beta(t_f)}}{}U(t_f,t_i)\ket{\alpha(t_i)}_{I}\nn
    %&\simeq \tensor[_{GB}]{\bra{\beta}}{}U(\infty,-\infty)\ket{\alpha}_{GB}
    &\simeq \tensor[_{0}]{\bra{\beta}}{}e^{-D_\beta-i\Theta^\rmout_\beta}U(\infty,-\infty)e^{C_\alpha+i\Theta^\rmin_\alpha}\ket{\alpha}_{0}
    \nn[1em]
    &=e^{-i\Theta_{\beta,\alpha}}\tensor[_{0}]{\bra{\beta}}{}e^{-D_\beta}U(\infty,-\infty)e^{C_\alpha}\ket{\alpha}_{0}\,,
    \end{align}
where the meaning of the approximation $\simeq$ is the same as in \eqref{eq:appSmatrix}, and we have introduced the following notation of phases:
\begin{align}
\Theta_{\beta,\alpha}:=\Theta^\rmout_{\beta}-\Theta^\rmin_{\alpha}. 
\end{align}

%%%%%%%%%%%%%%%%%%%%%%%%%%%%%%%%%%%%%%%%%%%%%%%%%%%%%%%
\subsection{Infrared factorization formula for general dressed states}
Now we study the condition for the dressing operators $C_{\alpha}\,, D_{\beta}$ to make the $S$-matrix elements \eqref{eq:gen_Smat} IR-safe.
Before showing the result, let us introduce the following inner product notation between two transverse vector fields $ C^A(\vk),  D^A(\vk)$ on the photon momentum space as (see appendix~\ref{app:transv} for details):
\begin{align}
\label{def:inpr}
    \ins{C}{D}:=\int_\text{soft}\widetilde{d^3k}\, C^A(\vk) D_A(\vk).
\end{align}
The subscript $\mathrm{s}$ represents that the integration is restricted to the soft momentum region.
The inner product $\ins{C}{D}$ defines the non-negative norm as \begin{align}
    \lVert C\rVert^2_\mathrm{s}:=\ins{C}{C^\ast}=\ins{C^\ast}{C}\geq 0
\end{align} 
for any complex functions $C^A(\vk)$ supported on the soft momentum region $\lambda \leq |\vk| \leq \Lambda_s$.

Using the soft photon theorem, we can show (see appendix~\ref{app:Ssoft} for the derivation) that the possibly IR-dangerous factor for the $S$-matrix elements \eqref{eq:gen_Smat} is factorized as
\begin{align}
\label{eq:general_factor}
    S_{\beta, \alpha}
    =S^{\mathrm{soft}}_{\beta,\alpha}(\lambda,\Lambda_s)
    S^{\mathrm{hard}}_{\beta, \alpha}(\Lambda_s)
    [1+(\text{subleading})].
\end{align}
In the above, 
$S^{\mathrm{hard}}_{\beta, \alpha}(\Lambda_s)$ denotes the conventional (undressed) $S$-matrix elements between the hard Fock states, $\ket{\alpha}_{0}$ and $\ket{\beta}_{0}$, with the effective IR-cutoff $\Lambda_{s}$, where $S^{\mathrm{hard}}_{\beta, \alpha}(\Lambda_s)$ is independent of the true IR regulator $\lambda$ and thus IR-safe.\footnote{
By dividing the Feynman loop integral with a different scale $\Lambda(\neq \Lambda_s)$ in the computation of $ S^{\mathrm{hard}}_{\beta, \alpha}$, we obtain
$S^{\text{hard}}_{\beta,\alpha}(\Lambda_s)= \left(\frac{\Lambda_s}{\Lambda}\right)^{A_{\beta,\alpha}} S^{\text{hard}}_{\beta,\alpha}(\Lambda)$ when the soft approximation is  valid for the range $|\vk| \leq \max(\Lambda_s,\Lambda)$, where $A_{\beta,\alpha}$ is a non-negative exponent which will be defined in \eqref{eq:IRsuppresfac}.}
The \textit{soft} $S$-matrix elements, $S^{\mathrm{soft}}_{\beta,\alpha}(\lambda,\Lambda_s)$, are given by
\begin{align}\label{eq:softSmat}
    S^{\mathrm{soft}}_{\beta,\alpha}(\lambda,\Lambda_s)=e^{-N_{D_\beta,C_\alpha}}
\end{align}
with the exponent
\begin{align}\label{eq:gen_soft_fact}
 &N_{D_\beta,C_\alpha}(\lambda,\Lambda_s)
   \nn[0.5em]
   &=\frac{1}{2}
   \lVert R_{\beta,\alpha}+D_\beta-C_\alpha\rVert^2_\mathrm{s}
   +i\Im\left[\ins{R^{\ast}_{\beta,\alpha}}{D_\beta+C_\alpha}
   +\ins{D^\ast_\beta}{C_\alpha}\right]
   +i\Theta_{\beta,\alpha}
   -i \Phi_{\beta,\alpha}\,,
\end{align}
where $R^A_{\beta, \alpha}(\vk)$ and $\Phi_{\beta,\alpha}$ are defined as follows\footnote{In the sum $n,l$ in \eqref{def:Phi_ba}, $n=l$ terms are removed.}:
\begin{align}
\label{def:R_beal}
    &R^A_{\beta, \alpha}(\vk):= R^A_{\beta}(\vk)- R^A_{\alpha}(\vk)
    = \sum_{n \in \alpha,\beta} \eta_{n}e_n\frac{p_n \cdot \epsilon^A(\hat{k})}{p_n\cdot k}\,,\\
    &\Phi_{\beta, \alpha}:= \sum_{n,l\in\alpha,\beta}\delta_{\eta_{n}\eta_{l}}\frac{  e_{n} e_{l} \eta_{n} \eta_{l}}{8\pi\beta_{n l}} \log \left(\frac{\lambda}{\Lambda_s}\right)\,. \label{def:Phi_ba}
\end{align}
In the above,  $R^A_{\beta}(\vk), R^A_{\alpha}(\vk)$ were defined in \eqref{def:RAal}, and $\eta_{n}$ denotes a sign factor which takes $+1$ for outgoing particles in  $\beta$ and $-1$ for incoming particles in $\alpha$, and $\beta_{n l}$ is the relativistic relative velocity of particles $n$ and $l$, which is given by $\beta_{n l} = \sqrt{1-\frac{m^{4}}{\left(p_{n} \cdot p_{l}\right)^{2}}}$.
%The phase $\Phi_{\beta,\alpha}$ is $\cdots$ appears in the following reason.
 We confine a brief derivation of \eqref{eq:general_factor} and  \eqref{eq:gen_soft_fact} in appendix~\ref{app:Ssoft}.
The phase $\Phi_{\beta, \alpha}$ can also be decomposed as
\begin{align}
    \Phi_{\beta, \alpha} =
    \Phi_{\beta}-\Phi_{\alpha} \qquad 
    \text{with }\quad 
    \Phi_{\psi}:= \sum_{n,l\in\psi}\frac{  e_{n} e_{l} }{8\pi\beta_{n l}} \log \left(\frac{\lambda}{\Lambda_s}\right) \quad (\psi = \beta, \alpha).
\end{align}

The subleading term in \eqref{eq:general_factor}  originates from the corrections 
to the  approximation \eqref{eq:asappcur} and also the subleading corrections of the soft photon theorem which is used in deriving \eqref{eq:general_factor} and  \eqref{eq:gen_soft_fact}.
Since our aim in this paper is to see how dressed states avoid the IR problem, 
the precise evaluation of the subleading terms is beyond the scope of our paper. 
Hence we will omit the subleading term for simplicity, and write \eqref{eq:general_factor} just as 
\begin{align}
\label{leading:general_factor}
    S_{\beta, \alpha}
    =S^{\mathrm{soft}}_{\beta,\alpha}(\lambda,\Lambda_s)
    S^{\mathrm{hard}}_{\beta, \alpha}(\Lambda_s).
\end{align}
This formula is a generalization of the infrared factorization formula of $S$-matrix in QED for general dressed states.
In the following, 
we  see that the formula reproduces the usual IR problem  for Fock states and also Chung's result in \cite{Chung:1965zza}. In the next subsection, we investigate the condition for general dresses to obtain the IR-safe $S$-matrix.

We also note that $S^{\mathrm{soft}}_{\beta,\alpha}$ is written in terms of the inner product with the typical form $\ins{X^\ast}{Y}$ as \eqref{eq:gen_soft_fact}.
This inner product is invariant under the transformation $\delta \epsilon_A^\mu \propto k^\mu$ as shown in appendix~\ref{app:transv}. Hence $S^{\mathrm{soft}}_{\beta,\alpha}$ is gauge invariant.

\paragraph{Undressed case reproduces the conventional IR divergence}\ \\
First, let us consider the undressed states, i.e., we take $\ket{\alpha}_{GB}, \ket{\beta}_{GB}$ as Fock states by setting  $C_{\alpha}=D_{\beta}=0$, and also $\Theta_\beta=\Theta_\alpha=0$. 
In this case, the exponent of the soft $S$-matrix in \eqref{eq:gen_soft_fact} is given by
\begin{align}\label{eq:underessN}
    N_{D_\beta=C_\alpha=0}(\lambda,\Lambda_s)
  =\frac{1}{2}
   \ins{R_{\beta, \alpha}}{R^{\ast}_{\beta, \alpha}}-i\Phi_{\beta,\alpha}.
\end{align}
The first term is given by
\begin{align}
\label{undressN}
    \frac{1}{2}
   \ins{R_{\beta, \alpha}}{R^{\ast}_{\beta, \alpha}}&=
   \frac{1}{2}
   \sum_{n,m \in \alpha,\beta}\eta_{n}\eta_{m}e_ne_m\int^{\Lambda_s}_{\lambda}\widetilde{d^3k} \frac{(p_n \cdot \epsilon^A)( p_m \cdot \epsilon^{\ast}_A)}{(p_n\cdot k) (p_m\cdot k)}\nn
   &=\frac{1}{2}\sum_{n,m \in \alpha,\beta}\eta_{n}\eta_{m}e_ne_m\int^{\Lambda_s}_{\lambda}\widetilde{d^3k} \frac{p_n \cdot p_m}{(p_n\cdot k) (p_m\cdot k)},
\end{align}
where we have used \eqref{def:R_beal} in the first equality, and the complete relation \eqref{eq:polcoml} and the global charge conservation $\sum_{n\in \alpha,\beta}\eta_{n}e_n=0$ in the second equality.
This is  the well-known IR divergent integral (see, e.g., \cite{Weinberg:1995mt}) for conventional $S$-matrix elements for Fock states. 
The integral is computed as 
\begin{align}\label{eq:IRsuppresfac}
   \frac{1}{2}
   \ins{R_{\beta, \alpha}}{R^{\ast}_{\beta, \alpha}} =\sum_{n,m\in\{\alpha,\beta\}}\frac{ e_{n} e_{m} \eta_{n} \eta_{m}}{8\pi^{2}\beta_{n m}} \log \left(\frac{1+\beta_{n m}}{1-\beta_{n m}}\right)\log\left(\frac{\lambda}{\Lambda_s}\right)
   =:-A_{\beta,\alpha}\log\left(\frac{\lambda}{\Lambda_s}\right)\,,
\end{align}
which causes a logarithmic divergence in the limit removing the IR cutoff $\lambda\to0$.
Plugging \eqref{eq:IRsuppresfac} into the soft $S$-matrix \eqref{eq:softSmat}, we obtain the following conventional IR suppression factor:
\begin{align}
     S^{\mathrm{soft}}_{\beta,\alpha}(\lambda,\Lambda_s)=e^{-N_{D_\beta=C_\alpha=0}}
     =\left(\frac{\lambda}{\Lambda_s}\right)^{A_{\beta,\alpha}}e^{i\Phi_{\beta,\alpha}}\,.%S^{\mathrm{hard}}_{\beta, \alpha}(\Lambda_s)
\end{align}
The IR divergence makes $\lim_{\lambda\to0}S^{\mathrm{soft}}_{\beta,\alpha}=0$ except for the identical scattering process $\alpha=\beta$ because $A_{\beta,\alpha}\ge 0$ and the case $A_{\beta,\alpha}=0$ is realized only for  the identical process \cite{Weinberg:1995mt, Carney:2017jut}.
This is the IR problem mentioned around \eqref{eq:zeroS}.
Here the identical scattering  process means that the information of external hard charged particles $(e_n, p_n)$ are the same for $\ket{\alpha}, \ket{\beta}$ although they can differ in the hard photon sector  (see \cite{Carney:2017jut} for detail). 

Recall that these ``undressed'' states contain the KF dress for finite time (see \eqref{eq:KFdressd} and \eqref{eq:realSmat}).
In this sense, the KF dresses obtained by evaluating the asymptotic dynamics are not enough to obtain IR-safe $S$-matrix elements \cite{Hirai:2019gio}, and we have to introduce additional dresses to cancel the divergences in \eqref{undressN} as pointed out in \cite{Hirai:2020kzx}.

\paragraph{Confirmation of Chung's result}\ \\
Next, we consider dressed states. 
As shown in \cite{Chung:1965zza}, 
Chung's dress, which is obtained by setting $C^A_{\alpha}=-R^A_{\alpha}$ and $D^A_\beta=-R^A_\beta$, is a way of resolving the IR problem.
In this case, the IR problematic function $N_{D_\beta,C_\alpha}$ in \eqref{eq:gen_soft_fact} becomes purely the phase term as  
\begin{align}
 N_{D_\beta,C_\alpha}\Big{|}_{D_\beta=-R_{\beta},C_\alpha=-R_{\alpha}}=\Im(R_{\alpha}^{\ast},R_{\beta})_s+i(\Theta_{\beta,\alpha}- \Phi_{\beta,\alpha})
 =i(\Theta_{\beta,\alpha}- \Phi_{\beta,\alpha})\,,
\end{align}
where we have used that $(R_{\alpha}^{\ast},R_{\beta})_s$ is real.\footnote{If we define $\eta^{T}_{\mu\nu}:=\eta_{T}^{AB}\epsilon_{A}^{\mu}(\hat{k})\epsilon_{B}^{\nu\ast}(\hat{k})$, then $\eta^{T}_{\mu\nu}$ is a real symmetric matrix because of \eqref{eq:polcoml}.
This fact immediately leads to the reality of $ (R_{\alpha}^{\ast},R_{\beta})_s$\,.
}
Thus, the factor $\left(\frac{\lambda}{\Lambda_s}\right)^{A_{\beta,\alpha}}$ which makes $S_{\beta,\alpha}=0$ in the limit $\lambda\to 0$ does not appear, and the $S$-matrix elements can take non-vanishing values for Chung's dressed states.
However, we still have the infinitely oscillating phase factor $e^{i \Phi_{\beta, \alpha}}$.
We can also remove them by taking $\Theta^\rmout_\beta=\Phi_\beta,\Theta^\rmin_\alpha=\Phi_\alpha$.
We then obtain $S^{\mathrm{soft}}_{\beta,\alpha}(\lambda,\Lambda_s)=1$. 
Hereafter, we suppose that Chung's dressed states contain these phases, that is, Chung's dressed states take the following forms 
\begin{align}
    \ket{\psi}=e^{-R_\psi +i \Phi_\psi}\ket{\psi}_0 \qquad (\psi=\alpha, \beta). 
\end{align}

To sum up, the $S$-matrix elements for Chung's dressed states are given by
\begin{align}
\label{eq:hardS}
    S_{\beta, \alpha}
    =
    S^{\mathrm{hard}}_{\beta, \alpha}(\Lambda_s)
\end{align}
which is clearly IR safe because it is independent of the IR cutoff $\lambda$.
Eq.~\eqref{eq:hardS} means that  the $S$-matrix elements for Chung's dressed states are approximately the same as the conventional $S$-matrix elements with an effective IR cutoff $\Lambda_s$ \cite{Mirbabayi:2016axw, Choi:2019rlz}.
We also note that \eqref{eq:hardS} has subleading corrections as we explained above \eqref{leading:general_factor}. We will comment on the subleading corrections in Sec.~\ref{sec:discuss}.

%%%%%%%%%%%%%%%%%%%%%%%%%%%%%%%%%%%%%%%%%%%%%%%%%%%%%%%
\subsection{General dress code for IR-safe $S$-matrix}\label{subsec:gene-dress-code}
%\paragraph{General dresses}
Here we investigate the possibility of more general dresses $C_\alpha, D_\beta$, not limited to Chung's dress, for IR-safe $S$-matrix elements. 
In general, the real part and imaginary part of the function $N_{D_\beta,C_\alpha}$ in \eqref{eq:gen_soft_fact} are written as
\begin{align}\label{eq:realN}
    &\Re(N_{D_\beta,C_\alpha})=
    \frac{1}{2}\lVert R_{\beta,\alpha}-C_\alpha+D_\beta\rVert^2_\mathrm{s}
    \ (\,\geq 0\,)\,,\\[1em]
    & \Im(N_{D_\beta,C_\alpha})=
    \Im\left[\ins{R^\ast_{\beta,\alpha}}{C_\alpha+D_\beta}+\ins{D^\ast_\beta}{C_\alpha}
    \right]
    +\Theta_{\beta,\alpha}
    -\Phi_{\beta,\alpha}\,.
    \label{eq:ImN}
\end{align}
%where we have used the fact that $R_{\beta,\alpha}^A$ is real because we have taken real polarization basis (see appendix \ref{app:transv}).
In particular, $\Re(N_{D_\beta,C_\alpha})$  is non-positive
and then the absolute value of the general soft factor  $S^{\mathrm{soft}}_{\beta,\alpha}=e^{-N_{D_\beta,C_\alpha}}$ is in the range 
\begin{align}
   0\leq |S^{\mathrm{soft}}_{\beta,\alpha}|\leq 1.
\end{align}
The vanishing soft factor $|S^{\mathrm{soft}}_{\beta,\alpha}|=0$ means that the IR problematic function  $N_{D_\beta,C_\alpha}$ has IR divergences.
To obtain the non-vanishing $S$-matrix elements, we have to choose the dresses so that $\Re(N_{D_\beta,C_\alpha})$ is finite:
\begin{align}\label{eq:Sdresscode}
    \lim_{\lambda\rightarrow 0}\Re\left[ N_{D_\beta,C_\alpha}(\lambda,\Lambda_s)\right]<\infty\,.
\end{align}
This leads to the following dress code:
\begin{align}
\label{cond:N=0}
   C_\alpha^A-D_\beta^A =R_{\beta,\alpha}^A+o(k^{-1})\,,
\end{align}
because  we have to cancel the singular function of $R_{\beta,\alpha} \simeq\mathcal{O}(k^{-1})$ which causes the logarithmic IR divergence as we saw in \eqref{eq:IRsuppresfac}.\footnote{Recall that $o(k^{-1})$ means that the singularity is weaker than $1/k$, i.e.,  we have $\lim_{k \to 0} k\, o(k^{-1})=0$.}
 
\paragraph{Memory effect from dress code}\ \\
An interpretation of the dress code \eqref{cond:N=0} is given as follows.
The coherent states created by $C_\alpha$ and $D_\beta$ correspond to the low-frequency parts of initial  and final electromagnetic waves. 
Eq.~\eqref{cond:N=0} states that the long-range parts of  the final radiation are completely fixed by the initial waves and the scattering data of the charged particles $\{e_n,\vp_n\}_{n\in \alpha} \to \{e_n,\vp_n\}_{n\in \beta}$.
This is nothing but 
the (leading) electromagnetic memory effect \cite{Bieri:2013hqa, Tolish:2014bka, Susskind:2015hpa}.
In fact, $R_{\beta,\alpha}^A$ is the same as the memory computed in \cite{Hamada:2018cjj} for classical point particles kicked as $\{e_n,\vp_n\}_{n\in \alpha} \to \{e_n,\vp_n\}_{n\in \beta}$.
It is shown that the memory effect is equivalent to the  conservation law of asymptotic symmetry \cite{Pasterski:2015zua, Mao:2017wvx, Hirai:2018ijc}. 
In fact, the dress code \eqref{cond:N=0} is related to the conservation law of the asymptotic symmetry in the following sense.
If states belong to different superselection sectors from each other, the transition between them should be prohibited by the superselection rule as mentioned in section~\ref{sec:sel_assym}. 
In other words,
$S$-matrix elements can take non-vanishing values only for states in the same superselection sector. 
To belong to the same sector, the states must satisfy the dress code \eqref{cond:N=0}, which leads to the memory effect.
Here we stress that the conservation law (or superselection rule) of the asymptotic symmetry is maintained \textit{thanks to the IR divergences} which set the amplitudes to zero for the prohibited transitions.
In this sense, we need IR divergences, and the IR problem concerning eq.~\eqref{eq:zeroS} is a consequence of the conservation law of the asymptotic symmetry, as emphasized in \cite{Kapec:2017tkm}.

Chung's dress, $C^A_{\alpha}=-R^A_{\alpha}$ and $D^A_\beta=-R^A_\beta$, satisfies the condition \eqref{cond:N=0}. 
We have other choices. 
For example, we can set the initial dress as $C_\alpha=0$ by taking the final dress as $D_\beta=-R_{\beta,\alpha}$, which corresponds to situations where we prepare initial states without any radiation.
This choice of dresses is the same as in \cite{Hirai:2020kzx}, and leads to the result \eqref{eq:hardS}.
This freedom of the dresses is the ``movability'' of photon clouds (dresses) discussed in \cite{Kapec:2017tkm, Choi:2017ylo}.\footnote{We cannot move the KF dress $R_{KF}$ in \eqref{eq:KFdressd} because of Gauss' law constraint as mentioned in \cite{Hirai:2020kzx}, i.e., any Fock state $\ket{\alpha}_0$ without $R_{KF}$ does not satisfy the physical state condition \eqref{eq:asBRSRcd}.} 
More generally, the dress code \eqref{cond:N=0} means that the $S$-matrix element is IR-safe if the finial state has a dressing $D_{\beta}$ depending on the initial dressing $C_{\alpha}$ up to subleading dressing as
\begin{align}\label{eq:dressmemory}
    D_{\beta}^{A}=C_{\alpha}^{A}-R_{\beta,\alpha}^A+o(k^{-1}),
\end{align}
where  $C_{\alpha}$ can contain singular terms than $\mathcal{O}(k^{-1})$ and also subleading corrections.

\paragraph{Superselection rule and the generalized Chung dress}\ \\
In the above, we have mentioned that the dress code \eqref{cond:N=0} allows the movability of clouds (dresses) which allows us to use the undressed states, for example, for initial states.
However, we can argue that the movability works only for  pure momentum eigenstates because the generic solutions of the dress code \eqref{cond:N=0} for arbitrary $\alpha$ and $\beta$ are essentially only Chung's dress, or more precisely only the generalized Chung's dress \eqref{int_gen_Chung} mentioned in the introduction.
We can also understand this fact from the asymptotic symmetry as follows.

As a warm-up, let us consider an undressed superposed incoming state
\begin{align}\label{eq:fsuperposed}
     \ket{f}_{GB}=f_1\ket{\alpha_1}_{0}+f_2\ket{\alpha_2}_{0}
\end{align} 
with arbitrary coefficients $f_1, f_2 \in \mathbb{C}$.
Here, we suppose that $\ket{\alpha_1}_{0}, \ket{\alpha_2}_{0}$ are different momentum eigenstates of hard states, and thus $ \ket{f}_{GB}$ is not momentum eigenstates.
Since $\ket{\alpha_1}_{0}, \ket{\alpha_2}_{0}$ are different hard states without soft parts, their asymptotic charges are different. 
Since the superselection rule of the asymptotic symmetry holds as argued in section~\ref{sec:sel_assym}, the two states $\ket{\alpha_1}_{0}, \ket{\alpha_2}_{0}$ cannot interfere with each other.
Any final state $\ket{\beta}_{GB}$ cannot have non-vanishing overlap both of $\ket{\alpha_1}_{0}$ and $\ket{\alpha_2}_{0}$. 
In fact, the superselection rule is executed by the infrared divergence as follows.
Since initial states are undressed states $(C_{\alpha_1}=C_{\alpha_2}=0)$, the final dress should be $D_\beta^A =-R_{\beta,\alpha_1}^A+o(k^{-1})$ or $D_\beta^A =-R_{\beta,\alpha_2}^A+o(k^{-1})$ by the dress code \eqref{cond:N=0} in order to have non-vanishing $S$-matrix elements $S_{\beta,f}\neq 0$.
Thus, if the final state $\ket{\beta}_{GB}$ is a momentum eigenstate, it can have non-vanishing overlap  with either $\ket{\alpha_1}_{0}$ or $\ket{\alpha_2}_{0}$.

In the above example, we started with the superposed state \eqref{eq:fsuperposed}. 
However, we cannot prepare such a superposition of states in different superselection sectors as in the statement ($\ast$) in subsec.~\ref{sec:sel_assym}, and we should consider superposition in the same superselection sector.
Without soft dressing, it is impossible to construct such superposition for general hard states.

On the other hand, two Chung's dressed states $ \ket{\alpha_i}=e^{-R_{\alpha_i}+i\Phi_{\alpha_i}}\ket{\alpha_i}_0$ ($i=1,2$) belong to the same superselection sectors.
It follows from the fact that there exists a final Chung's state $ \ket{\beta}=e^{-R_{\beta}+i\Phi_{\beta}}\ket{\beta}_0$ that have non-vanishing soft $S$-matrix elements for both states $\ket{\alpha_1}, \ket{\alpha_2}$ as we have seen in  \eqref{eq:hardS}.\footnote{We can also show that Chung's dressed states have the same asymptotic charges by explicitly computing the commutator between the asymptotic charges and Chung's dress \cite{Hirai:2020kzx}.}
Thus, superposition of Chung's dressed states makes sense. Here we consider the general superposition
\begin{align}\label{eq:chung_suppos}
    \ket{f}_{GB}=\sum_{i}f_i\,e^{-R_{\alpha_i}+i\Phi_{\alpha_i}}\ket{\alpha_i}_{0}.
\end{align}
This state has non-vanishing $S$-matrix elements with final dressed states 
\begin{align}
    &\ket{\beta}_{GB}=e^{D_{\beta}+i\Phi_{\beta}}\ket{\beta}_0
\end{align}
if and only if the final dresses are given by 
\begin{align}
    D_{\beta}^{A}=-R_{\beta}^A+o(k^{-1}).
\end{align}

The state \eqref{eq:chung_suppos} is essentially the general form of superposition such that the superposed states belong to the same superselection sectors. 
There are only two possible modifications of dresses.
One is just subleading changes of dresses that do not affect the leading asymptotic charges.
The other one is a common shift of dresses which may change the asymptotic charges but all of the superposed states are still in the same sector.
Thus, if we require the superposed states in the same superselection sector, general superposed states take the following form
\begin{align}
    \ket{f}_{GB}=\sum_{i}f_i\,e^{C_{\alpha_i}+i\Theta^\rmin_{\alpha_i}}\ket{\alpha_i}_{0},
    \qquad
    C_{\alpha_i}= -R_{\alpha_i} + \wC_{\alpha_i} + C,
    \label{C_gen_sup}
\end{align}
where $\wC_{\alpha_i}$ are subleading dresses $\wC_{\alpha_i}^A =o(k^{-1})$, and $C$ is arbitrary dresses independent of hard states $\alpha_i$.\footnote{Here we suppose that the common dress $C$ is more singular than $o(k^{-1})$. Indeed, if $C$ is $o(k^{-1})$, we can absorb it into $\wC_{\alpha_i}$.} 
The possible final states should have  similar dresses in order to have non-vanishing $S$-matrix elements with the above $\ket{f}_{GB}$. 
The general form is thus given by
\begin{align}
    \ket{g}_{GB}=\sum_{j}g_j\,e^{D_{\beta_j}+i\Theta^\rmout_{\beta_j}}\ket{\beta_j}_{0},
    \qquad
    D_{\beta_j}= -R_{\beta_j} + \wD_{\beta_j} + C,
    \label{D_gen_sup}
\end{align}
where $\wD_{\beta_j}$ are also $o(k^{-1})$ subleading dresses.
The initial and final dresses, \eqref{C_gen_sup},\eqref{D_gen_sup}, are indeed the general solution of the dress code \eqref{cond:N=0}. 
We will call this type of dresses,
\begin{align}\label{gen_Chung}
    \mathbf{C}_\alpha= -R_\alpha+\wC_\alpha+C, \quad  \mathbf{D}_\beta= -R_\beta+\wD_\beta+D,%\\
\end{align}
the {\it generalized Chung dress}, and represent them by bold symbols.
Here we allow the common part of final dress $D$ to be different from the initial one $C$ for later convenience.\footnote{We will see in the next section that the dress code for IR-safe inclusive cress-sections allows $C\neq D$ .}
Eqs.~\eqref{C_gen_sup},\eqref{D_gen_sup} indicate that we essentially do not have the movability of dresses because they are Chung's dressed states up to the subleading corrections and a common shift of dresses. 

\paragraph{$S$-matrix elements for the generalized Chung dressed states}\ \\
Let us compute the soft part of the $S$-matrix elements for the generalized Chung dressed states \eqref{gen_Chung}. 
As shown in \eqref{eq:softSmat}, it is given by 
\begin{align}
    S^{\mathrm{soft}}_{\beta,\alpha}(\lambda,\Lambda_s)=e^{-N_{\bD_\beta,\bC_\alpha}},
\end{align}
where $N_{\bD_\beta,\bC_\alpha}$ is given by \eqref{eq:gen_soft_fact} with setting $C_\alpha, D_\beta$ to the generalized Chung dresses $\bC_\alpha, \bD_\beta$ in \eqref{gen_Chung}. 
The real part of $N_{\bD_\beta,\bC_\alpha}$ is then given by
\begin{align}
    \Re(N_{\bD_\beta,\bC_\alpha})
    &=\frac{1}{2}\lVert \wD_{\beta}-\wC_\alpha
    +D-C
    \rVert^2_\mathrm{s}.
\end{align}
For general singular $C$ and $D$, this diverges, and it represents that if $C\neq D$, the initial and final states belong to different superselection sectors. The dress code \eqref{cond:N=0} requires $C=D$ as \eqref{D_gen_sup}. 
We will see in section~\ref{sec:reduced} that this restriction can be relaxed if we consider the inclusive cross-section. 

Here we concentrate on the case that  $C=D$. We then have a finite real part\footnote{
The expression \eqref{fin_real_DC} is just a formal one, because the subleading corrections to our approximation also contribute to the finite parts. 
We comment on the subleading corrections in section~\ref{sec:discuss}.}
\begin{align}
\label{fin_real_DC}
    \Re(N_{\bD_\beta,\bC_\alpha})
    &=\frac{1}{2}\lVert \wD_{\beta}-\wC_\alpha
    \rVert^2_\mathrm{s}.
\end{align}
The imaginary part of $N_{\bD_\beta,\bC_\alpha}$ with $C=D$ is also computed as
\begin{align}
\Im(N_{\bD_\beta,\bC_\alpha})=&\Im\left[
\ins{R^{\ast}_{\beta}}{\widetilde{D}_{\beta}+C}
-\ins{R^{\ast}_{\alpha}}{\widetilde{C}_{\alpha}+C}
+\ins{\widetilde{D}^{\ast}_{\beta}+C^\ast}{\widetilde{C}_{\alpha}+C}
\right]\nn &\quad
     +\Theta_{\beta,\alpha}-\Phi_{\beta,\alpha}.
     \label{ImN_genC=D}
\end{align}
This imaginary part generically diverges unless we choose appropriate phases $\Theta_{\alpha},\Theta_{\beta}$ 
as we did for original Chung's dressed states.
Indeed, we can make \eqref{ImN_genC=D} IR finite by setting
\begin{align}
\label{phase_gen_in}
    \Theta^\rmin_\alpha&= \Phi_\alpha -\Im\left[\ins{R^{\ast}_{\alpha}}{\widetilde{C}_{\alpha}+C}
-\ins{C^\ast}{\widetilde{C}_{\alpha}}
\right],
\\
 \Theta^\rmout_\beta&= \Phi_\beta -\Im\left[
\ins{R^{\ast}_{\beta}}{\widetilde{D}_{\beta}+C}
-\ins{C^{\ast}}{\widetilde{D}_{\beta}}
\right],
\label{phase_gen_out}
\end{align}
where we note that $\Theta^\rmin_\alpha$ does not depend on the quantities with index $\beta$, and also  $\Theta^\rmout_\beta$ does not on $\alpha$.
For these phases, the imaginary part of $N_{\bD_\beta,\bC_\alpha}$ reads
\begin{align}
\Im(N_{\bD_\beta,\bC_\alpha})=&\Im\left[
\ins{\widetilde{D}^{\ast}_{\beta}}{\widetilde{C}_{\alpha}}
\right]
\end{align}
which is IR finite because $\wC_{\alpha}$, $\wD_{\beta}$ are subleading dresses as  $\wC_{\alpha}^A(\vk), \wD_{\beta}^A(\vk)=o(k^{-1})$.
To sum up, in the above choice of phases, we have
\begin{align}
    S^{\mathrm{soft}}_{\beta,\alpha}(\lambda,\Lambda_s)=e^{-i\Im\left[
\ins{\widetilde{D}^{\ast}_{\beta}}{\widetilde{C}_{\alpha}}
\right]
-\frac{1}{2}\lVert \wD_{\beta}-\wC_\alpha
    \rVert^2_\mathrm{s}} = \bS^{\mathrm{soft}}_{\beta,\alpha}\,,
    \label{genCh_Ssoft}
\end{align}
where we have represented the soft $S$-matrix elements  $S^{\mathrm{soft}}_{\beta,\alpha}$ for the generalized Chung dress by $\bS^{\mathrm{soft}}_{\beta,\alpha}$, i.e., $\bS^{\mathrm{soft}}_{\beta,\alpha}:= e^{-N_{\bD_\beta,\bC_\alpha}}$.
The $S$-matrix elements between the generalized Chung dressed states are IR-safe as
\begin{align}
\label{N-fin_Smat}
    S_{\beta,\alpha}=\bS^{\mathrm{soft}}_{\beta,\alpha}S^{\mathrm{hard}}_{\beta,\alpha}=e^{-N_{\bD_\beta,\bC_\alpha}}S^{\mathrm{hard}}_{\beta,\alpha}.
\end{align}

The $S$-matrix element $S_{g,f}$ between the superposition of the generalized Chung dressed states $\ket{f}_{GB}, \ket{g}_{GB}$ given by \eqref{C_gen_sup}, \eqref{D_gen_sup} is 
\begin{align}
    S_{g,f}=\sum_{i,j}g^\ast_j f_i
    S_{\beta_j,\alpha_i}\,.
\end{align}
With the above choice of phases \eqref{phase_gen_in} and  \eqref{phase_gen_out}, the sot part of each term $S_{\beta_j,\alpha_i}$ is given by \eqref{genCh_Ssoft} with the replacement $\alpha\to \alpha_i, \beta\to \beta_j$.
We thus have 
\begin{align}\label{eq:IRsafeS}
    S_{g,f}=\sum_{i,j}g^\ast_j f_i
    e^{-i\Im\left[
\ins{\widetilde{D}^{\ast}_{\beta_j}}{\widetilde{C}_{\alpha_i}}
\right]
-\frac{1}{2}\lVert \wD_{\beta_j}-\wC_{\alpha_i}
    \rVert^2_\mathrm{s}}
    S^{\mathrm{hard}}_{\beta_j,\alpha_i}\,,
\end{align}
where each term is IR-safe.
That means that all the infrared divergences are removed at every term at any orders of perturbative expansion in the $S$-matrix \eqref{eq:IRsafeS}.
In particular, \eqref{eq:IRsafeS} yields $S_{g,f}=S^{\rmhard}_{g,f}$ for the case of no subleading dresses, $\wC_{\alpha_i}=\wD_{\beta_j}=0$.
Thus our analysis is a generalization of the agreement of the hard $S$-matrix and the $S$-matrix for Chung (Kulish-Faddeev) dressed states discussed in \cite{Chung:1965zza, Mirbabayi:2016axw, Choi:2019rlz} to the scatterings of the superposed states.
Because of the cancellation of IR divergences, 
\eqref{eq:IRsafeS} keeps all the interference effects without the decoherence unlike the undressed case \eqref{eq:fsuperposed}.

However, if we consider the inclusive cross-section, we may have further decoherence because we lose the information of final soft photons. 
We will see in section~\ref{sec:reduced} that the generalized Chung dress \eqref{gen_Chung} is also the unique dress that avoids all decoherence due to IR divergences in the inclusive cross-section.

%%%%%%%%%%%%%%%%%%%%%%%%%%%%%%%%%%%%%%%%%%%%%%%%%%%%%%%%%
\subsection{Soft photon theorem for dressed states}
We have seen that $S$-matrix elements between general dressed states \eqref{eq:dressdstates} are given by \eqref{eq:general_factor}. 
Here we consider the soft photon theorem for these states, which will be used to compute the inclusive cross-sections in section \ref{sec:reduced}. 
Let us add a soft photon with polarization $A$ to the initial dressed state as
\begin{align}
    \ket{\alpha+\gamma_\text{soft}}_{GB}=e^{C_\alpha+i\Theta_\alpha^{\rmin}} a^\dagger_A(\vk_\text{soft})\ket{\alpha}_{0}, 
\end{align}
and consider the $S$-matrix element between this initial state and the final state
\begin{align}
    \tensor[_{GB}]{\bra{\beta}}{}=
     \tensor[_{0}]{\bra{\beta}}{}e^{-D_\beta-i\Theta_\beta^{\rmout}}.
\end{align}
The $S$-matrix element is given by
\begin{align}\label{eq:softinsert}
    S_{\beta,\alpha+\gamma_\text{soft}}
    =e^{-i\Theta_{\beta,\alpha}}\tensor[_{0}]{\bra{\beta}}{}e^{-D_\beta}U(\infty,-\infty)e^{C_\alpha}a^\dagger_A(\vk_\text{soft})\ket{\alpha}_{0}.
\end{align}
When $a^\dagger_A(\vk_\text{soft})$ contracts with $U(\infty,-\infty)$, it reproduce the usual soft factor as in the standard soft theorem. 
In addition, $a^\dagger_A(\vk_\text{soft})$ also contracts with the dressing factors $e^{-D_\beta}, e^{C_\alpha}$. 
Thus, we obtain
\begin{align}
\label{dress-soft-thrm-in}
    S_{\beta, \alpha+\gamma_\text{soft}}=\left[-R_{\beta,\alpha A}(\vk_\text{soft})+C_{\alpha A}(\vk_\text{soft})-D_{\beta A}(\vk_\text{soft})\right]  S_{\beta, \alpha}\,,
\end{align}
at the leading order in the soft limit $\vk_\text{soft}\sim 0$, where $R_{\beta,\alpha A}$ is the standard soft factor defined in \eqref{def:R_beal}.
Similarly, if we add a soft photon to the final state as
\begin{align}
     \tensor[_{GB}]{\bra{\beta+\gamma_\text{soft}}}{} =
     \tensor[_{0}]{\bra{\beta}}{} a_A (\vk_\text{soft})e^{-D_\beta-i\Theta_\beta^\rmout} ,
\end{align}
we have 
\begin{align}
\label{dress-soft-thrm-out}
    S_{\beta+\gamma_\text{soft}, \alpha}= \left[R^{\ast}_{\beta,\alpha A}(\vk_\text{soft})-C^\ast_{\alpha A}(\vk_\text{soft})+D^\ast_{\beta A}(\vk_\text{soft})\right]S_{\beta, \alpha}.
\end{align}
Eqs.~\eqref{dress-soft-thrm-in} and \eqref{dress-soft-thrm-out} are the soft photon theorem for general dressed states.

If we use dressing factors satisfying the leading dress code $C_\alpha^A-D_\beta^A =R_{\beta,\alpha}^A$, like Chung's dress,
the soft factors in \eqref{dress-soft-thrm-in} and \eqref{dress-soft-thrm-out} vanish. 
As we saw above, $C_\alpha^A-D_\beta^A =R_{\beta,\alpha}^A$ means $\Re(N_{D_\beta,C_\alpha})=0$ or $|S^{\mathrm{soft}}_{\beta,\alpha}|=1$. 
Thus, when $S_{\beta, \alpha}$ is completely IR safe as $|S^{\mathrm{soft}}_{\beta,\alpha}|=1$, the soft theorem is trivial as 
\begin{align}
\label{soft-thrm_Chung}
   S_{\beta, \alpha+\gamma_\text{soft}}=  S_{\beta+\gamma_\text{soft}, \alpha}=0
\end{align}
at the leading soft order of $\vk_\text{soft}\sim 0$.\footnote{The zero in the right-hand side of \eqref{soft-thrm_Chung} just represents vanishing of  $\mathcal{O}(k_\text{soft}^{-1})$ terms which generally appear in the right-hand side of \eqref{dress-soft-thrm-in}. 
Thus, more precisely, we should write \eqref{soft-thrm_Chung} as
  $S_{\beta, \alpha+\gamma_\text{soft}}\sim  S_{\beta+\gamma_\text{soft}, \alpha} \sim  o(k_\text{soft}^{-1})$.}
In particular, Chung's dress states are orthogonal to other Chung's dressed states with additional soft photons at the leading order as shown in \cite{Gabai:2016kuf, Carney:2017oxp}.

Furthermore, the general soft photon theorem [\eqref{dress-soft-thrm-in} and \eqref{dress-soft-thrm-out}] leads to the following equation  for any dressed (or undressed) states $\alpha, \beta$:
\begin{align}
\label{soft-theorem=0}
   \lim_{\lambda \to 0}S_{\beta, \alpha+\gamma_\text{soft}}
   =\lim_{\lambda \to 0}S_{\beta+\gamma_\text{soft},\alpha}
   =o(\omega^{-1}),
\end{align}
i.e., 
   \begin{align}
\label{leading-soft-theorem=0}
    \lim_{\omega\to 0} \omega \times S_{\beta, \alpha+\gamma_\text{soft}}
    =\lim_{\omega \to 0} \omega \times S_{\beta+\gamma_\text{soft}, \alpha}=0 \qquad \text{in the limit $\lambda \to 0$},
\end{align} 
where $\omega= |\vk_\text{soft}|$. 
This is proved as follows.
If the dresses $C_\alpha, D_\beta$  satisfy the condition \eqref{cond:N=0}, we trivially obtain \eqref{soft-theorem=0}.
If the dresses $C_\alpha, D_\beta$ do not satisfy \eqref{cond:N=0}, we have $\lim_{\lambda\to 0}S_{\beta,\alpha}=0$. 
Thus, in this case, the right-hand sides of the soft photon theorem [\eqref{dress-soft-thrm-in} and \eqref{dress-soft-thrm-out}] vanish, and we obtain \eqref{soft-theorem=0}. 

The equation \eqref{soft-theorem=0} shows that the $\mathcal{O}(\omega^{-1})$  singular soft factor in the leading soft photon theorem does not make $S$-matrix elements singular, and does not contradict the unitarity constraint  $0\leq|S_{\beta, \alpha+\gamma_\text{soft}}|\leq1$.

It is easy to generalize the above soft photon theorem to the case with multiple soft photons.
For example, we consider the case where the final state has $N$ additional soft photons as
\begin{align}\label{eq:SbN+a}
    S_{\beta+N\gamma_\text{soft}, \alpha}=e^{-i\Theta_{\beta,\alpha}}
    \tensor[_{0}]{\bra{\beta}}{}
    \left( \prod_{\ell=1}^N a_{A_\ell}(\vk_\ell)\right)
    e^{-D_\beta}U(\infty,-\infty)e^{C_\alpha}\ket{\alpha}_{0},
\end{align}
where $\vk_\ell$ $(\ell=1,\dots, N)$ are momenta of soft photons.
As in \eqref{dress-soft-thrm-out}, we can evaluate this element, and obtain 
\begin{align}
    S_{\beta+N\gamma_\text{soft}, \alpha}
    &= \prod_{\ell=1}^N\left[R^{\ast}_{\beta,\alpha A_\ell}(\vk_\ell)-C^\ast_{\alpha A_\ell}(\vk_\ell)+D^\ast_{\beta A_\ell}(\vk_\ell)\right]
    S_{\beta, \alpha}
    \nn
     &= \prod_{\ell=1}^N\left[R^{\ast}_{\beta,\alpha A_\ell}(\vk_\ell)-C^\ast_{\alpha A_\ell}(\vk_\ell)+D^\ast_{\beta A_\ell}(\vk_\ell)\right]e^{-N_{D_\beta,C_\alpha}}
    S^{\mathrm{hard}}_{\beta, \alpha}
    .
    \label{dress-soft-thrm}
\end{align}
This is the multiple soft photon theorem for general dressed states. 
We will use this formula to compute the inclusive cross-sections in section \ref{sec:reduced}.

%%%%%%%%%%%%%%%%%%%%%%%%%%%%%%%%%%%%%%%%%%%%%%%%%%%%%%%%%%%%%%%%%%%%%%%%%%%%%%%%
\section{Dress code avoiding decoherence for inclusive cross-sections}
%\section{Inclusive computations with general dressed states and the decoherence problem}
\label{sec:reduced}

In a realistic setup, any detector has a minimum energy-resolution, which we represent by $\Lambda_s$, such that we cannot detect soft photons whose energy is less than $\Lambda_s$. 
We need to trace out soft photons which are invisible to the detector \cite{Bloch:1937pw}.
We perform these inclusive computations for general dressed states.

We consider the inclusive cross-section, which will be denoted by $\sigma^{\mathrm{inc}}(f^{\text{in}}\rightarrow g^{\text{out}})$, from general initial states $\ket{f^{\text{in}}(t_i)}$ to final hard states $\ket{g^{\text{out}}(t_f)}_H$. 
The hard states $\ket{g^{\text{out}}(t_f)}_H$ are states in the hard sector $\mathcal{H}^{\mathrm{hard}}$ obtained by tracing out soft photons from the entire physical Hilbert space.
This tracing-out procedure is the same as summing over all possible final states which cannot be distinguished from the detector in the conventional inclusive computations.
We suppose that initial and final time $t_i, t_f$ are sufficiently large so that the approximation \eqref{eq:conv_Smatrix} is valid. Thus, the entire Hilbert space is given by $\mathcal{H}_{\rmas}$ in sec.~\ref{sec:ASphys}.

We start from the initial density matrix associated with an initial state $\ket{f^{\text{in}}(t_i)} \in \mathcal{H}_{\rmas}$,
\begin{align}
    \rho(t_i)=\ket{f^{\text{in}}(t_i)}\bra{f^{\text{in}}(t_i)}\,.
\end{align}
Then the time evolved total density matrix at final time $t_f$ is given by
\begin{align}
    \rho(t_f)=U(t_f,t_i)\rho(t_i)U^{\dagger}(t_f,t_i)\,.
\end{align}
 The final hard density matrix $\rho^{\mathrm{hard}}(t_f)$ is obtained from $\rho(t_f)$ by taking the trace over the soft sector as $\rho^{\mathrm{hard}}(t_f)=``{\mathop{\mathrm{Tr}}_{\mathrm{soft}}}\rho(t_f)$''. 
However the definition of the trace has a subtlety. 
In the next subsection, we will give the precise definition of the reduced density matrix $\rho^{\mathrm{hard}}$ by taking care of the structure of the physical Hilbert space.

Our aim in this section is to see whether we can observe interference effects in the inclusive cross-section when we use general dressed states.
For this purpose, we consider general superposed states. 
First, initial states $\ket{f^{\text{in}}(t_i)} \in \mathcal{H}_{\rmas}$ are given by\footnote{This superposition can be continuous as $\sum \to \int$.} 
\begin{align}
\label{superposed_in}
\ket{f^{\text{in}}(t_i)}=\sum_{\alpha} f_\alpha \ket{\alpha(t_i)},
\end{align}
where $f_\alpha$ are complex coefficients and  $\ket{\alpha(t_i)}$ are general dressed states 
\begin{align}
\ket{\alpha(t_i)}=e^{R_{KF}(t_i)+i\Phi(t_i)} \ket{\alpha}_{GB} ,
\end{align}
where $\ket{\alpha}_{GB}$ are arbitrary dressed states as $\ket{\alpha}_{GB}=e^{C_\alpha+i\Theta^\rmin_\alpha} \ket{\alpha}_0$. We suppose that $\ket{\alpha}_0$ are hard momentum eigenstates in the Fock space. 

Since we consider sufficiently large initial and final time $t_i, t_f$ as we mentioned above, the time-dependent dress can be absorbed into the time-evolution operator $U(\infty,-\infty)$ as in \eqref{eq:conv_Smatrix}, and we just write the initial superposed states as
\begin{align}
\label{superposed_inGB}
\ket{f^{\text{in}}}_{GB}=\sum_{\alpha} f_\alpha \ket{\alpha}_{GB}.
\end{align}
Similarly, we consider general hard final states
\begin{align}
\label{superposed_out}
\ket{g^{\text{out}}}_H=\sum_{\beta} g_\beta \ket{\beta}_0\,,
\end{align}
with arbitrary complex coefficients $g_\beta$, 
where $\ket{\beta}_0$ are hard momentum eigenstates in the Fock space.
We will compute the inclusive cross-section which is the probability that we start from $\ket{f^{\text{in}}}_{GB}$ and obtain the final hard state $\ket{g^{\text{out}}}_H$ where final soft states are not specified. 
We represent the inclusive cross-section by $\sigma^{\mathrm{inc}}(f^{\text{in}}\rightarrow g^{\text{out}})$ which can be computed by the hard reduced density matrix $\rho^{\mathrm{hard}}$ as
\begin{align}
     &\sigma^{\mathrm{inc}}(f^{\text{in}}\rightarrow g^{\text{out}})
    =\tensor[_H]{\bra{g^{\text{out}}}}{}\rho^{\mathrm{hard}}\ket{g^{\text{out}}}_{H}.
\end{align}
Our task in the next subsection is to define  $\rho^{\mathrm{hard}}$.

%%%%%%%%%%%%%%%%%%%%%%%%%%%%%%%%%%%%%%%%%%%%%%%%%%%%%%%%%%%%%%%%%%%%
\subsection{Definition of trace over soft sector}
\label{sec:def-partial_trace}
Here we define a reduced density matrix on  the hard sector. 
Our detector cannot distinguish between an undressed state $\ket{\beta}_{0}$ and that to which with states added transverse soft photons added, \textit{e.g.}, $a^\dagger_A(\vk)\ket{\beta}_{0}$ with $|\vk|<\Lambda_s$.
Thus, the reduced density matrix obtained by taking the trace over these indistinguishable states is only relevant for observers using the detector. 
The indistinguishable states can be expanded by the following states 
\begin{align}
\label{photon-basis}
 \prod_{\ell=1}^N  a^\dagger_{A_\ell}(\vk_\ell) \ket{\beta}_{0}.
\end{align}
However, since our detector cannot distinguish soft sectors, 
we can use  another basis expanded by 
\begin{align}
\label{photon-basis_C}
 e^{D_\beta+i\Theta^\rmout_\beta}\prod_{\ell=1}^N  a^\dagger_{A_\ell}(\vk_\ell) \ket{\beta}_{0},
\end{align}
where $D_\beta$ is an arbitrary dressing operator defined in \eqref{eq:gen_dressC}.
To simplify the expression, we will sometime use the following sloppy notation that we write dressed states \eqref{photon-basis_C} as
\begin{align}
\label{sloppy-basis}
    \prod_{\ell=1}^N  a^\dagger_{A_\ell}(\vk_\ell) \widetilde{\ket{\beta}},
\end{align}
although we should note that $a^\dagger_{A_\ell}(\vk_\ell)$ acts on $\ket{\beta}_0$ before the dressing operators.
The reduced density matrix obtained by tracing out a subspace is usually independent of the choice of the basis of the subspace. 
However, we might have to take care because the space expanded by \eqref{photon-basis} can be orthogonal to the space expanded by \eqref{photon-basis_C} in the limit removing the IR cutoff $\lambda\to 0$ because of the incompleteness argued in section~\ref{sec:incomplete}.
This incompleteness follows from the superselection structure by the asymptotic symmetry. Formally, the entire physical Hilbert space is decomposed into the superselection sectors as
\begin{align}\label{eq:allphys}
   \mathcal{H}_\text{all phys} =\bigoplus_Q \mathcal{H}_Q
\end{align}
where $Q$ denotes the asymptotic charges and $\mathcal{H}_Q$ is a physical Hilbert space with the asymptotic charge fixed to $Q$.
Putting dresses that are singular in $\lambda \to 0$ like Chung's dress moves formally a state in a sector into another sector.
The orthogonality of different sets of bases (e.g., \eqref{photon-basis} and \eqref{photon-basis_C}) is due to the decomposition because they have different charges.
Each $\mathcal{H}_Q$ consists of hard and soft parts, and we consider a partial trace over the soft part.

As explained in subsec.~\ref{sec:sel_assym}, 
any quasi-local observables do not change sectors. 
In such a superselection structure, we have to take care of the definition of the partial trace.  
We encounter similar problems for the definition of entanglement entropy in gauge theories (see, e.g., \cite{Casini:2013rba}) and for the target space entanglement entropy (see, e.g., \cite{Mazenc:2019ety, Sugishita:2021vih}). 
In these situations, the key to avoiding the problems is demanding that the expectation values of operators are correctly reproduced from the reduced density matrix. We follow this approach here, and it leads us to take a partial trace on each superselection sector. 
We will explain the definition of the partial trace on a fixed sector $\mathcal{H}_Q$.

In $\mathcal{H}_Q$, the soft dresses depend on hard parts because of the condition that the total asymptotic charge must be a given $Q$. 
If a hard part is $\ket{\beta}_\text{hard}$, the soft part takes the form $\ket{\gamma;\beta}_\text{soft}:= e^{D_\beta}\ket{\gamma}_\text{soft}$ where $D_\beta$ is an  dressing operator defined in \eqref{eq:gen_dressC}, $\ket{\gamma}_\text{soft}$ represent Fock states of soft photons. 
The dressing operator $D_\beta$ must be chosen so that  the asymptotic charge of $\ket{\beta}_\text{hard}\ket{\gamma;\beta}_\text{soft}$  agrees with the given value $Q$.
In appendix \ref{app:Completeness}, we will explain that the basis \{$\ket{\beta}_\text{hard}\ket{\gamma;\beta}_\text{soft}$\} is complete in $\mathcal{H}_Q$ and also how to specify states in $\mathcal{H}_Q$.

We now discuss a generic situation such that states in the total Hilbert space are given by
\begin{align}\label{eq:Htot}
   \mathcal{H}_Q = \{\ket{\beta}_A \ket{\gamma;\beta}_{\overline{A}}\, |\,\ket{\beta}_A \in \mathcal{H}^{A},\,
   \ket{\gamma;\beta}_{\overline{A}} \in \mathcal{H}^{\overline{A}}_\beta\}.
\end{align}
Note that space $\mathcal{H}^{\overline{A}}_\beta$ depends on the states in sector $A$, but we assume that the degrees of freedom of $\gamma$ are independent of $\beta$ and thus the dimension of $\mathcal{H}^{\overline{A}}_\beta$  are the same as $\dim(\mathcal{H}^{\overline{A}}_\beta)=\dim(\mathcal{H}^{\overline{A}}_{\beta'})$ for any $\beta, \beta'$.
In our setup, $\mathcal{H}^{A}$ corresponds to the hard sector $\mathcal{H}^{\mathrm{hard}}$, and $\mathcal{H}^{\overline{A}}_\beta$ represents the space spanned by $\ket{\gamma;\beta}_\text{soft}$, i.e., space of soft photons with dress depending on hard state $\beta$.
We now consider the reduced density matrix on $\mathcal{H}^A$ by taking the partial trace over $\gamma$ (which corresponds to soft photons). 

For the total density matrix $\rho$ on $\mathcal{H}_Q$, we define the reduced density matrix $\rho^A$ on $\mathcal{H}^A$ whose matrix elements are given by\footnote{For simplicity, we represent  $\gamma$ (and also $\beta$ in the following) as the discrete variable although it should be written as integral if $\gamma$ (and also $\beta$) contains continuous variables.}
\begin{align}
\label{def:rhoA}
  \tensor[_{A}]{\bra{\beta}}{}
  \rho^A \ket{\beta'}_A:= \sum_\gamma 
  \tensor[_{A}]{\bra{\beta}}{}
    \tensor[_{\overline{A}}]{\bra{\gamma;\beta}}{}
    \rho \ket{\beta'}_A \ket{\gamma;\beta'}_{\overline{A}}.
\end{align}
This definition is reduced to the standard one if sector $\overline{A}$ is independent of sector $A$.
The generalized definition \eqref{def:rhoA} can be understood as follows.
Operators on $\mathcal{H}_Q$ that do not change the degrees of freedom $\gamma$ take the form
\begin{align}
    O=\sum_{\beta, \beta'}\sum_\gamma O^A_{\beta',\beta}\ket{\beta'}_A
    \ket{\gamma;\beta'}_{\overline{A}}
    \tensor[_{A}]{\bra{\beta}}{}\tensor[_{\overline{A}}]{\bra{\gamma;\beta}}{}.
\end{align}
This class of operators trivially acts on the degrees of freedom in $\mathcal{H}^{\overline{A}}_\beta$.
For any such operators, the reduced density  matrix $\rho^A$ should satisfy 
\begin{align}
\label{defining_rhoA}
    \Tr(O \rho)=\Tr_A (O^A \rho^A),
\end{align}
where $O^A$ corresponds to the sector $A$ part in $O$, that is, given by
\begin{align}
    O^A:=\sum_{\beta, \beta'} O^A_{\beta',\beta}\ket{\beta'}_A
    \tensor[_{A}]{\bra{\beta}}{}.
\end{align}
Here, the trace on $\mathcal{H}_Q$ is defined as 
\begin{align}
     \Tr(*) :=
     \sum_\beta \sum_\gamma
  \tensor[_{A}]{\bra{\beta}}{}
    \tensor[_{\overline{A}}]{\bra{\gamma;\beta}}{}
    \ast \ket{\beta}_A \ket{\gamma;\beta}_{\overline{A}}.
\end{align}
Then, we can show, by taking $O^A=\ket{\beta'}_A\tensor[_{A}]{\bra{\beta}}{}$ in \eqref{defining_rhoA}, that matrix elements of $\rho^A$ is given by \eqref{def:rhoA}. 

Another explanation of the definition \eqref{def:rhoA} is as follows.
Given a fixed way of dressing $e^{D[\beta]}$, the vector space 
$\mathcal{H}_{Q}=\{\ket{\beta}_A e^{D[\beta]}\ket{\gamma}_{\bar{A}}\}$ is formally isomorphic to a tensor product space
$\mathcal{H}^{A}_{Q}\otimes \mathcal{H}^{\overline{A}}_{Q}=\{\ket{\beta}_A \ket{\gamma}_{\bar{A}}\}$. 
Under this identification, the definition \eqref{def:rhoA} corresponds to that to be identified with the reduced density matrix $\rho_A$ given by the standard partial trace over $\mathcal{H}^{\overline{A}}_{Q}$. 

\paragraph{Crucial difference from naive computation of partial trace}\ \\
We have considered a generic situation and have defined the reduced density matrix on $\mathcal{H}^{A}$ as \eqref{def:rhoA}.
In our setup, $\mathcal{H}^{A}$ corresponds to the hard Fock space, and $\rho^A$ is the reduced density matrix for hard states.
What we should remark here is that the bra and ket states on the right-hand side of \eqref{def:rhoA} belong to the same $\mathcal{H}_Q$ and have the same asymptotic charges. 
Thus, the definition is based on the superselection structure, and this point is missed in the literature.  

Here we see in a simple example how our definition of $\rho^{A}$ gives a different result than $\rho^{A}$ computed by naive computation of trace.
Consider the case where $\rho$ is factorized as %$\rho=\sigma^{A}\otimes \sigma^{\overline{A}}$, e.g.,
$\rho=\ket{\beta}_{A\,A}\!\bra{\beta'}\otimes \ket{s;\beta}_{\overline{A}\,\overline{A}}\!\bra{s';\beta'}$.
In this case, $\rho^{A}$ defined  by \eqref{def:rhoA} is given by
\begin{align}\label{eq:tra_resl}
    \tensor[_{A}]{\bra{\beta}}{} \rho^A \ket{\beta'}_A
    = \sum_\gamma  \tensor[_{\overline{A}}]{\braket{\gamma;\beta}{s;\beta}_{\overline{A}\,\overline{A}}\!\braket{s';\beta'}{\gamma;\beta'}}{}_{\!\overline{A}}=\delta_{s,s'}
\end{align}
On the other hand, if we naively use the trace ``$\trc{\overline{A}}$ ''
as the trace over undressed soft states such as
\begin{align}
  \trc{\overline{A}} (\ast ) :=\sum_\gamma 
    \tensor[_{\overline{A}}]{\bra{\gamma}}{} (\ast)\ket{\gamma}_{\overline{A}},
\end{align}
we can define a naive reduced density matrix as
\begin{align}\label{wrong_tr}
   \tensor[_{A}]{\bra{\beta}}{} \rho^{A,\text{naive}} \ket{\beta'}_A :=\sum_\gamma 
  \tensor[_{A}]{\bra{\beta}}{}
    \tensor[_{\overline{A}}]{\bra{\gamma}}{}
    \rho \ket{\beta'}_A \ket{\gamma}_{\overline{A}}.
\end{align}
Then, we may calculate $ \rho^{A,\text{naive}}$ 
for the above $\rho=\ket{\beta}_{A\,A}\!\bra{\beta'}\otimes \ket{s;\beta}_{\overline{A}\,\overline{A}}\!\bra{s';\beta'}$ as follows:
\begin{align}\label{eq:naive_tr}
    \tensor[_{A}]{\bra{\beta}}{} \rho^{A,\text{naive}} \ket{\beta'}_A 
    =\trc{\overline{A}}\left(\ket{s;\beta}_{\overline{A}\,\overline{A}}\!\bra{s';\beta'}\right)
    \sim\tensor[_{\overline{A}}]{\braket{s';\beta'}{s;\beta}}{_{\overline{A}}}\, \overset{\lambda\to 0}{\longrightarrow}\,\delta_{\beta,\beta'}\delta_{s,s'}\,,
\end{align}
where we have used a naive  complete relation on the soft sector.
Compared to \eqref{eq:tra_resl}, another Kronecker delta $\delta_{\beta,\beta'}$ appears in \eqref{eq:naive_tr}.
This is due to the fact that, in \eqref{wrong_tr}, $\ket{\beta}_A \ket{\gamma}_{\overline{A}}$ and $\ket{\beta'}_A$  $\ket{\gamma}_{\overline{A}}$ belong to different superselection sectors unless $\beta=\beta'$. Therefore, a naive definition \eqref{wrong_tr} is not the partial trace on a superselection sector $\mathcal{H}_Q$ except for the diagonal components $\beta=\beta'$.
 The difference between \eqref{eq:tra_resl} and \eqref{eq:naive_tr} makes crucial differences in the result of inclusive cross-sections, as we will see in subsec.~\ref{subsec:resl}.

%%%%%%%%%%%%%%%%%%%%%%%%%%%%%%%%%%%%%%%%%%%%
\subsection{Hard density matrix for general dressed states}\label{subsec:Incout}

In our case, the sum over $\gamma$ in \eqref{def:rhoA} corresponds to taking the trace over the soft photons in the basis \eqref{photon-basis_C}.
That is, we have
\begin{align}
    &\sum_\gamma\ket{\beta'}_A
    \ket{\gamma;\beta'}_{\overline{A}}
    \tensor[_{A}]{\bra{\beta}}{}\tensor[_{\overline{A}}]{\bra{\gamma;\beta}}{}
  \nn
  &\to
  \sum_{N=0}^{\infty}\frac{1}{N!}
   \left( \prod_{\ell=1}^N 
    \int^{\Lambda_s}_{\lambda} \!\! 
    \widetilde{d^3k_\ell}\,
    \eta^{A_\ell B_\ell}_T\right)
    e^{D_{\beta'}+i\Theta^\rmout_{\beta'}}
    \left[\prod_{\ell=1}^N  a^\dagger_{B_\ell}(\vk_\ell)\right] \ket{\beta'}_{0}
\tensor[_{0}]{\bra{\beta}}{}
 \left[\prod_{\ell=1}^N
    a^\dagger_{A_\ell}(\vk_\ell)\right]
e^{-D_{\beta}-i\Theta^\rmout_{\beta}}.
\end{align}
Using the sloppy notation \eqref{sloppy-basis}, the reduced density matrix on hard sector is given by
\begin{align}
     \rho^{\mathrm{hard}}_{\beta, \beta'}&:=
    \sum_{N=0}^{\infty}\frac{1}{N!}
   \left( \prod_{\ell=1}^N 
    \int^{\Lambda_s}_{\lambda} \!\! 
    \widetilde{d^3k_\ell}\,
    \eta^{A_\ell B_\ell}_T\right)
    \widetilde{\bra{\beta}} \left[\prod_{\ell=1}^N a_{A_\ell}(\vk_\ell)\right]
    \rho
    \left[\prod_{\ell=1}^N
    a^\dagger_{B_\ell}(\vk_\ell)\right]
    \widetilde{\ket{\beta'}}\nonumber\\[1em]
    %\label{hard-off-diag}
%\end{align}
%\begin{align}
   %\rho^{\mathrm{hard}}_{\beta, \beta'} 
   &=\sum_{\alpha,\alpha'}f_\alpha f^\ast_{\alpha'}
   \sum_{N=0}^{\infty}\frac{1}{N!}
   \left( \prod_{\ell=1}^N 
  \int_{\text{soft}}\widetilde{d^3k_\ell}\, \eta^{A_\ell B_\ell}_T\right)
    S_{\beta+N\gamma_\text{soft}, \alpha}S^\dagger_{\alpha', \beta'+N\gamma_\text{soft}},
    \label{rhoH_sumN}
\end{align}
where $S_{\beta+N\gamma_\text{soft}, \alpha}$ was defined in
\eqref{eq:SbN+a}.
We have already computed the matrix element $S_{\beta+N\gamma_\text{soft}, \alpha}$ in \eqref{dress-soft-thrm}, 
and can also compute
$S^\dagger_{\alpha', \beta'+N\gamma_\text{soft}}$ in the same way.\footnote{
$S^\dagger_{\alpha', \beta'+N\gamma_\text{soft}}=\left(S_{ \beta'+N\gamma_\text{soft},\alpha'}\right)^{\ast}\,$.}
Taking the sum over $N$ in \eqref{rhoH_sumN}, we obtain the hard reduced density matrix as\footnote{
If we impose that the total energy of soft photon must be less than $\Lambda_s$ by considering the situation where a detector measures multi-photon simultaneously rather than detecting photons one by one, we should insert $\Theta(\Lambda_s-\sum_{l=1}^{N} \omega_{l})$ in \eqref{rhoH_sumN}.
It may give an additional factor in \eqref{rhoH-general} as in \cite{Weinberg:1995mt, Gomez:2018war, Choi:2019rlz}, and it  gives subleading contributions.
}
\begin{align}
\label{rhoH-general}
   \rho^{\mathrm{hard}}_{\beta, \beta'}=&  \sum_{\alpha,\alpha'}f_\alpha f^\ast_{\alpha'}\,
   e^{
   \ins{R^\ast_{\beta,\alpha }-C^\ast_{\alpha}+D^\ast_{\beta}}
   {R_{\beta',\alpha'}-C_{\alpha'}+D_{\beta'}}
   }
   e^{-N_{D_\beta,C_\alpha}-N^\ast_{D_{\beta'},C_{\alpha'}}}
   S^{\mathrm{hard}}_{\beta, \alpha} S^{\mathrm{hard}\dagger}_{\alpha',\beta'}\nn[1em]
   =&\sum_{\alpha,\alpha'}f_\alpha f^\ast_{\alpha'}\,e^{-N_{\alpha,\alpha'}^{\beta,\beta'}}S^{\mathrm{hard}}_{\beta, \alpha} S^{\mathrm{hard}\dagger}_{\alpha',\beta'}.
\end{align}
In the last equality we have defined 
\begin{align}\label{eq:Nfactor}
  N_{\alpha,\alpha'}^{\beta,\beta'}  :=N_{D_\beta,C_\alpha}+N^\ast_{D_{\beta'},C_{\alpha'}}- \ins{\Delta_{D_\beta,C_\alpha}^{\ast}}
   {\Delta_{D_{\beta'},C_{\alpha'}}}
\end{align}
with 
\begin{align}
    \Delta_{D_\beta,C_\alpha}:=R_{\beta,\alpha}-C_{\alpha}+D_{\beta}
    =(R_{\beta}+D_{\beta})-(R_{\alpha}+C_{\alpha})\,.
\end{align}
Since we have $\Re\left[N_{D_\beta,C_\alpha}\right]=\frac{1}{2}\lVert\Delta_{D_\beta,C_\alpha}\rVert_{\mathrm{s}}$ by \eqref{eq:realN}, the real part of \eqref{eq:Nfactor} is given by
\begin{align}\label{eq:RNalbe}
    \Re\left[N_{\alpha,\alpha'}^{\beta,\beta'}\right]
    &=\frac{1}{2}\ins{\Delta_{D_{\beta'},C_{\alpha'}}^{\ast}-\Delta_{D_{\beta},C_{\alpha}}^{\ast}}
   {\Delta_{D_{\beta'},C_{\alpha'}}-\Delta_{D_{\beta},C_{\alpha}}}\nn[1em]
   &=\frac{1}{2}\lVert\left(R_{\beta'}-R_{\beta}+D_{\beta'}-D_{\beta}\right)-\left(R_{\alpha'}-R_{\alpha}+C_{\alpha'}-C_{\alpha}\right)\rVert^2_\mathrm{s}\ (\ge 0)\,,
   \end{align}
   and the imaginary part of \eqref{eq:Nfactor} is given by
\begin{align}\label{eq:INalbe}
   \Im\left[N_{\alpha,\alpha'}^{\beta,\beta'}\right]
    =\Im\left[N_{D_\beta,C_\alpha}+N^\ast_{D_{\beta'},C_{\alpha'}}
    -\ins{\Delta_{D_\beta,C_\alpha}^{\ast}}
   {\Delta_{D_{\beta'},C_{\alpha'}}}\right]\,,
\end{align}
where $\Im\left[N_{D_\beta,C_\alpha}\right]$ was given by \eqref{eq:ImN}. 
%and 
%\begin{align}\label{eq:Imdeldel}
    %\Im\left[\ins{\Delta_{D_\beta,C_\alpha}^{\ast}}{\Delta_{D_{\beta'},C_{\alpha'}}}\right]
   % =\Im\left[(R^{\ast}_{\beta,\alpha},D_{\beta'}-C_{\alpha'})+(D^{\ast}_{\beta}-C^{\ast}_{\alpha},\Delta_{D_{\beta'},C_{\alpha'}})\right]\,.
%\end{align}

\paragraph{Generalized Chung dress}
We introduced the generalized Chung dress in \eqref{gen_Chung} as
\begin{align}\label{eq:gChung}
 &%C^{A}_{\alpha}=-R^{A}_{\alpha}+C^{\mathrm{sub},A}_{\alpha}+C^{A}\quad,\quad
 \bC^{A}_{\alpha}=-R^{A}_{\alpha}+\widetilde{C}^{A}_{\alpha}+C^{A}\quad,\quad
\bD^{A}_{\beta}=-R^{A}_{\beta}+\widetilde{D}^{A}_{\beta}+D^{A}\,.
\end{align}
%In \eqref{eq:gChung}, $\widetilde{C}^{A}_{\alpha}, \widetilde{D}^{A}_{\beta}$ are subleading dresses to Chung's dress, which behave as $\widetilde{C}^{A}_{\alpha}, \widetilde{D}^{A}_{\beta}=o(k^{-1})$ as $k\rightarrow 0$, and $C^{A}, D^{A}$ are arbitrary common dresses in the sense that they are independent of $\alpha$ and $\beta$. 
In the next subsection, we soon see that the above dress is also important for avoiding decoherence of interference effects.

Here we provide  $N_{\alpha,\alpha'}^{\beta,\beta'}$ defined in \eqref{eq:Nfactor} for the generalized Chung dress \eqref{eq:gChung}, which we write as  $\bN_{\alpha,\alpha'}^{\beta,\beta'}$. 
The real part of $\bN_{\alpha,\alpha'}^{\beta,\beta'}$ is given by
\begin{align}\label{eq:ReNcross}
    \Re\left[\bN_{\alpha,\alpha'}^{\beta,\beta'}\right]
   &=\frac{1}{2}\left\lVert\left(\widetilde{D}_{\beta'}-\widetilde{D}_{\beta}\right)-\left(\widetilde{C}_{\alpha'}-\widetilde{C}_{\alpha}\right)\right\rVert^2_\mathrm{s}\ (\ge 0)\,,
   \end{align}
   which is IR finite because $\widetilde{C}^{A}_{\alpha}, \widetilde{D}^{A}_{\beta}=o(k^{-1})$ as $k\rightarrow 0$.
For the generalized Chung dress \eqref{eq:gChung}, each term in $\Im\left[N_{\alpha,\alpha'}^{\beta,\beta'}\right]$ in \eqref{eq:INalbe} is given by plugging $\eqref{eq:gChung}$ into \eqref{eq:ImN} and \eqref{eq:INalbe} as
   \begin{align}   
   \label{eq:ImNDCgChung}
   &\Im\left[N_{\bD_\beta,\bC_\alpha}\right]
   =\Im\left[\ins{R^{\ast}_{\beta}}{\widetilde{D}_{\beta}+D}-\ins{R^{\ast}_{\alpha}}{\widetilde{C}_{\alpha}+C}+\ins{\widetilde{D}^{\ast}_{\beta}+D^\ast}{\widetilde{C}_{\alpha}+C}\right]
   %\\ &\hspace{7em}\quad
     +\Theta_{\beta,\alpha}-\Phi_{\beta,\alpha}\,,\\[1em]
 &\Im\left[\ins{\Delta_{\bD_\beta,\bC_\alpha}^{\ast}}
   {\Delta_{\bD_{\beta'},\bC_{\alpha'}}}\right]
    =\Im\left[\left(\widetilde{D}_{\beta}+D)^{\ast}-(\widetilde{C}_{\alpha}+C)^{\ast}\,,\,(\widetilde{D}_{\beta'}+D)-(\widetilde{C}_{\alpha'}+C)\right)_{\!\mathrm{s}}\right]\,,
    \label{eq:ImdeldelgChung}
\end{align}
where we have again used  $\Im\left[\ins{R^{\ast}_{m}}{R_{n}}\right]=0$ for $m,n=\alpha,\alpha',\beta,\beta'$.
Note that  $\Im\left[\bN_{\alpha,\alpha'}^{\beta,\beta'}\right]$ is also IR finite in the following sense.
For arbitrary functions, $C^{A}(\vk)$ and $D^{A}(\vk)$, the terms in \eqref{eq:ImNDCgChung} and \eqref{eq:ImdeldelgChung} can be IR divergent integrals.
However there is the term of $\Theta_{\beta,\alpha}=\Theta^\rmout_{\beta}-\Theta^\rmin_{\alpha}$ in \eqref{eq:ImNDCgChung}, where $\Theta^\rmout_{\beta},\Theta^\rmin_{\alpha}$  are the pure phases of Fock basis introduced in \eqref{eq:dressdstates}.
We can set $\Theta^\rmout_{\beta}$ and $\Theta^\rmin_{\alpha}$ as any functions of $\beta$ and $\alpha$ respectively because changing each phase of basis states is not relevant to any physical results.
Moreover potentially IR divergent phase terms are factorized into terms only with $\alpha$ index and terms only with $\beta$ index because there is no term having both indices of $\alpha^{(\prime)}$ and $\beta^{(\prime)}$ except for the IR finite parts: $\ins{\widetilde{D}^{\ast}_{\beta}}{\widetilde{D}_{\beta^{\prime}}}, \ins{\widetilde{D}^{\ast}_{\beta}}{\widetilde{C}_{\alpha^{(\prime)}}}, \ins{\widetilde{C}^{\ast}_{\alpha}}{\widetilde{D}_{\beta^{(\prime)}}}$ and $\ins{\widetilde{C}^{\ast}_{\alpha}}{\widetilde{C}_{\alpha^{\prime}}}$.
Therefore we can cancel the IR divergent terms in \eqref{eq:INalbe} by appropriately choosing $\Theta^\rmout_{\beta}$ and $\Theta^\rmin_{\alpha}$ (see Appendix~\ref{app:IRphase} for detail), and we find that
 $ \Im\left[\bN_{\alpha,\alpha'}^{\beta,\beta'}\right]$ in \eqref{eq:INalbe} can be rewritten as 
\begin{align}\label{eq:ImNalbegChung}
    %\Im\left[N_{\alpha,\alpha'}^{\beta,\beta'}\right]=
    \Im\left[\bN_{\alpha,\alpha'}^{\beta,\beta'}\right]
    =&
    \Im\left[\ins{\widetilde{D}^{\ast}_{\beta}}{\widetilde{C}_{\alpha}}-\ins{\widetilde{D}^\ast_{\beta'}}{\widetilde{C}_{\alpha'}}
    \right.\nn
    &\qquad\left.-\ins{\widetilde{D}^{\ast}_{\beta}}{\widetilde{D}_{\beta^{\prime}}}-\ins{\widetilde{C}^{\ast}_{\alpha}}{\widetilde{C}_{\alpha^{\prime}}}+\ins{\widetilde{D}^{\ast}_{\beta}}{\widetilde{C}_{\alpha^{\prime}}}+ \ins{\widetilde{C}^{\ast}_{\alpha}}{\widetilde{D}_{\beta^{\prime}}}\right]\nn[0.5em]
    =&\Im\left[\ins{\wD^{\ast}_{\beta'}}{\wD_{\beta}}+\ins{\wC^{\ast}_{\alpha'}}{\wC_{\alpha}}-\ins{\wD^{\ast}_{\beta'}-\wD^{\ast}_{\beta}}{\wC_{\alpha'}+\wC_{\alpha}}\right]
\end{align}
which obviously gives an IR finite phase.

%%%%%%%%%%%%%%%%%%%%%%%%%%%%%%%%%%%%%%
\subsection{Inclusive cross-sections and the dress code for no decoherence}\label{sebsec:IRdecoh}
In this subsection we study the inclusive cross-sections for general dressed states and show that the generalized Chung dress is special in the sense that it can avoid decoherence of all kinds of interference by IR divergences.

The probability for observing $ \ket{g^{\text{out}}}_{H}$ as outgoing hard particles for incoming state $\ket{f^{\text{in}}}_{GB}$ is given by the following transition probability:
\begin{align}\label{eq:inccross}
    &\sigma^{\mathrm{inc}}(f^{\text{in}}\rightarrow g^{\text{out}})\nn[1em]
    &=\tensor[_H]{\bra{g^{\text{out}}}}{}\rho^{\mathrm{hard}}\ket{g^{\text{out}}}_{H}
    =\sum_{\beta,\beta'}g_{\beta}g_{\beta'}^{\ast}\rho^{\mathrm{hard}}_{\beta\,\beta'}
    =\sum_{\alpha,\alpha'}\sum_{\beta,\beta'}g_{\beta}g_{\beta'}^{\ast}f_\alpha f^\ast_{\alpha'}\,\Gamma^{\beta,\beta'}_{\alpha,\alpha'}\,,
\end{align}
where we have used  \eqref{rhoH-general} and  defined ``the intensity factor'' $\Gamma^{\beta,\beta'}_{\alpha,\alpha'}$ in the final equality as
\begin{align}\label{eq:def_Gamma}
    \Gamma^{\beta,\beta'}_{\alpha,\alpha'}=e^{-N_{\alpha,\alpha'}^{\beta,\beta'}}S^{\mathrm{hard}}_{\beta, \alpha} S^{\mathrm{hard}\dagger}_{\alpha',\beta'}\,.
\end{align}
The above inclusive transition probability can be decomposed into the following four different terms: \vspace{0.5em}
\begin{align}\label{eq:cross_intf}
    \sigma^{\mathrm{inc}}(f^{\text{in}}\rightarrow g^{\text{out}})
    %=\sum_{\alpha,\alpha'}\sum_{\beta,\beta'}g_{\beta}g_{\beta'}^{\ast}f_\alpha f^\ast_{\alpha'}\,\Gamma^{\beta,\beta'}_{\alpha,\alpha'}\nonumber\\[1em]
    &=\underbrace{\sum_{\alpha,\beta}|g_{\beta}|^2|f_\alpha|^2\,\Gamma^{\beta,\beta}_{\alpha,\alpha}}_{\text{diagonal}}
    +\underbrace{\sum_{\alpha\neq\alpha'}\sum_{\beta}|g_{\beta}|^2f_\alpha f^\ast_{\alpha'}\,\Gamma^{\beta,\beta}_{\alpha,\alpha'}}_{\text{initial interference}}
    +\underbrace{\sum_{\alpha}\sum_{\beta\neq\beta'}g_{\beta}g_{\beta'}^{\ast}|f_\alpha|^2 \,\Gamma^{\beta,\beta'}_{\alpha,\alpha}}_{\text{final interference}}\nonumber\\[1em]
    &+\underbrace{\sum_{\alpha\neq\alpha'}\sum_{\beta\neq\beta'}g_{\beta}g_{\beta'}^{\ast}f_\alpha f^\ast_{\alpha'}\,\Gamma^{\beta,\beta'}_{\alpha,\alpha'}}_{ \text{initial/ final mixed interference}}\,.
\end{align}
To obtain a nonzero contribution from each  $\Gamma^{\beta,\beta'}_{\alpha,\alpha'}$ in the sum of \eqref{eq:cross_intf} for fixed $\alpha,\alpha',\beta,\beta'$, the dressing operators must satisfy at least the following dress code:
\begin{align}
\label{condGamma}
    0< \lim_{\lambda\rightarrow 0}\left|\Gamma^{\beta,\beta'}_{\alpha,\alpha'}\right|
    \quad \Rightarrow \quad
    \lim_{\lambda\rightarrow 0}\Re\left[N_{\alpha,\alpha'}^{\beta,\beta'}\right]<\infty\,.
\end{align}
Note that this condition is different from the condition \eqref{eq:Sdresscode} for obtaining IR-safe $S$-matrix elements.
We investigate the dresses satisfying the condition \eqref{condGamma} for each interference term in the following.

\paragraph{Diagonal terms $(\beta=\beta',\alpha=\alpha')$}\,\\
Eqs.~\eqref{eq:RNalbe} and \eqref{eq:INalbe} for $\beta=\beta',\alpha=\alpha'$ give
\begin{align}
     \Re\left[N_{\alpha,\alpha'}^{\beta,\beta'}\right]\Big{|}_{\beta=\beta',\alpha=\alpha'}
     =\Im\left[N_{\alpha,\alpha'}^{\beta,\beta'}\right]\Big{|}_{\beta=\beta',\alpha=\alpha'}
    =0\,,%\quad\Rightarrow\quad 
\end{align}
for {\it any} initial and final dressing operators, $C^{A}_{\alpha}, D^{A}_{\beta}$. 
In this case, $\Gamma_{\alpha,\alpha}^{\beta,\beta}$ defined in \eqref{eq:def_Gamma} is given by
\begin{align}\label{eq:diag_sur}
   \Gamma_{\alpha,\alpha}^{\beta,\beta}=\left|S^{\mathrm{hard}}_{\beta, \alpha}\right|^{2}\,,
\end{align}
which is always IR safe. This case is special in the sense that $ \left|\Gamma_{\alpha,\alpha}^{\beta,\beta}\right|$ is independent of the initial and final dressing operators, $C_{\alpha}, D_{\beta}$. 
In particular, $\left|\Gamma_{\alpha,\alpha}^{\beta,\beta}\right|$ is IR safe even for Fock states, i.e., $C_{\alpha}=D_{\beta}=0$, which is consistent with the result by Bloch-Nordsieck \cite{Bloch:1937pw}. 
This is the reason why the conventional computation using undressed Fock states works well for computing the cross-sections for scatterings from a momentum eigenstate to another momentum eigenstate.

\paragraph{Initial interference terms $(\beta=\beta',\alpha\neq\alpha')$}\,\\
Eq.~\eqref{eq:RNalbe} for $\beta=\beta'$ yields
\begin{align}\label{eq:ReNin}
     \Re\left[N_{\alpha,\alpha'}^{\beta,\beta'}\right]\Big{|}_{\beta=\beta'}
    =\frac{1}{2}\left\lVert R_{\alpha'}-R_{\alpha}+C_{\alpha'}-C_{\alpha}\right\rVert^2_\mathrm{s}\,.
   %= \Re(N_{C_{\alpha},C_{\alpha'}})\,.
\end{align}
Since $C_{\alpha}$ can only depend on $\alpha$, the condition, $ \Re(N_{\alpha,\alpha'}^{\beta,\beta'})\Big{|}_{\beta=\beta'}<\infty$, is solved as
\begin{align}\label{eq:CgenChung}
    C^{A}_{\alpha}
    =\bC^{A}_{\alpha}
    =-R^{A}_{\alpha}
    %+\mathcal{C}^{A}_{\alpha}
    +\widetilde{C}^{A}_{\alpha}+C^{A}\,,\quad D^{A}_\beta= \text{arbitrary}\,,
\end{align}
which means that we need to put the generalized Chung dress \eqref{eq:gChung} on the incoming states to avoid the decoherence of initial interference effects.
Putting the dress \eqref{eq:CgenChung} into the real part \eqref{eq:ReNin} and the imaginary part \eqref{eq:ImNalbegChung},
we then find 
\begin{align}
   \bN_{\alpha,\alpha'}^{\beta,\beta'}\Big{|}_{\beta=\beta'}
   =\frac{1}{2}\left\lVert \widetilde{C}_{\alpha'}-\widetilde{C}_{\alpha}\right\rVert^2_\mathrm{s}
   +\Im\left[\ins{\widetilde{C}^{\ast}_{\alpha'}}{\widetilde{C}_{\alpha}}\right]
   =N_{\mathbf{C}_{\alpha'},\mathbf{C}_{\alpha}}\,,
   \label{eq:NCC'}
\end{align}
where $N_{\bC_{\alpha},\bC_{\alpha'}}$ is given by \eqref{eq:gen_soft_fact}
with the replacement $C_\alpha \to \bC_{\alpha}$, and $D_\beta \to \bC_{\alpha'}$.
The detail of the final equality in \eqref{eq:NCC'} is given in Appendix~\ref{app:IRphase}.
We then find that the intensity factor in \eqref{eq:def_Gamma}  for the initial interference effect is given by 
\begin{align}\label{eq:fdaig_fact}
    \bGamma_{\alpha,\alpha'}^{\beta,\beta'}\Big{|}_{\beta=\beta'}
    =e^{-N_{\bC_{\alpha'},\bC_{\alpha}}}S^{\mathrm{hard}}_{\beta, \alpha} S^{\mathrm{hard}\dagger}_{\alpha',\beta}
    =\bS^{\mathrm{soft}}_{\alpha',\alpha}S^{\mathrm{hard}}_{\beta, \alpha} S^{\mathrm{hard}\dagger}_{\alpha',\beta},
\end{align}
where $\bS^{\mathrm{soft}}_{\alpha',\alpha}:=e^{-N_{\bC_{\alpha'},\bC_{\alpha}}}$ is the soft $S$-matrix   for generalized Chung dressed states for the process $\alpha \to \alpha'$.
Note that $D_\beta$ disappears in \eqref{eq:fdaig_fact}.
It means that the diagonal components of the reduced density matrix, $\rho^{\mathrm{hard}}_{\beta\,\beta}$ in \eqref{eq:inccross}, do not depend on the choice of dresses $D_\beta$ in the basis \eqref{photon-basis_C}. In other words, only the dressings of the initial states are relevant for $\rho^{\mathrm{hard}}_{\beta\,\beta}$. 
In particular, $\rho^{\mathrm{hard}}_{\beta\,\beta}$ can be correctly computed even if we use undressed final states $D_\beta=0$.

We note that  $S^{\mathrm{soft}}_{\alpha',\alpha}$, i.e. the soft $S$-matrix for the process $\alpha \to \alpha'$, generally appears in $\Gamma_{\alpha,\alpha'}^{\beta,\beta'}\Big{|}_{\beta=\beta'}$ for general dress $C_\alpha$ not restricted to the generalized Chung dress $\bC_\alpha$.
This is related to the superselection rule of the asymptotic symmetry as we will comment in section~\ref{sec:comme_resl}.

\paragraph{Final interference  terms $(\beta\neq\beta',\alpha=\alpha')$}\,\\
The same argument as above leads to the following result:
\begin{align}\label{eq:ReNout}
    & \Re\left[N_{\alpha,\alpha'}^{\beta,\beta'}\right]\Big{|}_{\alpha=\alpha'}
    =\frac{1}{2}\lVert R_{\beta'}-R_{\beta}+D_{\beta'}-D_{\beta}\rVert^2_\mathrm{s}
   = \Re(N_{D_{\beta},D_{\beta'}})<\infty\,\\[1em]
 \Rightarrow\quad &  C^{A}= \text{arbitrary},\quad D^{A}_{\beta}=\bD^{A}_{\beta}
 =-R^{A}_{\beta}+\widetilde{D}^{A}_{\beta}+D^{A}\,,\label{eq:DgenChung}
\end{align}
which means that we need to put the generalized Chung dress \eqref{eq:gChung} on the outgoing states to avoid the decoherence of final interference effects.
For the dress \eqref{eq:DgenChung}, we obtain
\begin{align}\label{eq:fin_NDD'1}
   \bN_{\alpha,\alpha'}^{\beta,\beta'}\Big{|}_{\alpha=\alpha'}
   &=\frac{1}{2}\left\lVert \widetilde{D}_{\beta'}-\widetilde{D}_{\beta}\right\rVert^2_\mathrm{s}+i\Im\left[\ins{\widetilde{D}^{\ast}_{\beta^{\prime}}}{\widetilde{D}_{\beta}}-2\ins{\wD^{\ast}_{\beta'}-\wD^{\ast}_{\beta}}{\widetilde{C}_{\alpha}}\right]\\[1em]
   &= N_{\bD_{\beta'},\bD_{\beta}}
     -2i\Im\left[\ins{\wD^{\ast}_{\beta'}-\wD^{\ast}_{\beta}}{\widetilde{C}_{\alpha}}\right]\,,\label{eq:fin_NDD'2}
\end{align}
where the detail of the final equality is given in Appendix~\ref{app:IRphase}.
For the generalized Chung dress, 
we then find that the intensity factor in \eqref{eq:def_Gamma} for the final interference term is given by %for $e^{C_{\alpha'}}\ket{\alpha'}$ to 
\begin{align}\label{eq:fin_int_fac}
    \bGamma_{\alpha,\alpha'}^{\beta,\beta'}\Big{|}_{\alpha=\alpha'}&=e^{-N_{\bD_{\beta'},\bD_{\beta}}-2i\Im\left[\ins{\wD^{\ast}_{\beta'}-\wD^{\ast}_{\beta}}{\widetilde{C}_{\alpha}}\right]}S^{\mathrm{hard}}_{\beta, \alpha} S^{\mathrm{hard}\dagger}_{\alpha',\beta'}\nn[1em] &=e^{-2i\Im\left[\ins{\wD^{\ast}_{\beta'}-\wD^{\ast}_{\beta}}{\widetilde{C}_{\alpha}}\right]}\bS^{\mathrm{soft}}_{\beta',\beta}(\lambda,\Lambda_s)S^{\mathrm{hard}}_{\beta, \alpha} S^{\mathrm{hard}\dagger}_{\alpha,\beta'}\,,
\end{align}
where $\bS^{\mathrm{soft}}_{\beta',\beta}(\lambda,\Lambda_s):=e^{-N_{\bD_{\beta'},\bD_{\beta}}}$ is the soft $S$-matrix element for the generalized Chung dress.
\paragraph{Initial/finial mixed interference terms $(\beta\neq\beta',\alpha\neq\alpha')$}\,\\
Since $C_{\alpha}$ and $D_{\beta}$ can only depend on $\alpha$ and $\beta$ respectively, in the same sense of what we explained around \eqref{eq:fsuperposed},
the condition, $ \Re(N_{\alpha,\alpha'}^{\beta,\beta'})<\infty$ for \eqref{eq:RNalbe}, leads to the following dresses:
\begin{align}\label{eq:gChungIF}
 &C^{A}_{\alpha}=\bC^{A}_{\alpha}=-R^{A}_{\alpha}+\widetilde{C}^{A}_{\alpha}+C^{A}\quad,\quad
D^{A}_{\beta}=\bD^{A}_{\beta}=-R^{A}_{\beta}+\widetilde{D}^{A}_{\beta}+D^{A}\,.
\end{align}
That is,  we need to put the generalized Chung dress \eqref{eq:gChung} on both the incoming state and the outgoing state to avoid the decoherence of initial/final mixed interference effects.
Note that the common parts $C^A$, $D^A$ can be different as $C^A \neq D^A$. This is a difference from the direct computations of $S$-matrix elements where the dress code \eqref{cond:N=0} requires $C^A=D^A$ as \eqref{D_gen_sup}.
We will comment on it in subsec.~\ref{sec:comme_resl}.
For the dress \eqref{eq:gChungIF}, the real and imaginary parts of $N_{\alpha,\alpha'}^{\beta,\beta'}$ are given by \eqref{eq:ReNcross} and  \eqref{eq:ImNalbegChung} respectively, and we obtain the following expression:
\begin{align}
    \bN_{\alpha,\alpha'}^{\beta,\beta'}
   &=N_{\bD_{\beta'},\bD_{\beta}}+N_{\bC_{\alpha'},\bC_{\alpha}}-2i\Im\left[\ins{\wD^{\ast}_{\beta'}-\wD^{\ast}_{\beta}}{\widetilde{C}_{\alpha}}\right]
  -\ins{\wD^{\ast}_{\beta'}-\wD^{\ast}_{\beta}}{\wC_{\alpha'}-\wC_{\alpha}}\,,
\end{align}
where all terms are IR finite. 
We then find that the intensity factor in \eqref{eq:def_Gamma}  is given by 
\begin{align}\label{eq:IF_int_fact}
    \bGamma_{\alpha,\alpha'}^{\beta,\beta'}
    %=e^{-N_{\bC_{\alpha'},\bC_{\alpha}}}S^{\mathrm{hard}}_{\beta, \alpha} S^{\mathrm{hard}\dagger}_{\alpha',\beta}
    =\bS^{\mathrm{soft}}_{\beta',\beta}\bS^{\mathrm{soft}}_{\alpha',\alpha}e^{2i\Im\left[\ins{\wD^{\ast}_{\beta'}-\wD^{\ast}_{\beta}}{\widetilde{C}_{\alpha}}\right]
   +\ins{\wD^{\ast}_{\beta'}-\wD^{\ast}_{\beta}}{\wC_{\alpha'}-\wC_{\alpha}}}
    S^{\mathrm{hard}}_{\beta, \alpha} S^{\mathrm{hard}\dagger}_{\alpha',\beta'}\,
\end{align}
for the generalized Chung dress.
This result reproduces the  intensity factors, \eqref{eq:fdaig_fact} and  \eqref{eq:fin_int_fac}, for initial and final interference effects just by setting $\beta=\beta'$ and $\alpha=\alpha'$, respectively.
\vspace{1em}

Here we summarize the results of the above analysis. 
In the inclusive cross-section, \eqref{eq:cross_intf}, for general superposed incoming and outgoing states, the decoherence of each interference term is related to the choices of initial and final dresses in the following manner:
\begin{itemize}
    \item The diagonal terms survive for arbitrary choice of dressing functions, $C_{\alpha},D_{\beta}$.
    Moreover the diagonal terms are independent of the choice of dress, so we can compute it 
    by using undressed states by setting $C_{\alpha}=D_{\alpha}=0$.
    \item The initial interference terms decohere due to IR divergence except when the initial dresses are chosen as
    $C_{\alpha}=\bC_{\alpha}=-R_{\alpha}+\widetilde{C}_{\alpha}+C$.
    \item The final interference terms decohere due to IR divergence except when the initial dresses are chosen as
    $D_{\beta}=\bD_{\beta}=-R_{\beta}+\widetilde{D}_{\beta}+D$
    \item The initial/final mixed interference terms vanish due to IR divergence except when the initial and final dresses are chosen as $C_{\alpha}=\bC_{\alpha}\ ,\ D_{\beta}=\bD_{\beta}$.
\end{itemize}
Therefore, the important conclusion is that {\it the generalized Chung dress \eqref{eq:gChung}, $C_{\alpha}=\bC_{\alpha}\,\ D_{\beta}=\bD_{\beta}$, is the only type of dress that all infrared divergences are completely canceled in any inclusive cross-sections and keeps all kinds of interference effects.} In particular, if initial/final mixed interference effects are observed in an experiment, the incoming and outgoing charged particles must be dressed by Chung's dress.
For the generalized Chung dress $C_{\alpha}=\bC_{\alpha}\ ,\ D_{\beta}=\bD_{\beta}$,
all the terms in \eqref{eq:cross_intf} can be expressed by the above analysis as 
\begin{align}\label{eq:cross_result}
    \sigma^{\mathrm{inc}}(f^{\text{in}}\rightarrow g^{\text{out}})%\nn[0.5em]
    %=\sum_{\alpha,\alpha'}\sum_{\beta,\beta'}g_{\beta}g_{\beta'}^{\ast}f_\alpha f^\ast_{\alpha'}\,\Gamma^{\beta,\beta'}_{\alpha,\alpha'}\nonumber\\[1em]
    &=\sum_{\alpha,\alpha'}\sum_{\beta,\beta'}g_{\beta}g_{\beta'}^{\ast}f_\alpha f^\ast_{\alpha'}\, S^{\mathrm{hard}}_{\beta, \alpha}\,S^{\mathrm{hard}\dagger}_{\alpha',\beta'}\,
    \bS^{\beta,\beta'}_{\alpha,\alpha'}\,,
    %\nn[0.5em]
    %&\times \bS^{\mathrm{soft}}_{\beta', \beta}\bS^{\mathrm{soft}}_{\alpha',\alpha}\exp{2i\Im\left[\ins{\wD^{\ast}_{\beta'}-\wD^{\ast}_{\beta}}{\widetilde{C}_{\alpha}}\right]
   %+\ins{\wD^{\ast}_{\beta'}-\wD^{\ast}_{\beta}}{\wC_{\alpha'}-\wC_{\alpha}}}\,.
\end{align}
where $\bS^{\beta,\beta'}_{\alpha,\alpha'}:=e^{- \bN_{\alpha,\alpha'}^{\beta,\beta'}}$ is the correction from subleading dresses $\wC_{\alpha},\wD_{\beta}$ given by
\begin{align}
    \bS^{\beta,\beta'}_{\alpha,\alpha'}=\bS^{\mathrm{soft}}_{\beta', \beta}\,\bS^{\mathrm{soft}}_{\alpha',\alpha}\,\exp{2i\Im\left[\ins{\wD^{\ast}_{\beta'}-\wD^{\ast}_{\beta}}{\widetilde{C}_{\alpha}}\right]
   +\ins{\wD^{\ast}_{\beta'}-\wD^{\ast}_{\beta}}{\wC_{\alpha'}-\wC_{\alpha}}}\,.
\end{align}
Thus the generalized Chung dress leads to completely IR safe results for all interference terms in \eqref{eq:cross_intf}.
We can forget about the constant dresses $C, D$ because they do not appear in the final result \eqref{eq:cross_result}.\footnote{The reason for the absence of $C, D$ was that they only appeared in the phases of transition probability and were absorbed into the phases of Fock basis in subsec.~\ref{subsec:Incout}.}

The result \eqref{eq:cross_result} includes the subleading correction $\bS^{\beta,\beta'}_{\alpha,\alpha'}$ from the subleading dresses $\wC_{\alpha},\wD_{\beta}$.
Since setting $\wC_{\alpha}=\wD_{\beta}=0$ leads to $\bS^{\beta,\beta'}_{\alpha,\alpha'}=0$, the leading approximation is just given by
\begin{align}\label{eq:lead_finresl}
    \sigma^{\mathrm{inc}}(f^{\text{in}}\rightarrow g^{\text{out}})
    =\sum_{\alpha,\alpha'}\sum_{\beta,\beta'}g_{\beta}g_{\beta'}^{\ast}f_\alpha f^\ast_{\alpha'}\, S^{\mathrm{hard}}_{\beta, \alpha}S^{\mathrm{hard}\dagger}_{\alpha',\beta'}
    =\left|\sum_{\alpha,\beta}g^\ast_{\beta}f_\alpha S^{\mathrm{hard}}_{\beta,\alpha}\right|^2\,,
\end{align}
which is just the transition probability of  the superposed states of hard particles with the IR cutoff $\Lambda_s$.

%%%%%%%%%%%%%%%%%%%%%%%%%%%%%%%%%%%%%%%%

%%%%%%%%%%%%%%%%%%%%%%%%%%%%%%%%%%%%%%%%
\subsection{Comments on the results}\label{sec:comme_resl}
%Relation to asymptotic symmetry

%%%%%%%%%%%%%%%%%%%%%%%%%%%%%%%%%%%%%%%%%%%%%%%%%%%%%%%%%%

\paragraph{Why undressed Fock states (partially) works well}\ \\
If the leading dress is not Chung's dress as $C^{A}_{\alpha}\neq-R^{A}_{\alpha}, D^{A}_{\beta}\neq-R^{A}_{\beta}$, the interference terms in  \eqref{eq:cross_intf} (or \eqref{eq:lead_finresl}) vanish by the infrared divergence as explained in subsec.~\ref{sebsec:IRdecoh}, and only the diagonal terms survive.
The important point is that the diagonal terms can be computed correctly even by undressed Fock states because the diagonal terms are independent of choices of dresses as explained around eq.~\eqref{eq:diag_sur}.
If we consider the scattering from a momentum eigenstate $\ket{\alpha}_{0}$ to another momentum eigenstate $\ket{\beta}_{0}$, the inclusive transition probability $\sigma^{\mathrm{inc}}(\alpha\ra\beta)$ is given only by the diagonal terms $\Gamma_{\alpha,\alpha}^{\beta,\beta}=\left|S^{\mathrm{hard}}_{\beta, \alpha}\right|^{2}$ in \eqref{eq:cross_intf}.
The result $\sigma^{\mathrm{inc}}(\alpha\ra\beta)=\left|S^{\mathrm{hard}}_{\beta, \alpha}\right|^{2}$ was obtained  for Chung (Kulish-Faddeev) dressed states in  \cite{Carney:2018ygh,Gomez:2018war,Choi:2019rlz}.
We have seen that this result is not limited to Chung's dress, and it holds for any dresses.
This is the reason why the standard computation using undressed states, where we only focus on $\sigma^{\mathrm{inc}}(\alpha\ra\beta)$ in most cases, works well.
However once we consider cross-sections for superposed states of different momentum eigenstates in  the undressed formalism, all the interference effects are set to zero by infrared divergence and also the zero-measure problem appears in the case of wave packets which we will consider in subsec.~\ref{subsec:resl}.
%The difference would be crucial when we consider spatially localized states as initial or final states.
%\begin{align}
    %\sigma^{\mathrm{inc}}(f^{\text{in}}\rightarrow g^{\text{out}})=\sum_{\alpha,\beta}|g_{\beta}|^2|f(\alpha)^2|\left|S^{\mathrm{hard}}_{\beta, \alpha}\right|^2\,.
%\end{align}
%The decoherence is related to the conservation law of the charge of the asymptotic symmetry as explained in the next.

\paragraph{Decoherence and asymptotic symmetry}\ \\
Here we point out the relation between  the asymptotic symmetry and the appearance of $S^{\mathrm{soft}}_{\alpha',\alpha}$ in the initial interference term  \eqref{eq:fdaig_fact}.
Although the result \eqref{eq:fdaig_fact} has been written down after putting the generalized Chung dress on hard states,
 we can show that the same relation holds even for arbitrary dresses $C_{\alpha}^{A}, D_{\beta}^{A}$ as
\begin{align}\label{eq:appr_softS}
    \Gamma_{\alpha,\alpha'}^{\beta,\beta'}\Big{|}_{\beta=\beta'}
    %=e^{-\bN_{C_{\alpha'},C_{\alpha}}}S^{\mathrm{hard}}_{\beta, \alpha} S^{\mathrm{hard}\dagger}_{\alpha',\beta}
    =S^{\mathrm{soft}}_{\alpha',\alpha}(\lambda,\Lambda_s)S^{\mathrm{hard}}_{\beta, \alpha} S^{\mathrm{hard}\dagger}_{\alpha',\beta}\,,
\end{align}
where $S^{\mathrm{soft}}_{\alpha',\alpha}(\lambda,\Lambda_s)=e^{-N_{C_{\alpha},C_{\alpha'}}}$ defined in \eqref{eq:softSmat} and $N_{C_{\alpha},C_{\alpha'}}$ is given by \eqref{eq:gen_soft_fact}
with the replacement, $\beta\to \alpha'$ and $D_\beta \to C_{\alpha'}$.
Then the diagonal component of the outgoing density matrix (= transition probability of $\ket{f^{\rmin}}\to \ket{\beta}$) is computed using \eqref{eq:appr_softS} as
\begin{align}
    \rho^{\mathrm{hard}}_{\beta, \beta}
    =\sum_{\alpha,\alpha'}f_\alpha f^\ast_{\alpha'}\,\Gamma^{\beta,\beta}_{\alpha,\alpha'}
    =\sum_{\alpha,\alpha'}f_\alpha f^\ast_{\alpha'}\,
    S^{\mathrm{soft}}_{\alpha',\alpha}(\lambda,\Lambda_s)S^{\mathrm{hard}}_{\beta, \alpha} S^{\mathrm{hard}\dagger}_{\alpha',\beta}\,.\label{eq:gen_iniS}
\end{align}
In section~\ref{subsec:gene-dress-code}, we saw that $S^{\mathrm{soft}}_{\alpha',\alpha}$ takes non-vanishing values only when $\alpha$ and $\alpha'$ have the same asymptotic charges. Thus, the appearance of $S^{\mathrm{soft}}_{\alpha',\alpha}$ in \eqref{eq:gen_iniS} reveals the connection between the decoherence and the asymptotic symmetry as we further discuss below.

If the initial states are undressed ($C_\alpha=0$ for all $\alpha$) as in the conventional inclusive computations,  $S^{\mathrm{soft}}_{\alpha',\alpha}$  becomes
\begin{align}
  S^{\mathrm{soft}}_{\alpha',\alpha}(\lambda,\Lambda_s) =e^{-\frac{1}{2}
  \ins{R_{\alpha'}-R_{\alpha}}{R_{\alpha'}-R_{\alpha}}
     },
\end{align}
which is the same as the standard IR problematic factor for the scattering between undressed states $\alpha \to \alpha'$. This factor vanishes in the limit removing the IR regulator $\lambda$ except for the cases $\alpha=\alpha'$\footnote{This equality means that only the information of hard charged particles is identical as explained below \eqref{undressN}. If we have states $\ket{\alpha}$ and  $\ket{\alpha'}$ such that they are different only in the hard photon sector, then $ S^{\mathrm{soft}}_{\alpha',\alpha}\neq 0$.} by
\begin{align}
\label{kronecker-delta}
 \lim_{\lambda \to 0}S^{\mathrm{soft}}_{\alpha',\alpha} =\delta^\text{hard charge}_{\alpha',\alpha},
\end{align}
where $\delta^\text{hard charge}_{\alpha',\alpha}$ represents a Kronecker delta such that it takes 1 if the information of hard charged particles $(e_n, p_n)_{n\in \alpha}$ agrees with $(e_n, p_n)_{n\in \alpha'}$ and otherwise zero (see \cite{Carney:2017jut} for details).
Thus, we find that only the identical terms $\alpha=\alpha'$ survive in the sum in \eqref{eq:gen_iniS} as
\begin{align}
\label{undress-no-diag}
   \rho^{\mathrm{hard}}_{\beta, \beta}
   =& \sum_{\alpha}|f_\alpha|^2\,
 |S^{\mathrm{hard}}_{\beta, \alpha} |^2.
\end{align}
This means the decoherence because the interference terms between $\alpha$ and $\alpha'\, (\alpha\neq \alpha')$ do not contribute, and the result is independent of  the relative phase between $\alpha, \alpha'$ existing in the initial state \eqref{superposed_in}.
The decoherence of initial interference for undressed superposed states is caused by the appearance of $\delta^\text{hard charge}_{\alpha',\alpha}$ as \eqref{kronecker-delta}.
This is a consequence of the conservation law of the asymptotic symmetry in the following sense. 
For undressed states, we do not have any soft radiation, which means that there are no soft charges. 
Then $\alpha$ and $\alpha'$  must have the same hard charges to belong to the same superselection sector of the asymptotic charges.
This means that a local electric current for state $\alpha$ agrees with that for state $\alpha'$ at any angle on the celestial sphere \cite{Carney:2018ygh}.  This indicates that $\alpha$ and  $\alpha'$ are the same hard state (up to neutral particles), that is, we have $\delta^\text{hard charge}_{\alpha',\alpha}$.

Even if we use dressed initial states, such a decoherence may occur. Indeed, we have  
\begin{align}
 \lim_{\lambda \to 0}S^{\mathrm{soft}}_{\alpha',\alpha} = 0
\end{align}
when the dressed states $\ket{\alpha}_{GB}, \ket{\alpha'}_{GB}$ belong to different sectors with respect to the asymptotic symmetry as we argued in subsec.~\ref{subsec:gene-dress-code}. 
We have to use appropriate dresses so that the dress code \eqref{cond:N=0}, in this case $C_\alpha^A-C_{\alpha'}^A =R_{\alpha',\alpha}^A+o(k^{-1})\,$, is satisfied in order to have non-vanishing $S^{\mathrm{soft}}_{\alpha',\alpha}$.
The initial interference term between $\alpha$ and $\alpha'$ does not decohere only when $\alpha$ and $\alpha'$ belong to the same superselection sector of the asymptotic symmetry. 
In other words, the superselection rule for the asymptotic symmetry holds for the initial interference even in the inclusive computations as similar to the $S$-matrix computations which do not involve the trace-out of soft photons.
In particular, if we want to keep all the initial interference terms in \eqref{eq:gen_iniS}, we have to put the (generalized) Chung's dress on hard states, as explained in subsec.~\ref{sebsec:IRdecoh}.
We will give a further comment on the superselection rule below.

Similarly, the final interference between the outgoing states $\beta, \beta'$ does not decohere only when they belong to the same sector of the asymptotic symmetry. 
This condition is also satisfied for (generalized) Chung's dress as explained in subsec.~\ref{sebsec:IRdecoh}.

%%%%%%%%%%%%%%%%%%%%%%%%%%%%%%%%%%%%%%%%%%%%%%%%%%%%%%%%%%
\paragraph{Subtlety of superselection rule in the inclusive computations}\ \\
It may sound trivial that the superselection rule for the asymptotic symmetry holds even for the inclusive computations.
However, we should be careful because we encounter the following problem.
If the superselection rule for the asymptotic charges holds, 
we naively expect that the conventional inclusive cross-section using the undressed Fock states is always zero except for the trivial process,
because the transition between different Fock states is not allowed by the conservation law of the asymptotic charges. 
On the other hand, we know that parts of the inclusive cross-section can  be nonzero even if we use Fock states. 
This is due to the order of the limit. 
As usual, we first introduce the IR regulator $\lambda$ which breaks the exact conservation law of the asymptotic charges, and take the sum over infinite soft photons, and finally remove the regulator $\lambda$. 
When we trace out soft photons, the information of soft parts of the final asymptotic charges might be lost. 
Hence we cannot trust the conservation law of the asymptotic charges in the order of the limit used in the inclusive computations, and we actually obtain the nonzero inclusive cross-section even for undressed states. 
Thus, it is dangerous to apply the superselection rule to the inclusive computations. 

We can also find a ``violation" of the conservation law of the asymptotic charges in our result for the initial/final mixed interference effect. 
We showed that if we use the generalized Chung dress \eqref{eq:gChungIF}, we obtain  non-zero cross-section. Below \eqref{eq:gChungIF}, we commented that $C^A$ can be different from $D^A$ unlike the direct computations of $S$-matrix elements $S_{\beta,\alpha}$. 
The difference $C^A\neq D^A$ means that the incoming states $\alpha$ have the different asymptotic charges from the outgoing states $\beta$.
This is a ``violation" of the conservation law in the inclusive computations.\footnote{Of course, this does not imply the true violation of the asymptotic symmetry. It just means that we cannot determine the precise values of the asymptotic charges when we cannot observe soft photons and we consider only inclusive quantities.}
That is, if we trace-out the soft photons, we lose the information of the relative shift of the asymptotic charges between the initial and final states.

Nevertheless, our result \eqref{eq:gen_iniS} concludes that we cannot still observe the interference among initial states in different sectors with respect to the asymptotic charges.
Similarly, we cannot observe the final interference in different sectors. 
Therefore, states in different sectors cannot interfere with each other. 
In this sense, a kind of the superselection rule still holds even for the inclusive computations.

%%%%%%%%%%%%%%%%%%%%%%%%%%%%%%%%%%%%%%%%
\subsection{Resolution of the zero-measure problem by dressed states}\label{subsec:resl}

A subtle but potentially critical problem about cross-sections for incoming wave packets was pointed out in \cite{Carney:2018ygh}.
Here we first review the problem  for undressed states and see that the problem can be avoided if we use proper dressed states, based on \cite{Carney:2018ygh}. 
Then we generalize the discussion to the case that both initial and final states are wave packets.
We show that we must put the generalized Chung dresses on both the initial and final wave packets to resolve  the zero-measure problem.
The definition of the reduced density matrix of hard sector \eqref{def:rhoA} plays a critical role in this resolution.

\paragraph{Resolving the zero-measure problem for initial wave packets}\ \\
Wave packets can be written as a continuous superposition of momentum eigenstates as
\begin{align}
 \ket{f}_{GB}=   \int d\alpha f(\alpha)\ket{\alpha}_{GB}.
\end{align}
This is just a continuous analog of \eqref{superposed_inGB}. 
Then the inclusive cross-section from this initial wave packet to a hard state $\beta$ is given by \eqref{eq:gen_iniS}  with the replacement $\Sigma\ra\int$ as
\begin{align}
     \rho^{\mathrm{hard}}_{\beta, \beta}
   =&  \int d\alpha  \int d\alpha'  f(\alpha) f^\ast(\alpha')  S^{\mathrm{soft}}_{\alpha',\alpha}
 S^{\mathrm{hard}}_{\beta, \alpha} S^{\mathrm{hard}\dagger}_{\alpha',\beta}\,.
 \label{hard_red_cont}
\end{align}
However, we encounter the following  zero-measure problem for the continuous case. 
If each $\ket{\alpha}_{GB}$ is an undressed state as $\ket{\alpha}_{GB}=\ket{\alpha}_{0}$, the soft factor $S^{\mathrm{soft}}_{\alpha',\alpha}$ is the Kronecker delta $\delta_{\alpha,\alpha'}^\text{hard charge}$ as \eqref{kronecker-delta}, not the delta function $\delta(\alpha, \alpha')$.  Then, plugging $S^{\mathrm{hard}}_{\beta, \alpha}=\delta(\beta-\alpha)-2\pi i T_{\beta,\alpha}$, where $\delta(\beta-\alpha)$ corresponds to the trivial (no scattering) process, into  \eqref{hard_red_cont}, we obtain
\begin{align}
     \rho^{\mathrm{hard}}_{\beta, \beta}
   &=  \int d\alpha  \int d\alpha'  f(\alpha) f^\ast(\alpha')  \delta_{\alpha,\alpha'}^\text{hard charge}
 \left(\delta(\beta-\alpha)-2\pi i T_{\beta,\alpha} \right) \left(\delta(\beta-\alpha')+2\pi i T^{\dagger}_{\beta,\alpha'} \right)\nn[0.5em]
 &=|f(\beta)|^2\,,\label{eq:Focknoscat}
\end{align}
where we have used that the integral $\int d\alpha  \int d\alpha'  f(\alpha) f^\ast(\alpha')  \delta_{\alpha,\alpha'}^\text{hard charge}T_{\beta,\alpha}T^{\dagger}_{\beta,\alpha'} $ vanishes because $T_{\beta,\alpha}$ is supposed to be a non-singular function
\footnote{Strictly speaking, $T_{\beta,\alpha}$ contains a delta function of the total momentum conservation.
However, it is not singular enough to save the integrals from the zero-measure problem because the number of integration variables is greater than the number of delta functions, as argued in \cite{Carney:2018ygh}.}
and the integrand is only supported on a zero measure set $\alpha=\alpha'$.
The result \eqref{eq:Focknoscat} means that not only the initial interference effects vanish but also any non-trivial  scatterings do not occur for undressed wave packets  \cite{Carney:2018ygh}.

This problem stems from the fact that we have used the wave packets composed of undressed states. 
However, we cannot construct such wave packets 
by the statement ($\ast$) in subsec.~\ref{sec:sel_assym} 
because undressed states with different hard momenta belong to different superselection sectors. 
We should use dressed states in $\int d\alpha f(\alpha)\ket{\alpha}_{GB}$ so that all 
$\ket{\alpha}_{GB}$ in this integral have the same asymptotic charges as \eqref{C_gen_sup}.
If we use the generalized Chung dressed states for all $\ket{\alpha}$, we have
$S^{\mathrm{soft}}_{\alpha',\alpha}=\bS^{\mathrm{soft}}_{\alpha',\alpha}=e^{-N_{\bC_{\alpha},\bC_{\alpha'}}}$ where $N_{\bC_{\alpha},\bC_{\alpha'}}$ is given by \eqref{eq:NCC'}.
Since $\bS^{\mathrm{soft}}_{\alpha',\alpha}$ is not a Kronecker delta, the zero-measure problem disappears.
In particular, $\bS^{\mathrm{soft}}_{\alpha',\alpha}= 1$ for the leading order for any $\alpha, \alpha'$.
Thus we need dressed states to construct the wave packets that give non-trivial cross-sections and this is an evidence of the necessity of using dressed states as argued in \cite{Carney:2018ygh}.

\paragraph{Resolving the zero-measure problem for initial/final wave packets}\ \\
Here we consider the zero-measure problem when both initial and final states are wave packets and see that the problem also disappears if we use the proper dressed states. We then compare our result with the one by \cite{Carney:2018ygh}.

The transition probability from an initial wave packet to a finial wave packet in general is given by \eqref{eq:inccross} with the replacement $\Sigma \ra \int$. %$\sum_{\alpha,\alpha'}\sum_{\beta,\beta'}\ra \int\!d\alpha\int\!d\alpha'\int\!d\beta\int\!d\beta'$. %and  we need to use dressed states to avoid the decoherence as argued in Subsec \ref{sebsec:IRdecoh}.
If we use undressed states for initial and final states, the intensity factor \eqref{eq:def_Gamma}  can be evaluated as
\begin{align}
    \left|\Gamma^{\beta,\beta'}_{\alpha,\alpha'}\right|%=\left|e^{-N_{\alpha,\alpha'}^{\beta,\beta'}}S^{\mathrm{hard}}_{\beta, \alpha} S^{\mathrm{hard}\dagger}_{\alpha',\beta'}\right|
    &\propto \lim_{\lambda\ra 0}\exp\left(-\Re\left[N_{\alpha,\alpha'}^{\beta,\beta'}\right]\right)\nn[0.5em]
    &=\lim_{\lambda\ra 0}\exp\left(-\frac{1}{2}\lVert\left(R_{\beta'}+R_{\alpha}\right)-\left(R_{\alpha'}+R_{\beta}\right)\rVert^2_\mathrm{s}\right)
    =\delta^\text{hard charge}_{\beta'\cup\alpha,\,\beta\cup\alpha'},
\end{align}
where we have used \eqref{eq:RNalbe} in the first equality and \eqref{kronecker-delta} in the second equality.
Again, the Kronecker-delta causes the zero-measure problem because only the trivial term, $\delta^\text{hard charge}_{\beta'\cup\alpha,\,\beta\cup\alpha'}\delta(\alpha-\beta)\delta(\alpha'-\beta')$, survives in  \eqref{eq:inccross}.
Thus any non-trivial scattering processes  are prohibited for undressed wave packets due to the infrared divergence, which is the same result as in \cite{Carney:2018ygh}.

This problem is also resolved by using suitable dressed states. If we put the generalized Chung dress on initial and final wave packets, the transition probability from an initial wave packet to a final wave packet in general is given by \eqref{eq:cross_result} with the replacement $\sum\ra \int$, which is free from the zero-measure problem.
In particular, the transition probability for leading Chung's dress $C^{A}_{\alpha}=-R^{A}_{\alpha}, D^{A}_{\beta}=-R^{A}_{\beta}$ is given by \eqref{eq:lead_finresl} with $\sum \ra \int$:
\begin{align}\label{eq:ourcross_Ch}
      \sigma^{\mathrm{inc}}(f^{\text{in}}\rightarrow g^{\text{out}})
    =\int\!d\alpha\!\int\!d\alpha'\!\int\!d\beta\!\int\!d\beta'\,g(\beta)g(\beta')f^{\ast}(\alpha) f^\ast(\alpha')\, S^{\mathrm{hard}}_{\beta, \alpha}S^{\mathrm{hard}\dagger}_{\alpha',\beta'}\,.
\end{align}
The result \eqref{eq:ourcross_Ch} keeps all kinds of interference effects.

On the other hand, the result of the hard density matrix for Chung's dress in \cite{Carney:2018ygh}, where the definition of the tracing out soft photons is different from ours as explained in subsec.~\ref{sec:def-partial_trace}, is given by\footnote{To rewrite Eq.\eqref{eq:Car_resl}  back in the notation of \cite{Carney:2018ygh}, we need the replacement,
$S^{\text{hard}}_{\beta,\alpha}\ra\mathbb{S}_{\beta,\alpha}$,  $\delta^\text{hard charge}_{\beta',\,\beta}\ra\langle W_{\beta}^{\dagger}W_{\beta'}\rangle$.
}
\begin{align}\label{eq:Car_resl}
    \rho^{\mathrm{hard}}_{\beta,\beta'}=\int\!d\alpha\int\!d\alpha'\, f(\alpha)f^{\ast}(\alpha')S^{\text{hard}}_{\beta,\alpha}S^{\text{hard}\dagger}_{\alpha',\beta'}\delta^\text{hard charge}_{\beta',\,\beta}\,.
\end{align}
If we use the density matrix  in the transition probability,
\begin{align}
    \sigma^{\mathrm{inc}}(f^{\text{in}}\rightarrow g^{\text{out}})
     =\int\!d\beta\!\int\!d\beta'\,g(\beta)g(\beta')\rho^{\mathrm{hard}}_{\beta,\beta'}\,,
\end{align}
 the off-diagonal components of $\rho_{\beta,\beta'}$ vanish due to the Kronecker delta in \eqref{eq:Car_resl}.
 This means that final (and initial/final) interference effects in \eqref{eq:cross_intf} vanish for  scatterings of wave packets dressed by Chung's one if we use a naive definition of the trace over soft photons \eqref{eq:naive_tr}.
 This is different from our result \eqref{eq:ourcross_Ch} in which all kinds of interference effects occur.
Moreover the Kronecker delta \eqref{eq:Car_resl} causes  the zero measure problem in $\beta,\beta'$ integrals, and then no scattering is concluded.
%Even if we ignore the problem, the off-diagonal components of $\rho_{\beta,\beta'}$ is zero
%\begin{align}
%\hspace{-1em}
     %\sigma^{\mathrm{inc}}(f^{\text{in}}\rightarrow g^{\text{out}})
     %&=\int\!d\beta\!\int\!d\beta'\,g(\beta)g(\beta')\rho^{\mathrm{hard}}_{\beta,\beta'}\nn[0.5em]
    %&=\int\!d\alpha\!\int\!d\alpha'\!\int\!d\beta\!\int\!d\beta'\,g(\beta)g(\beta')f^{\ast}(\alpha) f^\ast(\alpha')S_{\beta,\alpha}S^{\ast}_{\beta',\alpha'}\delta^\text{hard charge}_{\beta',\,\beta}.
%\end{align}
This should be not correct because in  realistic experiments we can measure only cross-sections for  scatterings of wave packets, not the strict  momentum eigenstates, and we observe nontrivial scatterings.
Our result does not lead to this problem as explained above, and seems to be more reasonable.
The difference originates from our definition of a reduced density matrix based on a superselection structure of the Hilbert space, which is explained in subsec.~\ref{sec:def-partial_trace}.
The result supports our definition of the reduced density matrix \eqref{def:rhoA}.

%Thus, we think that our result is more natural.

%%%%%%%%%%%%%%%%%%%%%%%%%%%%%%%%%%%%%%%%%%%%%%%%%%
\section{Discussions}\label{sec:discuss}

\paragraph{Application of dressed states}
We have seen that dressed states are needed to obtain IR-safe (i.e., non-vanishing in the limit removing the IR cutoff $\lambda\to 0$) $S$-matrix elements. 
They are also necessary even for  the inclusive computations which only observe hard particles,  because without appropriate dresses we cannot see correct interference between hard particles.
Thus, we should investigate the technique of computing various quantities in the dressed state formalism.
In particular, dressing is a necessary treatment when we consider wave packets. 
We need the generalized Chung dresses so that each momentum state in wave packets has the same total asymptotic charges. 

We should also understand the physical meanings of the generalized Chung states \eqref{gen_Chung}.
We have seen that the generalized Chung dressed states are special in the sense that they are necessary to compose superposed states and also to avoid the decoherence and the zero-measure problem.
In \cite{Arkani-Hamed:2020gyp}, it is argued that Chung's dress\footnote{More precisely, the natural dress in \cite{Arkani-Hamed:2020gyp} is the one obtained by enlarging $k$-integral in Chung's dress  \eqref{eq:Chung-dress} to the entire momentum not restricted to the soft region. In our analysis, we have a freedom to add such hard dresses because hard photons are not relevant to IR behaviors.} is also a natural one for celestial amplitudes. 
It sounds interesting to understand this argument independently from the celestial amplitudes.

%\paragraph{Cancellation of IR divergent phase factors}\cite{Laddha:2018myi}

\paragraph{Subleading corrections}
We have seen that $S$-matrix elements between  appropriate dressed states are IR-safe.
Although our analysis is enough to see whether the $S$-matrix elements are IR-safe or not, we need to perform a more careful analysis to evaluate the precise values of the $S$-matrix elements. 
For example, the IR-safe result
\eqref{N-fin_Smat} is precisely written as
\begin{align}
    S_{\beta, \alpha}
    = e^{-N_{\bD_\beta,\bC_\alpha}(\lambda,\Lambda_s)}
    S^{\mathrm{hard}}_{\beta, \alpha}(\Lambda_s)[1+(\text{subleading})]
\end{align}
like \eqref{eq:general_factor}.
Because the subleading term which we ignore in our analysis is also finite, it would be the same order as the finite soft factor $e^{-N_{\bD_\beta,\bC_\alpha}}$.
As written above \eqref{leading:general_factor}, the subleading term comes from the subleading corrections in the large time expansion \eqref{eq:asexcurt} and also the subleading corrections to the soft photon theorem.
It is interesting to evaluate these corrections in a simple setup, and we leave it to the future work.
It is also interesting to study how to fix the subleading dresses $\wC_\alpha, \wD_\beta$ in the generalized Chung dress which appear in $e^{-N_{\bD_\beta,\bC_\alpha}}$. The results in \cite{Choi:2019rlz} might be helpful for this future work.
Moreover we have implicitly assumed that the upper energy scale of soft photons contained in the dresses is the same as the energy resolution scale of our detector.
However it may be better to set the two scales differently as discussed in \cite{Gomez:2018war}, and it may also give some subleading corrections.

\paragraph{Subleading soft theorem}
The conventional leading  soft photon theorem seems to say that $S$-matrix elements containing an external soft photon with energy $\omega$ are singular as $1/\omega$. 
One might think that it contradicts the unitarity constraint. 
However, as we have seen, we always have  \eqref{soft-theorem=0}, and the unitarity constraint is  kept at least at the leading $1/\omega$ order. 
In \cite{Laddha:2018myi, Sahoo:2018lxl, Saha:2019tub}, it is argued that there are subleading $\mathcal{O}(\log \omega)$ corrections to the soft theorem. 
The corrections also seem to violate the unitarity constraint.
Thus, the $\mathcal{O}(\log \omega)$ singularity also should be absent as  
\begin{align}
\label{absence-log}
   \lim_{\lambda \to 0}S_{\beta, \alpha+\gamma_\text{soft}}
   =\lim_{\lambda \to 0}S_{\beta+\gamma_\text{soft},\alpha}
   =o(\log \omega)
\end{align}
for any dressed (and also undressed) states $\alpha, \beta$.
The non-existence of the $1/\omega$ singularity is related to the asymptotic charge conservation, which means that $S$-matrix elements vanish unless the condition \eqref{cond:N=0} is satisfied.
Similarly, we expect that the non-existence of the $\log \omega$ singularity is related to the subleading asymptotic charge conservation derived in \cite{Hirai:2018ijc}, where it was shown that subleading behaviors\footnote{The subleading terms are roughly $(\log r)/r$ corrections to the dominant $1/r$ terms at the asymptotic region.} of electromagnetic fields should satisfy a constraint to maintain the asymptotic charge conservation.

So far we have discussed that adding a single soft photon does not produce a  singular change of $S$-matrix elements.
Furthermore, it might be natural to speculate that a finite number of soft photons do not change $S$-matrix elements in the limit that the energy of soft photons is strictly zero.
That is, we conjecture the following equation with any finite number $N$ 
\begin{align}
\label{conjecture}
    \lim_{\omega\to 0} S_{\beta, \alpha+N\gamma_\text{soft}}
    =\lim_{\omega\to 0} S_{\beta+N\gamma_\text{soft}, \alpha}=S_{\beta, \alpha} \qquad \text{in the limit $\lambda \to 0$},
\end{align}
for any states $\alpha$ and $\beta$. 
In order to check it, we have to investigate the $\mathcal{O}(\omega^{0})$ corrections to the soft theorem. 
Such $\mathcal{O}(\omega^{0})$ corrections for undressed states are given by Low's soft photon theorem \cite{Low:1954kd, Low:1958sn, Burnett:1967km, GellMann:1954kc} (see also \cite{Sahoo:2020ryf} for  another sub-subleading proposal), which is also related to the asymptotic symmetry \cite{Lysov:2014csa, Campiglia:2016hvg, Conde:2016csj, Hirai:2018ijc}. 
It is not clear now whether \eqref{conjecture} is correct or not, because $\mathcal{O}(\omega^{0})$ terms do not conflict with the unitarity constraint. 
To check \eqref{absence-log} and \eqref{conjecture}, we have to develop the subleading soft photon theorem for general dressed states.

\paragraph{Averaging over the initial  soft sector}
In section~\ref{sec:reduced}, we have considered an inclusive process because 
we cannot distinguish final soft photons by realistic detectors. 
We also do not have the ability to prepare the desired initial dressed states because we do not have the fine resolution to distinguish soft dresses.
It implies that we should also take the average over the soft sector of the initial states
to compare experimental results.
We know that the average does not play any role in the conventional inclusive computations because of the IR factorization of QED.
Schematically, if we decouple the hard and soft sectors,  the average of the initial soft photons and the summation of the final ones can be written as
\begin{align}
    \frac{1}{\mathrm{dim}\mathcal{H}_\text{soft}}\sum_{i:\text{soft}}\sum_{f:\text{soft}}|\bra{f}S^\text{soft} \ket{i}|^2 S^\text{hard}_{\beta,\alpha}S^{\text{hard} \dagger}_{\alpha',\beta},
\end{align}
where $\mathrm{dim}\mathcal{H}_\text{soft}$ is a formal dimension of the soft sector given by $\mathrm{dim}\mathcal{H}_\text{soft}=\sum_{i:\text{soft}}$.
The soft part can be computed schematically by the completeness relation in the soft sector as
\begin{align*}
    &\frac{1}{\mathrm{dim}\mathcal{H}_\text{soft}}\sum_{i:\text{soft}}\sum_{f:\text{soft}}|\bra{f}S^\text{soft} \ket{i}|^2
    \nn
    &=\frac{1}{\mathrm{dim}\mathcal{H}_\text{soft}}\sum_{i:\text{soft}}\bra{i}S^{\text{soft}\dagger} S^\text{soft}\ket{i}
    =\frac{1}{\mathrm{dim}\mathcal{H}_\text{soft}}\sum_{i:\text{soft}}1=1.
\end{align*}
Thus, we only have the transition amplitude for hard sectors as in the case without the average.
For general dressed states, it could be expected that the average of initial soft photons gives us the same result as that without the average.

\paragraph{Collinear divergences}
In massless QED, collinear divergences arise from on-shell and off-shell photons that are emitted in a direction parallel to the momentum of one of external hard electrons. 
Both soft and hard collinear photons give such divergences.
The divergences arising from soft collinear photons can be removed by generalized Chung's dress because the dress cancels all contributions from soft photons in the inclusive cross-section as shown in section \ref{sebsec:IRdecoh}.
 We need  further considerations to confirm whether collinear divergence caused by hard collinear photons can also be eliminated by using appropriate hard dressing. 
The difficulty of dressed states for massless QED is also discussed recently in ref.\cite{Prabhu:2022mcj}.

\paragraph{Extension to gravitons}
One of the interesting future directions is extending our analysis to gravity.
IR divergences by soft gravitons are almost the same as those by soft photons. 
Hence, we also need dressed states to obtain non-vanishing $S$-matrix elements in gravity \cite{Ware:2013zja, Choi:2017bna, Choi:2017ylo, Wilson-Gerow:2018egh, Choi:2019rlz}. 
At least for the linearized gravity, the dressed states are given by coherent states of soft gravitons as in QED.
It is interesting to see whether the simple coherent states are enough even in nonlinear gravity to obtain  IR-safe $S$-matrix elements.
As we have seen that our condition for dressed states, \eqref{cond:N=0}, to have
non-vanishing $S$-matrix elements can be interpreted as the memory effect in QED.
We expect that this is also the case for nonlinear gravity. 
Thorne showed in \cite{Thorne:1992sdb} that nonlinear gravitational memory can be written as linear memory by including hard gravitons into matter fields. 
This implies that the change of soft graviton  dresses (corresponding to $D_\beta-C_\alpha$ in our QED case) is fixed only by hard information.

We finally remark on the superselection rule in quantum gravity.
As in QED, gravitational theories also have the asymptotic symmetry and the associated superselection rule. 
This indicates that we cannot observe any interference effects among different superselection sectors. 
Recently it has been proposed to observe some decoherence effects induced by gravitons to establish the quantum nature of gravity \cite{Parikh:2020nrd, Kanno:2020usf} (see also \cite{Danielson:2022tdw}).
However, as we saw in QED, for naive superposed states we always encounter decoherence because of the superselection rule.
To discuss relative phases between superposed states, these states should have the same asymptotic charges of gravity.
Such states must be dressed states as in QED.

%%%%%%%%%%%%%%%%%%%%%%%%%%%%%%%%%%%%%%%%%%%%%%%%%%%%%%%%%%%%%%
%%%%%%%%%%%%%%%%%%%%%%%%%%%
%%%%%%%%%%%%%%%%%%%%%%%%%%%%%%%%%%%%%%%%%%%%%%%%%%%%%%%%%%%%%%
%%%%%%%%%%%%%%%%%%%%%%%%%%%%%%%%%%%%%%%%%%%%%%%%%%%%%%%%%%%%%%
\section*{Acknowledgement}
We thank Hideo Furugori, Asuka Ito,  Yoshio Kikukawa, Toshifumi Noumi, Teruaki Suyama,  Hidetoshi Taya, Seiji Terashima, Junsei Tokuda, and Masahide Yamaguchi for the discussions.
SS also thanks the organizers and participants of the seminars at Kobe University, Yukawa Institute, Tokyo Institute of Technology, University of Tokyo (Komaba), and RIKEN (iTHEMS) for  opportunities to present and refine the results of this paper.
HH acknowledges support from JSPS KAKENHI Grant Number JP 21K13925.
SS acknowledges support from JSPS KAKENHI Grant Number JP 21K13927 and 22H05115.

%%%%%%%%%%%%%%%%%%%%%%%%%%%%%%%%%%%%%%%%%%%%%%%%%%%%%%%%%%%%%%
%%%%%%%%%%%%%%%%%%%%%%%%%%%%%%%%%%%%%%%%%%%%%%%%%%%%%%%%%%%%%%
\appendix
\section{Notation of transverse directions}
\label{app:transv}
We represent the two transverse polarization vectors by $\epsilon_A^\mu(\hat{k})=(0,\vec{\epsilon}_A(\hat{k}))$ ($A=1,2$) where $\vec{\epsilon}_A$ depends only on the direction of photon momentum $\vk$. 
The polarization vectors  satisfy the transverse condition $\vk \cdot \vec{\epsilon}_A(\hat{k})=0$. 
We also represent the inner product of two polarization vectors by $\eta^{T}_{AB}$
as $\eta^{T}_{AB}(\hat{k})=\eta_{\mu\nu}\epsilon_A^{\mu\ast}(\hat{k})\epsilon_B^{\nu}(\hat{k})$.
We assume that our basis is taken so that $\eta^{T}_{AB}$ is symmetric\footnote{For general complex basis, $\eta^{T}_{AB}$ is Hermitian but not needed to be symmetric.} as $\eta^{T}_{AB}=\eta^{T}_{BA}$ to avoid computational confusion due to non-symmetric $\eta^{T}_{AB}$.
We also suppose that $\eta^{T}_{AB}$ is non-negative definite.
For example, real orthogonal polarization bases and circular polarization bases give real symmetric non-negative definite $\eta^{T}_{AB}$.
The matrix $\eta^{T}_{AB}$ is invertible, and we represent the inverse matrix by $\eta^{AB}_{T}$. 
We have the completeness relation 
\begin{align}\label{eq:polcoml}
    \eta^{\mu\nu}
    =\eta_{T}^{AB}\epsilon_{A}^{\mu}(\hat{k})\epsilon_{B}^{\nu\ast}(\hat{k})
    -\frac{k^\mu \tilde{k}^\nu+\tilde{k}^\mu k^\nu}{2\omega^2},
\end{align}
where $\tilde{k}^\mu$ is the spatially reflected vector of $k^\mu$ as  $\tilde{k}^\mu=(\omega, -\vk)$ for $k^\mu=(\omega, \vk)$. Note that we have $k\cdot \tilde{k}=-2\omega^2$.

The mode expansion of the gauge field (in the interaction picture) is given by
\begin{align}
    A_\mu(x)=\int\widetilde{d^3k}\,\left[a_\mu(\vk) e^{ik \cdot x}+a_\mu^\dagger(\vk) e^{-ik \cdot x}\right].
\end{align}
The ladder operators satisfy the following commutation relation
\begin{align}
    [a_\mu(\vk), a_\nu^\dagger(\vk')]=(2\omega)(2\pi)^3 \eta_{\mu\nu}\delta^3 (\vk-\vk').
\end{align}
The ladder operators of the transverse photons  are given by
\begin{align}
 a_A(\vk):= \epsilon_A^{\mu\ast} (\hat{k})  a_\mu(\vk),
 \quad   
 a_A^\dagger(\vk):= \epsilon_A^{\mu} (\hat{k})  a_\mu^\dagger(\vk).
\end{align}

In this paper, indices $A, B$ are lowered or raised by $\eta^{T}_{AB}$ or $\eta_{T}^{AB}$, \textit{e.g.}, 
$C_A=\eta^{T}_{AB} C^B$.\footnote{We have assumed that $\eta_{T}^{AB}$ is symmetric. 
For general complex basis $\epsilon^\mu_A$ of transverse polarization, $\eta_{T}^{AB}$ is not symmetric and we have to take care of the definition of the lower or raised indices as $C_A= C^B\eta^T_{BA},\, C^\ast_A=\eta^{T}_{AB}C^{B \ast}$ to ensure $C_A=(C_A^\ast)^\ast$, where note that $(\eta^T_{AB})^\ast=\eta^T_{BA}$.}
The inner product that we introduced in \eqref{def:inpr} is written as 
\begin{align}\label{def:inpr:app}
    \ins{C}{D}=\int^{\Lambda_s}_{\lambda}\widetilde{d^3k}\,\eta^{T}_{AB}(\vk)  C^A(\vk) D^B(\vk).
\end{align}

A choice of the transverse vectors is 
$\vec{\epsilon}_A=\frac{\partial\hat{k}}{\partial \Omega^A}$ where $\Omega^A$ are (arbitrary) coordinates of the unit two-sphere parameterized by unit vector $\hat{k}$, 
\textit{e.g.}, 
$\Omega^1=\theta, \Omega^2=\phi$ in the polar coordinate system. 
Then, $\eta^{T}
_{AB}$ are nothing but the metric components of the unit two-sphere in the  $\Omega$-coordinates. 

In our analysis, the soft part of $S$-matrix elements will be written in terms of the above inner products with a typical form $\ins{X^\ast}{Y}$.
We can confirm that this inner product $\ins{X^\ast}{Y}$ is independent of the polarization vectors $\epsilon_A^\mu$ as follows.
Indeed, transverse vectors $X_A, Y_A$ can be written by the transverse projection of four-vectors $X_\mu, Y_\mu$ as $X_A=\epsilon_A^\mu X_\mu$, $Y_A=\epsilon_A^\mu Y_\mu$. 
Then we have
\begin{align}
   \eta_{T}^{AB} X_A^\ast Y_B=\eta_{T}^{AB} \epsilon_A^{\mu\ast}\epsilon_B^\nu X^\ast_\mu\, Y_\nu
   = \left(\eta^{\mu\nu}
    +\frac{k^\mu \tilde{k}^\nu+\tilde{k}^\mu k^\nu}{2\omega^2}\right)X^\ast_\mu\, Y_\nu,
\end{align}
and it leads to 
\begin{align}
    \ins{X^\ast}{Y}=\int^{\Lambda_s}_{\lambda}\widetilde{d^3k}\,\left(\eta^{\mu\nu}
    +\frac{k^\mu \tilde{k}^\nu+\tilde{k}^\mu k^\nu}{2\omega^2}\right)X^\ast_\mu(\vk)\, Y_\nu(\vk).
\end{align}
Thus, $\ins{X^\ast}{Y}$ is independent of the polarization vectors $\epsilon_A^\mu$.
This equation also shows that $\ins{X^\ast}{Y}$ is invariant under longitudinal changes of $X^\ast_\mu$ or $Y_\mu$ like $\delta X^\ast_\mu \propto k^\mu$, $\delta Y_\mu \propto k^\mu$.

%%%%%%%%%%%%%
%%%%%%%%%%%%%
%%%%%%%%%%%%%
\section{Asymptotic current for massive fermions}\label{sec:ascurrent}
In this section we evaluate the electric current of massive Dirac fields for a large time region and derive the result \eqref{eq:asexcurt}.
The following analysis is a more detailed version of the analysis in \cite{Kulish:1970ut}, which includes an order estimation of a subleading correction.

The mode expansion of Dirac fields in the interaction picture is given by
\begin{align}
%&A^{I}_\mu(\vx)
%=\intkw A(t,\vk)e^{i\kx}
%=\intkw \left[a_\mu(\vk)e^{ikx} +a_\mu^{\dagger}(\vk) e^{-ikx}\right]\,,\\
&\psi(x)
=\sum_{s=\pm}\intp \left[b_s(\vp)u_s(\vp)e^{ipx}+d^{\dagger}_s(\vp)v_s(\vp)e^{-ipx}\right]\,,\\
&\overline{\psi}(x)
=\sum_{s=\pm}\intp \left[d_s(\vp)\overline{v}_s(\vp)e^{ipx}+b^{\dagger}(\vp)\overline{u}_s(\vp)e^{-ipx}\right]\,,
\end{align}
where the commutation relations are 
\begin{align}
    %& [\,a_\mu(\vk),a_\nu^{\dagger}(\vkp)\,]=(2\omega)\eta_{\mu\nu}(2\pi)^3  \delta^{(3)}(\vk-\vkp),\\
    & \{\,b_s(\vp),b_{s'}^{\dagger}(\vpp)\,\}=(2E_{p})\delta_{ss'}(2\pi)^3  \delta^{(3)}(\vp-\vpp),\\
    & \{\,d_s(\vp),d_{s'}^{\dagger}(\vpp)\,\}=(2E_{p})\delta_{ss'}(2\pi)^3  \delta^{(3)}(\vp-\vpp)\,.
\end{align}
In momentum space, the current density operator in the normal ordered form is given by 
\begin{align}\label{eq:cur_in_mom}
    &:j^{\mu}(t,\vk):
    =i\intx e^{-i\vk\vdot\vx}:\overline{\psi}(t,\vec{x})\gamma^{\mu}\psi(t,\vec{x}):\nn[0.5em]
    %&= \sum_{s,s'}\intx \intq \intp e^{-i\vk\vdot\vx}\non
 %&\quad\left[d_s(\vq)\overline{v}_s(\vq)e^{iqx}+b^{\dagger}(\vq)\overline{u}_s(\vq)e^{-iqx}\right]
 %\gamma^{\mu}\left[b_{s'}(\vp)u_{s'}(\vp)e^{ipx}+d^{\dagger}_{s'}(\vq)v_{s'}(\vp)e^{-ipx}\right]\non
  &=i\sum_{s,s'=\pm} \intp\frac{1}{2E_q}
 \bigg{(}b^{\dagger}_s(\vq)b_{s'}(\vp)\overline{u}_s(\vq)\gamma^{\mu}u_{s'}(\vp)e^{i(E_q-E_{p})t}\Big{|}_{\vq=\vp-\vk}\nn[0.5em]
   &\quad -d^{\dagger}_{s'}(\vp)d_s(\vq)\overline{v}_s(\vq)\gamma^{\mu}v_{s'}(\vp)e^{-i(E_q-E_{p})t}\Big{|}_{\vq=\vp+\vk}
   %\non
   %&\quad
   +d_s(\vq)b_{s'}(\vp)\overline{v}_{s}(\vq)\gamma^{\mu}u_{s'}(\vp)e^{-i(E_q+E_{p})t}\Big{|}_{\vq=-\vp+\vk}\nn[0.5em]
   &\quad +b^{\dagger}_s(\vq)d^{\dagger}_{s'}(\vp)\overline{u}_{s}(\vq)\gamma^{\mu}v_{s'}(\vp)e^{i(E_q+E_{p})t}\Big{|}_{\vq=-\vp-\vk}
\ \ \bigg{)}.
\end{align}
For large time $t$, stationary points of phases $e^{\pm i (E_q \pm E_p)t}$ dominate in the $\vp$-integral.
The ``stationary'' points of each phase function in varying $p$ are given by
\begin{align}
     &0=\frac{\partial}{\partial p^i}\left(E_{p}+E_{p\pm k}\right)=\frac{p^{i}}{E_{p}}+\frac{p^{i}\pm k^{i}}{E_{p\pm k}}\quad \Rightarrow \quad \vec{p}=\mp\frac{1}{2}\vec{k}\,,\label{eq:collsp}\\[1em]
      &0=\frac{\partial}{\partial p^i}\left(E_{p}-E_{p\pm k}\right)=\frac{p^{i}}{E_{p}}-\frac{p^{i}\pm k^{i}}{E_{p\pm k}}\quad \Rightarrow \quad %\left\{
      \begin{array}{l}
      \text{any}\  \vec{p}\ \  \text{for}\ \  \vec{k}=\vec{0}\\
      (\text{no solution for}\ \  \vec{k}\neq\vec{0}\,)\,.
      \label{eq:softpt}\end{array}
      %\right.
\end{align}
The first stationary point \eqref{eq:collsp} corresponds to the process in which a collinear   particle-antiparticle pair changes to a photon, or vice versa.
The second solutions \eqref{eq:softpt} are not stationary points, but rather a value of $\vec{k}$ where the phase vanishes.   This case corresponds to the emission or absorption of a soft photon. 
In these cases, the phases take the following values:
\begin{align}
\left.\left(E_{p}+E_{p\pm k}\right)\right|_{\vec{p}=\mp\frac{1}{2}\vec{k}}
    =\sqrt{4m^2+|\vec{k}|^2}\,,\quad
    \left.\left(E_{p}-E_{p\pm k}\right)\right|_{\vec{k}=\vec{0}}
    =0\,.
\end{align}
Since  the case \eqref{eq:collsp} is  the only set of stationary points, the last two terms in \eqref{eq:cur_in_mom} at the stationary points \eqref{eq:collsp} may dominate the integral for large $t$.
However we should separately evaluate the integral of the first terms in \eqref{eq:cur_in_mom} for  $\vec{k}\sim 0$ because the stationary phase approximation is not applicable for the  phases  for $\vec{k}\sim 0$ where the corresponding phases are not large.
More precisely, since the phases of the first two terms in \eqref{eq:cur_in_mom} can be expanded around $k=0$ as\footnote{The higher order term $\mathcal{O}({\vk}^2)$ is more precisely $E_{p}\times \mathcal{O}(\frac{{\vk}^2}{E_p^2},\frac{(\vp\vdot\vk)^2}{E_{p}^4})$.}
\begin{align}
    (E_{p}-E_{p\pm k})t=\left\{\pm \frac{\vp\cdot \vk}{E_{p}}+\mathcal{O}({\vk}^2)\right\}t\,,
\end{align}
the stationary approximation for the $\vp$-integral in \eqref{eq:cur_in_mom} is not valid if $|\vk t|$ is not large.
Therefore we cannot discard the contribution from the non-stationary points of $\vp$ when $\vec{k}\sim \vec{0}$. Rather
we evaluate the two cases, \eqref{eq:collsp} and \eqref{eq:softpt}, separately in the following.
\\[1em]
%%%%%%%%%%%%%%%%%%%%%%%%
{\bf Contributions from collinear emission/absorption stationary points ($\vec{p}=\mp\frac{1}{2}\vec{k}$)}\\
The Hessian matrix of the corresponding phases functions are given by
\begin{align}
    \left.\frac{\partial^2}{\partial p^i \partial p^j}(E_{p}+E_{p\pm k})\right|_{\vec{p}=\mp\frac{1}{2}\vec{k}}
    %=\left.\frac{2\delta_{ij}}{E_{p}}-\frac{p^ip^j+(p\pm k)^i(p\pm k)^{j}}{E_{p}^3}\right|_{\vec{p}=\mp\frac{1}{2}\vec{k}}
    %=\frac{2\delta_{ij}}{E_{p}}-\frac{2p^{i}p^{j}}{E_{p}^3}
    =\left.\frac{2}{E_p}\left(\delta_{ij}-\frac{p^ip^j}{E_{p}^2}\right)\right|_{\vec{p}=\mp\frac{1}{2}\vec{k}}
    =2H_{ij}(\vec{p})\Big{|}_{\vec{p}=\mp\frac{1}{2}\vec{k}}
\end{align}
where  $H_{ij}:=\frac{1}{E_p}\left(\delta_{ij}-\frac{p^ip^j}{E_{p}^2}\right)$ has eigenvalues $\frac{m^2}{E_{p}^3},\frac{1}{E_{p}}, \frac{1}{E_{p}}$.
Using the stationary phase approximation with
%\begin{align}
    $\det(H)=\left.\frac{m^2}{E_{p}^5}\right|_{\vec{p}=\mp\frac{1}{2}\vec{k}}
    =\frac{m^2}{E^5_{\vec{k}/2}}$
    where
    $ E_{\vec{k}/2}
    =\sqrt{m^2+\frac{\vec{k^2}}{4}}=\frac{1}{2}\sqrt{4m^2+\vec{k}^2}$,
    %=\frac{m^2}{(m^2+\frac{\vec{k}^2}{4})^{\frac{5}{2}}}=\frac{2^5m^2}{(4m^2+\vec{k}^2)^{\frac{5}{2}}}\,.
%\end{align}
one finds that the integration of the phase function in the last two terms in \eqref{eq:cur_in_mom} with a sufficiently  smooth arbitrary function $f(\vp)$ is evaluated as
\begin{align}
\int\!\frac{d^3p}{(2\pi)^3} e^{i(E_{p}+E_{p\pm k})t}f(\vp)
&= e^{i\sqrt{4m^2+\vec{k}^2}\,t}\int \frac{d^{3}p}{(2\pi)^3}e^{it H_{ij}p^{i}p^{j}}f(\mp \vk/2)
+\mathcal{O}(t^{-2})\nn[1em]
&=e^{i\sqrt{4m^2+\vec{k}^2}\,t+i\frac{3}{4}\pi}\frac{(E_{\vec{k}/2})^{\frac{5}{2}}}{8\pi^{\frac{3}{2}}m t^{\frac{3}{2}}}f(\mp \vk/2)+\mathcal{O}(t^{-2})\,.
%+\text{subleading}
\end{align}
Thus the stationary point approximation at $\vp=\mp\frac{1}{2}\vk$ gives the following result:
\begin{align}
 :j^{\mu}(t,\vk):\ 
 =\ &i\frac{(E_{ \vec{k}/2})^{\frac{1}{2}}}{2^5\pi^{\frac{3}{2}}m t^{\frac{3}{2}}}\bigg{(}
 d_s(\vk/2)b_{s'}(\vk/2)\overline{v}_{s}(\vk/2)\gamma^{\mu}u_{s'}(\vk/2)e^{-i\sqrt{4m^2+\vec{k}^2}\,t+i\frac{3}{4}\pi}\nn
 &+d^{\dagger}_s(-\vk/2)b^{\dagger}_{s'}(-\vk/2)\overline{u}_{s}(-\vk/2)\gamma^{\mu}v_{s'}(-\vk/2)e^{i\sqrt{4m^2+\vec{k}^2}\,t-i\frac{3}{4}\pi}\bigg{)} +\mathcal{O}(t^{-2})\,,\label{eq:jas_coll}
\end{align}
which falls off as $t^{-\frac{3}{2}}$ for $|t|\rightarrow \infty$.
\\[1em]
%%%%%%%%%%%%%%%%%%%%%%%%
%%%%%%%%%%%%%%%%%%%%%%%%
{\bf Contributions from soft emission/absorption  ($\vec{k}\sim 0$)}\\
Using the following spinor property:
\begin{align}
    \overline{u}_{s'}(\vec{p})\gamma^\mu u_s(\vec{p})
    =\overline{v}_{s'}(\vec{p})\gamma^\mu v_s(\vec{p})=-2ip^{\mu}\delta_{s',s}\,,
\end{align}
one finds as written in \cite{Kulish:1970ut} that the first two terms in \eqref{eq:cur_in_mom} are given as follows:
\begin{align}
    %:j^{\mu}(t,\vk):
    &\text{(First two terms in \eqref{eq:cur_in_mom})}\nn
    &=%\delta_{\mathrm{soft}}(\vk)
    \sum_{s,s'=\pm} \intp\frac{1}{2E_q}
 \bigg{(}
    b^{\dagger}_s(\vq)b_{s'}(\vp)\overline{u}_s(\vq)\gamma^{\mu}u_{s'}(\vp)e^{i(E_q-E_{p})t}\Big{|}_{\vq=\vp-\vk}\nn[0.5em]
   &\hspace{10em} 
   -d^{\dagger}_{s'}(\vp)d_s(\vq)\overline{v}_s(\vq)\gamma^{\mu}v_{s'}(\vp)e^{-i(E_q-E_{p})t}\Big{|}_{\vq=\vp+\vk}
\ \ \bigg{)}\nn[0.5em]
&=
\int\!d^3p\frac{p^{\mu}}{E_p}e^{-i\frac{\vec{p}\cdot\vec{k}}{E_p}t}\rho(p)+\text{(subleading)}\,,
%\,,\quad\text{for $\vec{k} 0$}
\label{eq:ascur_mom}
\end{align}
where $\rho(p)$ is the density operator defined by
\begin{align}
    \rho(p)=\frac{1}{2E_p(2\pi)^3}\sum_{s=\pm}\left(b^{\dagger}_s(\vq)b_{s}(\vp)-d^{\dagger}_s(\vq)d_{s}(\vp)\right)\,.
\end{align}
The subleading term in \eqref{eq:ascur_mom}, comes from higher order terms in the expansion of the integrand around $\vk=\vec{0}$, namely the soft expansion.
Since the approximation in \eqref{eq:ascur_mom} is valid  around $\vec{k}\sim \vec{0}$, we define 
\begin{align}
j^{\mu}_{\rmas}(t,\vk)    :=\int\!d^3p\frac{p^{\mu}}{E_p}e^{-i\frac{\vec{p}\cdot\vec{k}}{E_p}t}\rho(p)\,\widetilde{\delta}_{\rmsoft}(\vk)
\end{align}
where $\widetilde{\delta}_{\rmsoft}(\vk)$ is a function which gives one around $\vec{k}\sim \vec{0}$ and zero otherwise.
One natural choice of  ``$\vec{k}\sim \vec{0}$'' is  the region where $|\vk t|\lesssim 1$, but we do not precisely determine the region because the result of $S$-matrix will turn out to be independent of the detail of the region at least at the leading order of the large time and the soft approximation.

Comparing the results \eqref{eq:jas_coll} and \eqref{eq:ascur_mom}, given that $j_{\rmas}(t,\vk)$  gives non-zero contribution in the $S$-matrix calculation, we obtain the following result:
\begin{align}
    :j^{\mu}(t,k):\ \rightarrow\  j^{\mu}_{\rmas}(t,\vk)\quad (\,|t|\rightarrow \infty\,)\label{eq:asj_mom_lim}
\end{align}
where the contribution from $\vp=\mp\frac{1}{2}\vk$ has been dropped because \eqref{eq:jas_coll}
falls off by $t^{-\frac{3}{2}}$ as $|t|\rightarrow \infty$.
The Fourier transformation of \eqref{eq:asj_mom_lim} in the momentum variable is given by
\begin{align}
    :j^{\mu}(t,\vec{x}):&\rightarrow\  \int\!\frac{d^3k}{(2\pi)^3}e^{i\vk\cdot\vx}j^{\mu}_{\rmas}(t,\vk)\quad (\,|t|\rightarrow \infty\,)\nn
    &= \intpw\,\rho(\vp)\, \frac{p^\mu}{E_p}\, \delta^{(3)}_{\rmsoft}(\vx -\vp t/E_p) 
    =j^{\mu}_{\rmas}(t,\vx)\,,
\end{align}
where $\delta^{(3)}_{\rmsoft}(\vx)$ is defined as
\begin{align}
    \delta^{(3)}_{\rmsoft}(\vx):=\int\!\frac{d^3k}{(2\pi)^3}e^{i\vk\cdot\vx}\,\widetilde{\delta}_{\rmsoft}(\vk)
    =\int_{\vec{k}\sim \vec{0}}\frac{d^3k}{(2\pi)^3}e^{i\vk\cdot\vx}\,.
\end{align}
We thus have obtained \eqref{eq:asexcurt}.

We introduce the $i\epsilon$ prescription to the asymptotic current.
The KF dressing operator with $i\epsilon$ prescription given by \eqref{def:RKFi} can be traced back to the regularization to the asymptotic current as
\begin{align}
j^{\mu}_{\rmas}(t,\vk)    =\int\!d^3p\frac{p^{\mu}}{E_p}e^{-i\frac{\vec{p}\cdot\vec{k}-i\eta(t)\epsilon}{E_p}t}\rho(p)\,\widetilde{\delta}_{\rmsoft}(\vk)
=e^{-\epsilon| t|}\int\!d^3p\frac{p^{\mu}}{E_p}e^{-i\frac{\vec{p}\cdot\vec{k}}{E_p}t}\rho(p)\,\widetilde{\delta}_{\rmsoft}(\vk)\,,
\end{align}
which is equivalent to regularizing $j^{\mu}_{\rmas}(t,\vx)$ as
\begin{align}
\label{reg-current}
    j^{\mu}_{\rmas}(t,\vx)
    =e^{-\epsilon |t|}\intpw\,\rho(\vp)\, \frac{p^\mu}{E_p}\, \delta^{(3)}_{\rmsoft}(\vx -\vp t/E_p) \,.
\end{align}

%%%%%%%%%%%%%%%%%%
\subsection*{Kulish-Faddeev phases}
We evaluate the phase \eqref{eq:defphase},
\begin{align}
\label{KFphase-commu}
    \Phi_{KF}(t,t')&=\frac{i}{2}\int^{t}_{t'}dt_1 \int^{t_1}_{t'} dt_2 [V_{\rmas}(t_1),V_{\rmas}(t_2)]
\end{align}
using the regularized asymptotic current \eqref{reg-current}.

The commutator $[V_{\rmas}(t_1),V_{\rmas}(t_2)]$ takes the form
\begin{align}
   &[V_{\rmas}(t_1),V_{\rmas}(t_2)]
   \nn
   &=\int_{\vec{k}\sim\vec{0}}\widetilde{d^3k}\,\intpw\,\intqw \,\rho(\vp)\rho(\vq)\,\frac{p\cdot q}{E_p E_q}
    e^{-\epsilon (|t_1|+|t_2|)}\left[e^{i \frac{\vp \cdot \vk}{E_p}t_1-i \frac{\vq \cdot \vk}{E_q}t_2-i\omega (t_1-t_2)}-\mathrm(c.c.)\right]
   \nn
   &= \int_{\vec{k}\sim\vec{0}}\widetilde{d^3k}\,\intpw\,\intqw \,\rho(\vp)\rho(\vq)\,(v_p\cdot v_q)\,
   e^{-\epsilon (|t_1|+|t_2|)}\left[e^{iv_p \cdot k t_1-i v_q \cdot k t_2}-\mathrm(c.c.)\right],
\end{align}
where we have introduced four-velocity vectors  $v_p^\mu=p^\mu/E_p=(1,\vp/E_p)$ and $v_q^\mu=q^\mu/E_q$.

We are interested in $\Phi_{KF}(t_i, T_i)$ and $\Phi_{KF}(T_f, t_f)$ in the limit $T_i\to -\infty$, $T_f\to \infty$.
Because of the $i\epsilon$ prescription, the phases are simplified in the limit as
\begin{align}
    &\lim_{T_i \to -\infty}\Phi_{KF}(t_i, T_i)
    =\Phi(t_i),
    \qquad
    \lim_{T_f \to \infty}\Phi_{KF}(T_f, t_f)
    =-\Phi(t_f)
\end{align}
with
\begin{align}
    \Phi(t):=&\frac{i}{2}
    \int_{\vec{k}\sim\vec{0}}\widetilde{d^3k}\,\intpw\,\intqw\,\rho(\vp)\rho(\vq)\,(v_p \cdot v_q)\, \frac{e^{-\epsilon|t|}}{2} 
    \nn
    &\times\left[\frac{e^{i(v_p-v_q)\cdot k t}}{(v_{p}-v_{q})\cdot k+i\eta(t)\epsilon}\left(\frac{1}{v_p\cdot k+i \eta(t)\epsilon}
    +\frac{1}{v_q\cdot k-i \eta(t)\epsilon}\right)-(c.c.)
    \right],
    \label{phiKF-soft}
\end{align}
where $\eta(t):=\mathrm{sgn}(t)$.

Although it is difficult to precisely evaluate $k$-integral in \eqref{phiKF-soft} because of the restriction $\vk\sim 0$, we can find the qualitative behavior of the phase $\Phi(t)$ just by removing the restriction as the conventional evaluation of IR divergences (see \cite{Weinberg:1995mt}).
Then, the phase is given by 
\begin{align}
    \Phi(t)
   \sim& -\intpw\,\intqw \,\rho(\vp)\rho(\vq)\,(v_p\cdot v_q)\,
   \nn
   &\qquad\times\int^{t}_{-\infty}dt_1 \int^{t_1}_{-\infty} dt_2
      e^{-\epsilon (|t_1|+|t_2|)}
   \int\widetilde{d^3k}\,
   \sin\left[k_\mu(v_p^\mu t_1 -v_q^\mu t_2)  \right].
\end{align}
The computation of this integral is almost the same as in \cite{Kulish:1970ut}, and the result is as follows:
\begin{align}
    \Phi(t)=\frac{1}{8\pi}
    \intpw\,\intqw \,\rho(\vp)\rho(\vq)\,\frac{p\cdot q}{\sqrt{(p\cdot q)^2-m^4}}\Gamma(0,\epsilon |t|),
\end{align}
where $\Gamma(s,x)$ is the incomplete gamma function, $\Gamma(s,x)=\int^{\infty}_{x}dt\, t^{s-1} e^{-t}$.
Note that, if $\epsilon |t| \ll 1$, we have  $\Gamma(0,\epsilon |t|) :=-\gamma -\log(\epsilon |t|)+ \mathcal{O}(\epsilon |t|)$ where $\gamma$ is Euler's constant.\footnote{If we are interested in the infinite time $|t|\to \infty$, we should take the limit $|t|\to \infty$ with fixed $\epsilon$. Then, $\lim_{|t|\to \infty}\Phi(t)=0$.} 
Thus, if we consider finite $t$ with $\epsilon \to 0$, the phase $\Phi(t)$ diverges as $\log \epsilon$.
However, the phase always appears in the combination $e^{-\Phi(t_f)+\Phi(t_i)}$ in $S$-matrix elements.
In this combination, we can take $\epsilon \to 0$ and obtain
\begin{align}
    \lim_{\epsilon\to 0}[-\Phi(t_f)+\Phi(t_i)]=\frac{1}{8\pi}
    \intpw\,\intqw \,\rho(\vp)\rho(\vq)\,\frac{p\cdot q}{\sqrt{(p\cdot q)^2-m^4}} \log\left(\frac{|t_f|}{|t_i|}\right).
\end{align}

%%%%%%%%%%%%%%%%%%%%%%%%%%%
%%%%%%%%%%%%%%%%%%%%%%%%%%%
\section{IR factorization for general dresses}\label{app:Ssoft}
In this appendix, we provide a derivation of \eqref{eq:general_factor}, \textit{i.e.}, we show the following IR factorization 
\begin{align}
\tensor[_{0}]{\bra{\beta}}{}e^{-D_\beta-i\Theta_\beta}U(\infty,-\infty)e^{C_\alpha+i\Theta_\alpha}\ket{\alpha}_{0}
=S^{\mathrm{soft}}_{\beta,\alpha}(\lambda,\Lambda_s)
    S^{\mathrm{hard}}_{\beta, \alpha}(\Lambda_s)
    [1+(\text{subleading})]
\end{align}
with the soft factor 
$S^{\mathrm{soft}}_{\beta,\alpha}=e^{-N_{D_\beta,C_\alpha}}$ given by 
\begin{align}
\label{eq_app:gen_soft_fact}
&N_{D_\beta,C_\alpha}(\lambda,\Lambda_s)
   \nn[0.5em]
   &=\frac{1}{2}
   \lVert R_{\beta,\alpha}+D_{\beta}-C_{\alpha}\rVert^2_\mathrm{s}
   +i\Im\left[\ins{R^{\ast}_{\beta,\alpha}}{D_\beta+C_\alpha}
   +\ins{D^\ast_\beta}{C_\alpha}\right]
   +i\Theta_{\beta,\alpha}
   -i \Phi_{\beta,\alpha}\,.
\end{align}

First, if we do not have the dressings $C_\alpha, D_\beta$ in the $S$-matrix elements, we have the standard IR factorization (see e.g. \cite{Weinberg:1995mt}) as
\begin{align}
\tensor[_{0}]{\bra{\beta}}{}U(\infty,-\infty)\ket{\alpha}_{0}
=e^{-\frac{1}{2}
   \ins{R_{\beta, \alpha}}{R^{\ast}_{\beta, \alpha}}+i \Phi_{\beta,\alpha}}
    S^{\mathrm{hard}}_{\beta, \alpha}(\Lambda_s)
    [1+(\text{subleading})].
\end{align}
Here we briefly remind how the exponent, $\frac{1}{2}\ins{R_{\beta, \alpha}}{R^{\ast}_{\beta, \alpha}}-i \Phi_{\beta,\alpha}$, appears.
In any $S$-matrix for Fock states, the IR divergent part of loop corrections by soft photons can be exponentiated and the exponent is given by 
\begin{align}\label{eq:IRexponent}
   \frac{1}{2}\sum_{n,m \in \alpha,\beta}\int_{\lambda\le |\vk|\le\Lambda_s}\frac{d^4k}{(2\pi)^4} \frac{\eta_{n}\eta_{m}e_ne_m(p_n \cdot  p_m) }{(k^2-i\epsilon)(p_n\cdot k-i\eta_{n}\epsilon)) (p_m\cdot k-i\eta_{m}\epsilon))}\,,
\end{align}
where $\eta_{n}$ denotes a sign factor which takes $+1$ for outgoing particles in  $\beta$ and $-1$ for incoming particles in $\alpha$.
Performing the $k^{0}$-integral  in \eqref{eq:IRexponent}  gives $-\frac{1}{2}\ins{R_{\beta, \alpha}}{R_{\beta, \alpha}}+i \Phi_{\beta,\alpha}$ by closing the contour appropriately and picking up the poles (see \cite{Weinberg:1995mt} for details).

Next, by the Baker–Campbell–Hausdorff formula, 
the dressing factors $e^{C_\alpha}, e^{-D_\beta}$ can be written as
\begin{align}
\begin{split}
    e^{C_\alpha}
    &=e^{-\frac{1}{2}\int_\text{soft}\widetilde{d^3k} C_\alpha^A(\vk)C^\ast_{\alpha A}(\vk)}
     e^{ -\int_\text{soft}\widetilde{d^3k} 
   C^{A\ast}_\alpha(\vk) a_{A}^{\dagger}(\vk)}
     e^{ \int_\text{soft}\widetilde{d^3k} 
  C^A_\alpha(\vk)  a_{A}(\vk)},
  \\[1em]
    e^{-D_\beta}
    &=e^{-\frac{1}{2}\int_\text{soft}\widetilde{d^3k} D^A_\beta(\vk)D^\ast_{\beta A}(\vk)}
     e^{ \int_\text{soft}\widetilde{d^3k} 
   D^{A\ast}_\beta(\vk) a_{A}^{\dagger}(\vk)}
     e^{ -\int_\text{soft}\widetilde{d^3k} 
  D^A_\beta(\vk)  a_{A}(\vk)}.
  \end{split}
\end{align}
The dressing factors act on hard states as
\begin{align}
\label{eq:CDadded}
\begin{split}
   \tensor[_{0}]{\bra{\beta}}{}e^{-D_\beta}
   &=e^{-\frac{1}{2}\ins{D_\beta}{D_{\beta}^\ast}}
    \tensor[_{0}]{\bra{\beta}}{}e^{ -\int_\text{soft}\widetilde{d^3k} 
  D^A_\beta(\vk)  a_{A}(\vk)},
  \\[1em]
    e^{C_\alpha}\ket{\alpha}_0
   &=e^{-\frac{1}{2}\ins{C_\alpha}{C_{\alpha}^\ast}}
 e^{ -\int_\text{soft}\widetilde{d^3k} 
  C^{A\ast}_\alpha(\vk)  a_{A}^\dagger(\vk)}
  \ket{\alpha}_0.
\end{split}  
\end{align}
Thus, the dressing factors play the role of adding soft photons to the hard states with the IR factor $e^{-\frac{1}{2}\ins{C_\alpha}{C_{\alpha}^\ast}-\frac{1}{2}\ins{D_\beta}{D_{\beta}^\ast}}$. 
The added soft photons in \eqref{eq:CDadded} also give an additional leading IR factor 
$e^{\ins{R_{\beta,\alpha}}{C^\ast_\alpha}-\ins{R^{\ast}_{\beta,\alpha}}{D_\beta}}$ 
because the soft photon theorem tells us that the contraction between a soft photon and $U(\infty,-\infty)$ gives a soft factor $R_{\beta,\alpha}$.
The soft photons themselves also contract as
\begin{align}
   \tensor[_{0}]{\bra{\beta}}{}e^{ -\int_\text{soft}\widetilde{d^3k} 
  D^A_\beta(\vk)  a_{A}(\vk)} e^{ -\int_\text{soft}\widetilde{d^3k} 
  C^{A\ast}_\alpha(\vk)  a_{A}^\dagger(\vk)}
  \ket{\alpha}_0
  =e^{\ins{D_\beta}{C_\alpha^\ast}}
  \tensor[_{0}]{\braket{\beta}{\alpha}}{_0}.
\end{align}
Combining all of the IR factors, we find that the exponent of the soft factor is given by 
\begin{align}
\label{eq_app:gen_soft_fact2}
 N_{D_\beta,C_\alpha}(\lambda,\Lambda_s)
   %\nn
   &=\frac{1}{2}
   \left[
   \ins{R_{\beta, \alpha}}{R^{\ast}_{\beta, \alpha}}
   +\ins{C_\alpha}{C_{\alpha}^\ast}
   +\ins{D_\beta}{D_{\beta}^\ast}\right]\nn
   &\qquad-\ins{R_{\beta,\alpha}}{C^\ast_\alpha}
   +\ins{R^{\ast}_{\beta,\alpha}}{D_\beta}
   -\ins{D_\beta}{C_\alpha^\ast}
   +i\Theta_{\beta,\alpha}
   -i \Phi_{\beta,\alpha}\,.
\end{align}
We can easily check that \eqref{eq_app:gen_soft_fact} is equivalent to \eqref{eq_app:gen_soft_fact2}.

%%%%%%%%%%%%%%%%%%%%%%%%%%%%%%%%%%%%%%%%%%
\section{Complete basis in $\mathcal{H}_{Q}$}
\label{app:Completeness}

Here we consider the following states discussed in section \ref{sec:def-partial_trace},
\begin{align}\label{eq:basisinQ}
\ket{\beta}_\text{hard}\ket{\gamma;\beta}_\text{soft}:=\ket{\beta}_\text{hard}e^{D_\beta}\ket{\gamma}_\text{soft}, 
\end{align} 
where $\ket{\beta}$ represents a Fock state of hard particles, $D_\beta$ is an arbitrary dressing operator defined in \eqref{eq:gen_dressC}, and $\ket{\gamma}_\text{soft}$ represents a Fock state of soft photons.  
We argue that the set of states in the form of \eqref{eq:basisinQ} 
can be a complete set of physical states with a fixed asymptotic charge $Q$, i.e. $\mathcal{H}_Q$, if the dressing $D_\beta$ is appropriately chosen.

The space $\mathcal{H}_Q$ consists of the hard and soft parts. 
We suppose the hard part can be dealt with as the standard Fock space. Thus, the hard states are spanned by the Fock basis $\{\ket{\beta}_\text{hard}\}$. This Fock space $\mathcal{H}^{\text{hard}}$ corresponds to $\mathcal{H}^{A}$ in \eqref{eq:Htot}. 
The state $\ket{\beta}_\text{hard}$ has a hard asymptotic charge $Q_\text{hard}[\beta]$ determined by the hard information of $\beta$ (the electric charges and momenta in $\ket{\beta}$). 
The soft sector must have the soft asymptotic charge $Q_\text{soft}=Q-Q_\text{hard}[\beta]$ so that the total asymptotic charge agrees with the given $Q$. %It depends on the information of the hard state $\beta$.
Thus, the soft sector is not independent of the hard sector.
We need a dressing operator so that the soft sector has the soft asymptotic charge $Q_\text{soft}$ because the Fock states with finite soft particles do not carry the soft asymptotic charges (see ref.~\cite{Hirai:2019gio}). 
Let $e^{D[Q_\text{soft}]}$ be such a dressing operator that $e^{D[Q_\text{soft}]} \ket{0}_\text{soft}$ has the soft asymptotic charge $Q_\text{soft}$. 
Then, any finite excited states of soft photons $e^{D[Q_\text{soft}]} \ket{\gamma}_\text{soft}$ also have the same charge $Q_\text{soft}$ where $\ket{\gamma}_\text{soft}$ are arbitrary (finite) Fock states with soft photons. 
Thus, a set of soft states with charge $Q_\text{soft}$ is
\begin{align}
\label{Hsoft}
   \mathcal{H}_{Q_\text{soft}}= \left\{\,e^{D[Q_\text{soft}]} \ket{\gamma}_\text{soft}\,|\,\ket{\gamma}_\text{soft} \in \mathcal{H}_\text{soft}^\text{Fock}\,\right\}.
\end{align}
This space corresponds to what we wrote by $\mathcal{H}^{\overline{A}}_\beta$ in \eqref{eq:Htot} and $e^{D[Q_\text{soft}]} \ket{\gamma}_\text{soft}$ does to $\ket{\gamma;\beta}_{\bar{A}}$.
The degrees of freedom $\gamma$ is independent of $\beta$, and we take the trace over them when we say the trace over soft photons. 

We now explain that the above $\mathcal{H}_{Q_\text{soft}}$ with a fixed dress $e^{D[Q_\text{soft}]}$ is a complete set of the soft sector with the soft asymptotic charge $Q_\text{soft}$. 
There is an ambiguity in the choice of the dressing operator $e^{D[Q_\text{soft}]}$ because we can add the subleading corrections as $e^{D[Q_\text{soft}]+D^{sub}}$ which does not change the asymptotic charges.
However, such states $e^{D[Q_\text{soft}]+D^{sub}}\ket{\gamma}_\text{soft}$ can have nonzero overlaps with states in the above $\mathcal{H}_{Q_\text{soft}}$, and can be expressed in terms of them. 
Thus, we think that $\mathcal{H}_{Q_\text{soft}}$ is complete and it is enough to sum over the degrees of freedom $\gamma$ for the trace over the soft photons.

To summarize, the way to specify states is as follows. First, we choose a superselection sector by fixing $Q$. This is a choice of a vacuum in the infinitely degenerated vacua. This is almost the same as a choice of a vacuum in theories with spontaneous symmetry breaking.
Next, we specify the hard sector. Then, depending on the hard asymptotic charges $Q_\text{hard}$, the soft sector is determined so that the soft asymptotic charge is $Q_\text{soft}=Q-Q_\text{hard}$, and takes the form of $\mathcal{H}_{Q_\text{soft}}$ on \eqref{Hsoft}. Since, up to the dress $e^{D[Q_\text{soft}]}$, the soft states in  $\mathcal{H}_{Q_\text{soft}}$ are usual Fock states, the specification is the same as the standard one.

%%%%%%%%%%%%%%%%%%%%%%%%%%%%%%%%%%%%%%%%%%
\section{Some computations of IR phase factors}
\label{app:IRphase}

%\paragraph{Ambiguity of the phase of transition probability}\ \\[0.5em]
\subsection*{Derivation of \eqref{eq:ImNalbegChung}}%\label{subsec:subphase}
We compute each term in \eqref{eq:INalbe} for the generalized Chung dress \eqref{eq:gChung}, i.e., 
\begin{align}\label{app:INalbe}
   \Im\left[N_{\alpha,\alpha'}^{\beta,\beta'}\right]
    &= \Im\left[N_{\bD_\beta,\bC_\alpha}+N^\ast_{\bD_{\beta'},\bC_{\alpha'}}
    -\ins{\Delta_{\bD_\beta,\bC_\alpha}^{\ast}}
   {\Delta_{\bD_{\beta'},\bC_{\alpha'}}}\right]\,.
\end{align}

The first term, $\Im\left[N_{\bD_\beta,\bC_\alpha}\right]$, can be written as follows:
 \begin{align}   
   %\begin{split}
   &\Im\left[N_{\bD_\beta,\bC_\alpha}\right]\nn
   &=\Im\left[\ins{R^{\ast}_{\beta}}{\widetilde{D}_{\beta}+D}-\ins{R^{\ast}_{\alpha}}{\widetilde{C}_{\alpha}+C}+\ins{\widetilde{D}^{\ast}_{\beta}+D^\ast}{\widetilde{C}_{\alpha}+C}\right]
     +\Theta_{\beta,\alpha}-\Phi_{\beta,\alpha}\nn
     %&=\Im\left[\ins{R^{\ast}_{\beta}}{\widetilde{D}_{\beta}+D}-\ins{\widetilde{D}^{\ast}_{\beta}}{C}-\ins{R^{\ast}_{\alpha}}{\widetilde{C}_{\alpha}+C}-\ins{D^{\ast}}{\widetilde{C}_{\alpha}}-\ins{\widetilde{D}^{\ast}_{\beta}}{\widetilde{C}_{\alpha}}+\ins{D^{\ast}}{C}\right]\nn
     %&\qquad+\Theta_{\beta,\alpha}-\Phi_{\beta,\alpha}\nn
     &=P^{\rmout}_\beta-P^{\rmin}_\alpha+\Theta^{\rmout}_{\beta}-\Theta^{\rmin}_{\alpha}+\Im\left[\ins{\widetilde{D}^{\ast}_{\beta}}{\widetilde{C}_{\alpha}}+\ins{D^{\ast}}{C}\right]\,, \label{eq:ImNfac}
\end{align}
where we have defined
\begin{align}
\begin{split}
    &P^{\rmout}_\beta=\Im\left[\ins{R^{\ast}_{\beta}}{\widetilde{D}_{\beta}+D}+\ins{\widetilde{D}^{\ast}_{\beta}}{C}\right]-\Phi_{\beta}\,,\\[1em]
    &P^{\rmin}_\alpha=\Im\left[\ins{R^{\ast}_{\alpha}}{\widetilde{C}_{\alpha}+C}+\ins{\widetilde{C}^{\ast}_{\alpha}}{D}\right]-\Phi_{\alpha}\,.
   \label{eq:fac_f}
   \end{split}
\end{align}
Then $\Im\left[N^{\ast}_{\bD_{\beta'},\bC_{\alpha'}}\right]$ is given by flipping the sign of each term in \eqref{eq:ImNfac} and replacing  $\alpha, \beta$ with $\alpha', \beta'$. 
The final term in \eqref{app:INalbe} 
 is given by \eqref{eq:ImdeldelgChung} and it can be expressed as 
\begin{align}
   \Im\left[\ins{\Delta_{\bD_\beta,\bC_\alpha}^{\ast}}
   {\Delta_{\bD_{\beta'},\bC_{\alpha'}}}\right]
    =&\Im\left[\ins{\wD^{\ast}_{\beta}}{\wD_{\beta'}}+\ins{\wC^{\ast}_{\alpha}}{\wC_{\alpha'}}-\ins{\wD^{\ast}_{\beta}}{\wC_{\alpha'}}-\ins{\wC^{\ast}_{\alpha}}{\wD_{\beta'}}\right]\nn[0.5em]
    %&\qquad
    &\qquad+\widetilde{P}^{\rmout}_\beta-\widetilde{P}^{\rmout}_{\beta'}-\widetilde{P}^{\rmin}_\alpha+\widetilde{P}^{\rmin}_{\alpha'}
    \label{eq:Imdel_fac}
\end{align}
where we have defined
\begin{align}
    \widetilde{P}^{\rmout}_\beta=\Im\left[\ins{\wD^{\ast}_{\beta}}{D-C}\right]\quad,\quad
    \widetilde{P}^{\rmin}_\alpha=\Im\left[\ins{\wC^{\ast}_{\alpha}}{D-C}\right]\,.
    \label{eq:delfac_f}
\end{align}
Using \eqref{eq:ImNfac} and \eqref{eq:Imdel_fac}, we can write \eqref{app:INalbe} as
 \begin{align}
 \Im\left[N_{\alpha,\alpha'}^{\beta,\beta'}\right]
    &= \Im\left[N_{\bD_\beta,\bC_\alpha}+N^\ast_{\bD_{\beta'},\bC_{\alpha'}}
    -\ins{\Delta_{\bD_\beta,\bC_\alpha}^{\ast}}
   {\Delta_{\bD_{\beta'},\bC_{\alpha'}}}\right]\\[1em]
  &= \Im\left[\bN_{\alpha,\alpha'}^{\beta,\beta'}\right]
  +\left(P^{\rmout}_\beta-\widetilde{P}^{\rmout}_\beta+\Theta^{\rmout}_{\beta}\right)
  -\left(P^{\rmin}_\alpha-\widetilde{P}^{\rmin}_\alpha+\Theta^{\rmin}_{\alpha}\right)\nn[1em]
  &\hspace{6em}-\left(P^{\rmout}_{\beta'}-\widetilde{P}^{\rmout}_{\beta'}+\Theta^{\rmout}_{\beta'}\right)
  +\left(P^{\rmin}_{\alpha'}-\widetilde{P}^{\rmin}_{\alpha'}+\Theta^{\rmin}_{\alpha'}\right),
  %+P^{\rmout\ast}_{\beta'}-P^{\rmin\ast}_{\alpha'}-\Theta^{\rmout}_{\beta'}+\Theta^{\rmin}_{\alpha'}
  %+\widetilde{P}^{\rmout\ast}_{\beta'}-\widetilde{P}^{\rmin\ast}_{\alpha'}
  %+P^{\rmout}_\beta-P^{\rmin}_\alpha+\Theta^{\rmout}_{\beta}-\Theta^{\rmin}_{\alpha}
  %+P^{\rmout\ast}_{\beta'}-P^{\rmin\ast}_{\alpha'}-\Theta^{\rmout}_{\beta'}+\Theta^{\rmin}_{\alpha'}
  %+\widetilde{P}^{\rmout}_\beta+\widetilde{P}^{\rmout\ast}_{\beta'}-\widetilde{P}_{\rmin}(\alpha)-\widetilde{P}^{\rmin\ast}_{\alpha'}
\end{align}
where $\Im\left[\bN_{\alpha,\alpha'}^{\beta,\beta'}\right]$ is defined in \eqref{eq:ImNalbegChung}.

The terms involving $C$ or $D$ in \eqref{eq:fac_f} and \eqref{eq:delfac_f} can be IR divergent because $C,D$ are supposed to be singular functions of $k$ as $k\rightarrow 0$  (otherwise they can be absorbed into $\widetilde{C}_{\alpha}$ and $\widetilde{D}_{\beta}$).
On the other hand, $\Theta_{\psi}\ (\psi=\alpha,\beta)$ is a pure phase of Fock basis $\ket{\psi}_{0}$, introduced in \eqref{eq:dressdstates}.
We can set $\Theta^{\rmin}_{\psi}, \Theta^{\rmout}_{\psi}$ as any function of $\psi$ and the choice does not affect any physical results.
Therefore the IR divergent terms in the phase \eqref{app:INalbe} do not cause any problem if they can be canceled by appropriate $\Theta_{\psi}$.
By choosing $\Theta^{\rmout}_{\beta}, \Theta^{\rmin}_{\alpha}$ as
\begin{align}
\begin{split}\label{eq:Phasechoice}
    &\Theta^{\rmout}_{\beta}=\widetilde{P}^{\rmout}_\beta-P^{\rmout}_\beta
    =\Im\left[\ins{\wD^{\ast}_{\beta}}{D-2C}-\ins{R^{\ast}_{\beta}}{\widetilde{D}_{\beta}+D}\right]+\Phi_{\beta}\,,\\[1em]
    &\Theta^{\rmin}_{\alpha}=\widetilde{P}^{\rmin}_\alpha-P^{\rmin}_\alpha
    =-\Im\left[\ins{\wC^{\ast}_{\alpha}}{C}+\ins{R^{\ast}_{\alpha}}{\widetilde{C}_{\alpha}+C}\right]+\Phi_{\alpha}\,,
    \end{split}
\end{align}
we obtain $\Im\left[N_{\alpha,\alpha'}^{\beta,\beta'}\right]
  = \Im\left[\bN_{\alpha,\alpha'}^{\beta,\beta'}\right]$, which is \eqref{eq:ImNalbegChung}.

%%%%%%%%%%%%%%%%%%%%%%%%%%%%%%%%%%%%%%%%%%%%%%%%
\subsection*{Derivation of \eqref{eq:NCC'} }%\label{subsec:proof_NCC'}
Here we derive the final equality of \eqref{eq:NCC'}, namely $\bN_{\alpha,\alpha'}^{\beta,\beta'}\Big{|}_{\beta=\beta'}=N_{\bC_{\alpha'},\bC_\alpha}$, by showing  the agreement of the real and imaginary parts of both sides.
Here $\bC^{A}_{\alpha}, \bC^{A}_{\alpha^{\prime}}$ are the generalized Chung dress:
\begin{align}\label{eq:gchungCC}
\bC^{A}_{\alpha}=-R^{A}_{\alpha}+\widetilde{C}^{A}_{\alpha}+C^{A}\quad,\quad
    \bC^{A}_{\alpha^{\prime}}=-R^{A}_{\alpha^{\prime}}+\widetilde{C}^{A}_{\alpha^{\prime}}+C^{A}.
\end{align}
$N_{\bC_{\alpha'},\bC_{\alpha}}$ is given by \eqref{eq:gen_soft_fact}
with the insertion $C_\alpha \to \bC_{\alpha}$ and $D_\beta \to \bC_{\alpha'}$. 
Then the real part of $N_{\bC_{\alpha'},\bC_{\alpha}}$ is given by
\begin{align}
    \Re(N_{C_{\alpha'},C_\alpha})
    =\frac{1}{2}\lVert \wC_{\alpha'}-\wC_\alpha\rVert^2_\mathrm{s}
    =\Re\left[\bN_{\alpha,\alpha'}^{\beta,\beta'}\right]\Big{|}_{\beta=\beta'}
\end{align}
where the final equality follows from the first equality in \eqref{eq:NCC'}.

The imaginary part of $N_{\bC_{\alpha'},\bC_{\alpha}}$  is given by \eqref{eq:ImNfac} with the replacement $\bD_\beta \to \bC_{\alpha'}$ as
\begin{align}
     \Im\left[N_{C_{\alpha'},C_\alpha}\right]
   =&\Im\left[\ins{R^{\ast}_{\alpha'}}{\widetilde{C}_{\alpha'}+C}-\ins{R^{\ast}_{\alpha}}{\widetilde{C}_{\alpha}+C}+\ins{\widetilde{C}^{\ast}_{\alpha'}+C^\ast}{\widetilde{C}_{\alpha}+C}\right]\nn &\quad
   %\\ &\hspace{7em}\quad
     +\Theta_{\alpha',\alpha}-\Phi_{\alpha',\alpha}
     \label{eq:ImC'C_ch}
\end{align}
where $\Theta_{\alpha',\alpha}$ is given by \eqref{eq:Phasechoice} with the same replacement as
\begin{align}
    \Theta_{\alpha',\alpha}
    &=\Theta^{\rmin}_{\alpha'}
    -\Theta^{\rmin}_{\alpha}\nn[1em]
    &=\Im\left[\ins{\wC^{\ast}_{\alpha'}}{-C}-\ins{R^{\ast}_{\alpha'}}{\widetilde{C}_{\alpha'}+C}+\ins{\wC^{\ast}_{\alpha}}{C}+\ins{R^{\ast}_{\alpha}}{\widetilde{C}_{\alpha}+C}\right]+\Phi_{\alpha',\alpha}.
    \label{eq:PC'C}
\end{align}
  Plugging \eqref{eq:PC'C} into \eqref{eq:ImC'C_ch}, we obtain
  \begin{align}
      \Im\left[N_{\bC_{\alpha'},\bC_\alpha}\right]=\Im[\ins{\wC^{\ast}_{\alpha'}}{\wC_\alpha}]
      %=-\Im[\ins{\wC^{\ast}_{\alpha}}{\wC_{\alpha'}}]
      =\Im\left[\bN_{\alpha,\alpha'}^{\beta,\beta'}\right]\Big{|}_{\beta=\beta'}\,,
  \end{align}
 where the final equality follows from the first equality in \eqref{eq:NCC'}.
  
\subsection*{Derivation of \eqref{eq:fin_NDD'2}}%\label{sec:proof_NDD'}
 Here we derive the equality of \eqref{eq:fin_NDD'2} by showing  the agreement of the real part and imaginary part of both sides.
$N_{\bD_{\beta'}, \bD_{\beta}}$ with the generalized Chung dress
\begin{align}\label{eq:gchungDD}
\bD^{A}_{\beta}=-R^{A}_{\beta}+\widetilde{D}^{A}_{\beta}+D^{A}\quad,\quad
    \bD^{A}_{\beta^{\prime}}=-R^{A}_{\beta^{\prime}}+\widetilde{D}^{A}_{\beta^{\prime}}+D^{A}
\end{align} 
is computed similar to $N_{\bC_{\alpha'},\bC_\alpha}$ in the above.
The real part is 
  \begin{align}
    \Re(N_{\bD_{\beta'},\bD_{\beta}})
    =\frac{1}{2}\lVert \wD_{\beta'}-\wD_{\beta}\rVert^2_\mathrm{s}
    =\Re\left[\bN_{\alpha,\alpha'}^{\beta,\beta'}\right]\Big{|}_{\alpha=\alpha'}\,,
\end{align}
where the final equality follows from \eqref{eq:fin_NDD'1}.
The imaginary part is 
\begin{align}
     &\Im\left[N_{\bD_{\beta'},\bD_{\beta}}\right]\nn[0.5em]
    &=\Im\left[\ins{R^{\ast}_{\beta'}}{\widetilde{D}_{\beta'}+D}-\ins{R^{\ast}_{\beta}}{\widetilde{D}_{\beta}+D}+\ins{\widetilde{D}^{\ast}_{\beta'}+D^\ast}{\widetilde{D}_{\beta}+D}\right]
   %\\ &\hspace{7em}\quad
     +\Theta_{\beta',\beta}-\Phi_{\beta',\beta},
     \label{eq:ImDD'_ch}
\end{align}
where
\begin{align}
    \Theta_{\beta',\beta}
    &=\Theta^{\rmout}_{\beta'}-\Theta^{\rmout}_{\beta}\nn[1em]
    &=\Im\left[\ins{\wD^{\ast}_{\beta'}}{-D}-\ins{R^{\ast}_{\beta'}}{\widetilde{D}_{\beta'}+D}+\ins{\wD^{\ast}_{\beta}}{D}+\ins{R^{\ast}_{\beta}}{\widetilde{D}_{\beta}+D}\right]+\Phi_{\beta',\beta}.
    \label{eq:PDD'}
\end{align}
  Plugging \eqref{eq:PDD'} into \eqref{eq:ImDD'_ch}, we obtain
  \begin{align}
      \Im\left[N_{\bD_{\beta'},\bD_{\beta}}\right]
      =\Im[\ins{\wD^{\ast}_{\beta'}}{\wD_\beta}]\,.
  \end{align}
We then find 
 \begin{align}
     \bN_{\alpha,\alpha'}^{\beta,\beta'}\Big{|}_{\alpha=\alpha'}
     =N_{\bD_{\beta'},\bD_{\beta}}
     +2\Im\left[\ins{\widetilde{D}_{\beta'}-\wD_{\beta}}{\widetilde{C}^{\ast}_{\alpha}}\right]
 \end{align}
 where the final equality follows from  \eqref{eq:fin_NDD'1}.
    
%%%%%%%%%%%%%%%%%%%%%%%%%%%%%%%%%%%%%%%%%%%%%%%%%%%%%%%%%%%%%%

%%%%%%%%%%%%%%%%%%%%%%%%%%%%%%%%%%%%%%%%%%%%%%%%%%%%%%%%%%%%%%
\bibliographystyle{utphys}
\bibliography{ref_dress}
\end{document}